\documentclass[12pt]{article}
\usepackage[margin=1in]{geometry}
\usepackage{amsthm}
\usepackage{amsmath}
\usepackage{amssymb}
\usepackage[hidelinks]{hyperref}
\hypersetup{
    colorlinks=true,
    linkcolor=blue,
    citecolor=blue
}

\usepackage{graphicx, caption, subcaption}
\usepackage{tikz}
\usepackage{setspace}
\usepackage{multirow}
\usepackage[capposition=top]{floatrow}
\usepackage{epigraph}
\usepackage{eufrak}
\usepackage{verbatim}
\usepackage{etoolbox}
\usepackage{epstopdf}
{

  }
\usepackage{tablefootnote}

\newcommand{\thetav}{\mbox{\boldmath$\theta $}}

\newcommand{\Lambdav}{\mbox{\boldmath$\Lambda $}}
\newcommand{\Gammav}{\mbox{\boldmath$\Gamma $}}

\newcommand{\muxz}{a}
\newcommand{\muzy}{b}

\usepackage[
backend=bibtex,natbib=true,
style=numeric,
citestyle=bwl-FU%authoryear
]{biblatex}
\newtheorem{defn}{Definition}

\newtheorem{theorem}{Theorem}

\usepackage{bbm}
\newtheorem{assumption}{Assumption}

\newtheorem{Algorithm}{Algorithm}

\input{tcilatex}

\begin{document}
\title{\vspace{-2cm}Matching Function Equilibria with Partial Assignment: Existence, Uniqueness and Estimation}
\author{\vspace{-0.5cm}Liang Chen\thanks{School of Economics, Zhejiang University. Email: \tt liang.chen@zju.edu.cn}
\and Eugene Choo\thanks{Division of Social Science, Yale-NUS College. Email: \tt eugene.choo@yale-nus.edu.sg}
\and Alfred Galichon\thanks{New York University and Courant Institute. Email: \tt ag133@nyu.edu or galichon@cims.nyu.edu. Support from ERC CoG-866274 EQUIPRICE is acknowledged}
\and Simon Weber\thanks{University of York. Email: \tt simon.weber@york.ac.uk}}
\maketitle
\vspace{-0.5cm}
\begin{abstract}

%%% LIMIT OF 150 WORDS

We argue that models coming from a variety of fields, such as matching models and discrete choice models among others, share a common structure that we call {\it matching function equilibria with partial assignment}. This structure includes an aggregate matching function and a system of nonlinear equations. We provide a proof of existence and uniqueness of an equilibrium and propose an efficient algorithm to compute it.  For a subclass of matching models, we also develop a new parameter-free approach for constructing the counterfactual matching equilibrium.  It has the advantage of not requiring parametric estimation when computing counterfactuals. We use our procedure to analyze the impact of the elimination of the Social Security Student Benefit Program in 1982 on the marriage market in the United States. We estimate several candidate models from our general class of matching functions and select the best fitting model using information based criterion.


\end{abstract}

\clearpage
\newpage

\section{Introduction}

Social scientists and demographers have long been fascinated with the workings of the marriage market and the important roles it plays in issues such as migration, fertility, intergenerational transmission of wealth, among others.\footnote{
For example,  \cite{eika2019educational} found educational assortative mating accounts for a nonnegligible
part of the cross-sectional inequality in household income in countries, such as the United States, Denmark, Germany, the United Kingdom, and Norway.}
%In particular, how does the distribution of marriages between heterogeneous agents get affected by demographic changes in the number of available men and women, or some exogenous changes that affects the desirability of certain matches relative to remaining single.
Confronted with aggregate data about who marries whom\footnote{While there are some examples where preference ordering are directly solicited from marriage market participants, these cases are rare. See \cite{HitschHortacsuAriely2010} for example.}, demographers and economists have heavily relied on \emph{%
aggregate matching functions}. Such functions allow the researcher
to relate the number of matches between two partners of given
characteristics (such as age or education) to their respective supply of single individuals in the
population.\footnote{A standard matching function is the harmonic marriage
matching function used by demographers \parencite[see e.g.][]{QianPreston1993, Schoen1981}.}
One drawback of this earlier approach is the absence of behavioral foundation for these matching functions. Consequently, it is difficult to interpret the model parameters or think about estimation.

\cite{choo2006who} proposed an approach to estimate the aggregate matching surplus using an equilibrium transferable utility model of marriage where agents have unobserved and heterogeneous taste for partners of known types. Using only aggregate data, their approach rationalizes a matching function where preference parameters are primitives of the behavioral model. Unlike some matching markets where preference data may be directly collected, it is not immediately clear how data on final matches could be used to identify preferences. The approach borrows many ideas from the structural industrial organization literature where estimating model primitives from equilibrium models using limited aggregate data is a norm. These identified primitives are important inputs when considering counterfactual experiments such as how the distribution of marriage responds to exogenous demographic shocks or government interventions. \cite{dupuy2014personality}, \cite{ChiapporiSalanieWeiss2017} and \cite{galichon2012}, among others, build on this and relax some of the assumptions in the \cite{choo2006who} model.
This paper unifies the themes in these papers and makes four contributions.
%The main contribution in this paper is three-fold.

%XXX
%First, we show that many recent contributions to the matching
%literature share a common structure with a behaviorally coherent aggregate matching function that relates the mass of matches to the mass of unmarried individuals in its
%center, and a system of nonlinear equations that balances available individuals for each type.
%We show that this system can be solved using an Iterative Projective Fitting Procedure (IPFP), and provide a general proof of existence and uniqueness of a solution under mild condition on the matching function.
%This result is important for policy analysis as it ensures a unique counterfactual equilibrium when considering the effects of policy changes.
%We argue that our approach is not anecdotal.
%It is striking to see that many
%settings in the matching literature can be characterized by an aggregate matching function and a system of nonlinear equations.\footnote{
%This includes the search and matching model of \textcite{shimer2000},
%Non Transferable Utility models of \textcite{Dagsvik2000} and \textcite{menzel2015large},
%Transferable Utility models of \textcite{choo2006who} and \textcite{Siow2008},
%Imperfectly Transferable Utility models of \textcite{GalichonKominersWeber2019},
%demand models of \textcite{BerryLevinsohnPakes1995},
%bilateral trade models of \textcite{HeadMayer2014}, and among others.}
%XXX

Our first contribution is to define a class of models, that we call \emph{matching function equilibrium models}, that share a common structure with, in its center, a behaviorally coherent aggregate matching function and a system of nonlinear equations that balances available individuals for each type. It is striking that our characterization captures a large variety of settings in the literature, including transferable,  non-transferable, imperfectly transferable utility, static and dynamic models, among others.\footnote{
This includes partial assignment models such as
Non Transferable Utility models of \textcite{Dagsvik2000} and \textcite{menzel2015large},
 Static Transferable Utility models of \textcite{choo2006who},
Dynamic Transferable Utility model of \cite{choo2015dynamic},
Imperfectly Transferable Utility models of \textcite{GalichonKominersWeber2019},
and models with full assignment such as
differentiated goods demand models of \textcite{BerryLevinsohnPakes1995},
bilateral trade models of \textcite{HeadMayer2014}, non additive random utility models, models of the labor market with full assignment, among others.
}
In this paper, we focus on models with partial assignment where some agents can be  unmatched in equilibrium.\footnote{
In a companion paper, \textcite{ChenChooGalichonEtAl2020}, we focus on a competitive equilibrium with substitutes,
and discuss as an application matching models with full assignment where all agents are matched in equilibrium.}
We provide general results on the proofs of equilibrium existence and uniqueness for the matching function equilibrium
models with partial assignment. Our results slightly extends those of \textcite{GalichonKominersWeber2019} and are closely related to the existence and uniqueness results found in \textcite{mourifie2019marriage}. These results are important for policy analysis as they ensure a unique counterfactual equilibrium when considering the effects of policy changes (see our third and fourth contributions below).
Moreover, we propose a new algorithm which provides
an efficient way of computing the unique equilibrium. 
This is quite useful in practice.

Our second contribution is in the estimation of structural parameters in the matching functions. We propose a general maximum likelihood strategy that estimates a number of candidate  matching models consistent with our  characterization of matching functions. The final estimates are obtained using a model selection criterion.
Existing methodology arbitrarily estimates the model of choice without entertaining the possibility that other models might be more consistent with the data generating process.
Our characterization of this general class of models allows us to consider a set of models
from which we select the one that best fit the data generating process.
We propose two computational approaches: (i) the nested approach; and (ii) the  Mathematical Program with Equilibrium Constraints (MPEC) approach.\footnote{We provide analytical expressions of the gradient of the log-likelihood for both approaches for efficient estimation.
Furthermore, we show that for models featuring a constant return to scale matching function, one can obtain close-form formulas for the confidence intervals of estimates.}
%{\color{red} Our characterization of this general class of models allows us to select a model, that best represents the data generating process, among a set of candidate models.}
%Our computational approaches uses  the  mathematical program with equilibrium constraints (MPEC) procedure.

%Finally, we conduct simulation experiments to illustrate the advantages and disadvantages of both approaches.
%The simulation results show that although the number of iterations is relatively similar across the two approaches, performing one iteration can be computationally more intensive in the MPEC approach. It is because the MPEC approach requires the calculation of the Jacobian matrix, which is computationally burdensome.

For our third contribution, we show that for a subclass of models, we can conduct counterfactual experiments without estimating the model parameters. We call this the
{\it parameter-free} approach.
All the papers in the literature have taken the standard approach where we first estimate the structural parameters followed by computation of the counterfactual equilibrium using the estimated parameters. We show that for a subclass of matching function equilibrium models whose matching functions are
(i) of the Cobb-Douglas form, and (ii) multiplicatively separable in parameters,
we can compute the counterfactual equilibrium without knowledge of the estimated parameters.
These two properties allow us to
reformulate these equilibrium matching models into a system of nonlinear equations, with ratios of counterfactual to observed equilibrium matching as unknowns. The new system of non-linear equations is free from structural parameters. We show that this new system of equations has a unique solution in terms of these ratios which allows us to compute the counterfactual equilibrium accordingly.
%While this parameter-free approach has limitations in terms of scope of models it applies to, it has the computational advantage of not requiring the estimation of the structural parameters in the matching functions.
Our new approach has the computational advantage of not requiring the estimation of the structural parameters in the matching functions.

Our final contribution is to apply our proposed estimation approaches
to analyze the impact of the elimination of the Social Security Student Benefit Program in 1982 on the marriage market.
%using the Transferable Utility model in \textcite{choo2006who} (CS hereafter).
In the United States, the Social Security Student Benefit Program established in 1965 provided financial aid for children of deceased, disabled or retired workers to attend college.\footnote{Recipients need to be not married and enrolled in college fulltime. The financial aid is provided up until the
semester the recipient turns  22 years of age.} The elimination of this benefit program
in 1982 is one of the largest policy change in college students' financial aid in the United States.
The number of college beneficiaries is estimated to have dropped from about 600,000 in 1981 to 66,000 in 1986.
The average total monthly payment fell substantially from about \$196 million in 1981 to \$26 million in 1986.
\textcite{dynarski2003} show that this policy change had a significant causal effect on reducing students' college attendance and completion among eligible students.

Our goal in this empirical application is to estimate the impact of the policy change on
the marriage distribution by computing the counterfactual marriage distribution in 1987/88 had the Social Security Student Benefit Program not been eliminated.
%by using our parametric and the parameter-free approaches.
Using causal effect estimates from the education literature, and our estimates of aid eligibility computed from the 1986 Current Population Survey, we first
construct a counterfactual distribution of available single men and women by age and education. We parametrically estimate the matching surpluses from the Exponential Transferable Utility (ETU)  matching model.
Special cases of this model include the Transferable Utility (TU) model of \textcite{choo2006who} (CS hereafter),
Non Transferable Utility (NTU) model of \textcite{GalichonHsieh2017},
and  the Harmonic Mean Matching function of \textcite{Schoen1981}.\footnote{ These ETU models are also a subset of Imperfectly Transferable Utility (ITU) models.}  We subsequently conduct a model selection using information based criterion  to pick the model that best fits the data.
Using the estimates in the selected model, we compute the marriage distributions in the observed scenario and counterfactual scenario had the Social Security Student Program not been eliminated.
We show that in the counterfactual scenario, there will be around 17,150 (3.00\%) fewer marriages among high school graduates, 8,154 (2.68\%) more marriages among college graduates and a negligible 11 (0.03\%) less marriages among above college graduates in 1987/88.
Interestingly, marriages between college educated men and high school educated women also increase by around 1,528 (2.04\%) while the marriages between college educated women and high school educated men would decrease.
This is probably due to the social norms of men preferring spouses who
are not more educated than themselves, as embedded in the preference parameters.

Our paper relates to the literature on matching functions used by demographers, e.g. \textcite{QianPreston1993, Schoen1981}. However, as we argue above, these matching functions lack behavioral micro-foundation. We look for such foundations in the two-sided matching literature, starting from the canonical models of matching with transferable utility, e.g. \textcite{KoopmansBeckmann1957, ShapleyShubik1971, Becker1973}, and non transferable utility, e.g. \textcite{GaleShapley1962}. In particular, we rely on recent advances on the structural estimation of these models, which exploits heterogeneity in tastes for identification purposes, e.g. \textcite{Dagsvik2000,menzel2015large,choo2006who,choo2015dynamic}. We argue that these models share a common structure, that revolves around an aggregate matching function and a system of nonlinear equations. These models, and their variants, provide us with a behaviorally coherent matching function, as pointed out in \textcite{choo2006who}, \textcite{choo2006estimating}, \textcite{ChiapporiSalanieWeiss2017} and \textcite{mourifie2019marriage}.

However, we also go beyond the TU and NTU cases. We argue in Section \ref{sec: examples} that this structure can be found in a variety of models. Therefore, our paper also relates to the literature on matching with imperfectly transferable utility, e.g. \textcite{Schoen1981}, \textcite{Qian1998}, \textcite{GalichonKominersWeber2019} and \textcite{GayleShephard2019}.
%, BLP style models of differentiated good demand and supply, e.g. \textcite{BonnetGalichonOHaraEtAl2018}, and trade models, e.g. \textcite{HeadMayer2014}.
Moreover, most existing matching functions in the literature restricts the equilibrium masses of matched agents to depend only on the equilibrium masses of unassigned own-type agents,
which suffers from the ``Independence of Irrelevant Alternatives (IIA)'' property \parencite{Gualdani2023-ds}.
In this paper, we investigate general matching functions which allow for 
the equilibrium masses of matched agents to depend on the equilibrium masses of unassigned agents of all types.
To the best of our knowledge, ours is the first paper to point out that the matching function structure is found in a surprisingly large number of models.

We provide the complete econometric toolbox to take these models to the data. In particular, we show how to estimate parametric versions of these models using a nested or MPEC approach. Thus our paper relates to the MPEC literature, e.g. \textcite{DubeFoxSu2012}, \textcite{SuJudd2012} and \textcite{PangSuLee2015}. We show how to conduct counterfactual experiments by using both the parametric and parameter-free approaches, and provide a number of computational techniques to ensure tractability.

Our empirical application contributes to the large body of literature on the impacts of financial aid programs on education.
Existing papers have primarily focused on the effects of aid on education attainments.
 For example, see
\textcite{manski1983college} and \textcite{kane1994college} on the Pell grant introduced in 1973, \textcite{reyes1997educational} on the Middle Income
Student Assistance Act, and  \textcite{angrist1993effect} on the War II G.I. Bills.
\textcite{dynarski2003} showed that the elimination of the Social Security Student Benefit Program has had a large significant causal effect on students' college attendance and completion.
It is well known that education is a primary attribute with which individuals match on in the marriage market. It has important implications on fertility and population growth, labor-force
participation, income inequality, etc \parencite[see e.g.][]{grossbard1988women,Becker1991,doepke2019bargaining,eika2019educational}.
However, there has been little research on how financial aid program affects the marriage distributions. Our paper contributes to the growing literature that uses structural techniques to evaluate the effect of policy changes on the marriage market \parencite[e.g.][]{adda2020there, schulz2020marriage}.

\noindent
\textbf{Organization of the paper.} Section \ref{sec:MFE} introduces and
characterizes matching function equilibrium models with partial assignment.
Section \ref{sec:MLE} outlines the
estimation by maximum likelihood and provides an analytic expression for the
gradient of the log-likelihood. We also provide the formulas for computing the confidence intervals of
the parameter estimates in models with homogeneous matching functions.
Section \ref{sec:counterfactuals} outlines the parameter-free approach for estimating the counterfactual equilibrium for a subclass of matching function equilibrium models.
Section \ref{sec:application} illustrates our approaches by investigating the impact of the elimination of the Social Security Student Benefit Program in 1982 on the marriage market in the United States. Section \ref{sec:con} concludes.
All proofs of our main results can be found in the Appendix. Additional results, numerical experiments, details about the data construction and additional figures are available in the Online Appendix.

%\bigskip
% \textbf{Notations}. In the following, we still use the marriage market as
% our main example and continue to use its particular terminology. Namely, we
% let $n_{x}$ be the mass of men of type $x$, and $m_{y}$ be the mass of women
% of type $y.$ The mass of matches between men of type $x$ and women of type $%
% y $ is the vector $(\mu _{xy})_{x\in \mathcal{X},y\in \mathcal{Y}}$, while
% the vectors $(\mu _{x0})_{x\in \mathcal{X}}$ and $(\mu _{0y})_{y\in \mathcal{%
% Y}}$ denote the mass of single men of type $x$ and single women of type $y$,
% respectively. The sets of types are denoted $\mathcal{X}$ for men and $%
% \mathcal{Y}$ for women, while $\mathcal{X}_{0}=\mathcal{X}\cup \mathcal{\{}0%
% \mathcal{\}}$ and $\mathcal{Y}_{0}=\mathcal{Y}\cup \mathcal{\{}0\mathcal{\}}$
% introduce singlehood as an option. We introduce the set of married pairs $%
% \mathcal{XY}=\mathcal{X\times Y}$ and the set of household types, $\mathcal{%
% XY}_{0}=\mathcal{X\times Y}\cup \mathcal{X\times }\left\{ 0\right\} \cup
% \left\{ 0\right\} \times \mathcal{Y}$. Finally, the number of households is $%
% N=\sum_{xy\in \mathcal{XY}_{0}}\mu _{xy}$. Note that our terminology can
% easily be adapted to fit any two-sided matching framework (e.g replacing men
% by workers and women by firms).

\section{Matching Function Equilibrium Models with Partial Assignment}\label{sec:MFE}

In this section, we show that many behavioural models in the matching
literatures can be
characterized by an aggregate matching function and a system of nonlinear
equations. We focus on situations in which we allow for unassigned agents, and provide a formal definition for this class of matching
models, that we label \emph{matching function equilibrium} models with partial assignment.\footnote{
In another paper,  \textcite{ChenChooGalichonEtAl2020}, we consider the full assignment case where all agents are matched
in equilibrium.}

We first provide a general proof of existence and uniqueness of the equilibrium in these models.
Consider a general marriage market where two populations of men
(indexed by $i\in I$)
and women
(indexed by $j\in J$)
meet and may form heterosexual pairs.\footnote{The setup
 and notations can be adapted to any two-sided one-to-one matching
markets, e.g., replacing men by workers and women by firms.}
We assume that men (resp. women) can be gathered in groups of
similar characteristics, or types, $x\in \mathcal{X}$ (resp. $y\in \mathcal{Y%
}$), with $|\mathcal{X}|$ (resp. $|\mathcal{Y}|$) denoting the number of
types for men (resp. women). The total mass of men of
type $x$ (resp. women of type $y$) is denoted $n_x$ (resp. $m_y$).
We define the sets, $\mathcal{X}_{0}=\mathcal{X}%
\cup \mathcal{\{}0\mathcal{\}},$ and $\mathcal{Y}_{0}=\mathcal{Y}\cup
\mathcal{\{}0\mathcal{\}},$ which include singlehood as an option. The set of married pairs is given by $\mathcal{XY}=\mathcal{X\times Y},$ and the set of household types by $\mathcal{XY}_{0}=\mathcal{X\times Y}\cup
\mathcal{X\times }\left\{ 0\right\} \cup \left\{ 0\right\} \times \mathcal{Y}
$. The mass of marriages between men of type $x$ and women of type $y$ is
denoted by $\mu _{xy}$, while the mass of single men of type $x$ and the mass
of single women of type $y$ are denoted by $\mu _{x0}$ and $\mu _{0y}$, respectively.

Let $\mu = \{\mu_{xy}\}_{xy \in \mathcal{XY}}$,
$\muxz = \{\mu_{x0}\}_{x \in \mathcal{X}}$,
and $\muzy = \{\mu_{0y}\}_{y \in \mathcal{Y}}$
denote the vectors of married and unmarried populations.
We can now define matching function equilibrium models with partial assignment below.

\subsection{Definition}

In this partial assignment case, we assume that $\mu _{xy}$ is a
deterministic function of the masses of unassigned agents of all types, denoted  by $a$ and $b$,
respectively.

\begin{defn}{\bf Matching function equilibrium model with partial
assignment}\label{def: AMF}\\ A matching function equilibrium model with partial
assignment determines the mass of
$(x,y) \in \mathcal{XY}$-type matches$,$ $\mu_{xy},$ and the mass vectors  of
unassigned agents of all types$,$ $\muxz=\{\mu_{x0}\}_{x \in \mathcal{X}}$ and $\muzy =\{\mu_{0y}\}_{y \in \mathcal{Y}}$,  by an aggregate matching
function which relates the former to the latter by
\begin{equation}
\mu _{xy}=M_{xy}(\muxz,\muzy),  \label{eq: matching_function_def}
\end{equation}%
where $\muxz$ and $\muzy$ are determined by a system of
nonlinear accounting equations$,$%
\begin{equation}
\QATOPD\{ . {n_{x}=\mu_{x0}+\sum_{y\in \mathcal{Y}}M_{xy}(\muxz,\muzy),~\forall x\in \mathcal{X}}
{m_{y}=\mu _{0y}+\sum_{x\in \mathcal{X}%
}M_{xy}(\muxz,\muzy),~\forall y\in \mathcal{Y}},\label{eq: nonlinear_system}
\end{equation}
in equilibrium.
\end{defn}

The matching equilibrium in these models is fully characterized by
the system of nonlinear equations \eqref{eq: nonlinear_system} with unknown
masses of unassigned agents, $(\muxz,\muzy)$.
The matching function defined in \eqref{eq: matching_function_def} describes the simultaneous relationship that exists in equilibrium between the masses of unassigned agents and the masses of matched agents.
It allows for the equilibrium masses of matched agents not only depend on the equilibrium masses of unassigned agents of own types, but also depend on the masses of unassigned agents of other types.

In the matching literature, researchers often utilize the logit specification on utility shocks to derive the matching functions, which restricts that the equilibrium masses of matches to depend only 
on the equilibrium masses of unassigned own-type agents (see examples a) to e) in section \ref{sec: examples}).
This restriction implies that the substitution patterns suffer from the implausible ``Independence of Irrelevant Alternatives (IIA)'' property, which may be potentially limit the scope of applications for these models in the literature.
Our framework also allows for the equilibrium masses of matches depend 
on the equilibrium masses of unassigned other-type agents,
which does not suffer from the IIA substitution restriction.
We now provide some examples to show that the defined matching function equilibrium
model encompasses many behavioural models in the matching
literature.

\subsection{Examples} \label{sec: examples}

\noindent\textbf{a) Models with Transferable Utility (TU).}\\
\textcite{choo2006who} develops a TU model where agents have
heterogenous preferences over their potential partners and face a
choice problem with random utility shocks.
It delivers a well-known aggregate matching function.
Let $\alpha _{xy}$ and $\gamma _{xy}$
represent pre-transfer utilities that a type $x$ man and $y$ woman receives from a $(x,y)$ type
 % \in \mathcal{XY}$
 match respectively.
The equilibrium payoff for
man $i$ of type $x,$ who marries with a type $y$ woman, is
$u_{iy}=\alpha_{xy}-\tau _{xy}+\epsilon _{iy}$.
It is the sum of a
systematic component ($U_{xy}\equiv \alpha _{xy}-\tau _{xy}$, where $\tau
_{xy}$ is the transfer from the type $x$ man to the type $y$ woman) and an idiosyncratic part ($\epsilon_{iy}$).
Similarly, the payoff for woman $j$ of type $y$ who marries
a type $x$ man is
$v_{xj}=\gamma _{xy}+\tau _{xy}+\eta _{xj}$,
where the systematic part is denoted
by $V_{xy} \equiv \gamma _{xy}+\tau _{xy}$.

Following \textcite{McFadden1974}, \textcite{choo2006who} assume the random utility shocks have Extreme Value Type I (logit) distributions.
The authors show that the equilibrium systematic payoffs can be recovered from the observed matching probabilities,
that is $U_{xy}=\log \frac{\mu _{xy}}{\mu
_{x0}}$ and $V_{xy}=\log \frac{\mu _{xy}}{\mu _{0y}}$. Let $\Phi _{xy}=\alpha _{xy}+\gamma _{xy}$ denote the total
systematic
surplus from a
$(x,y) $ % \in \mathcal{XY}$-type
match. It follows from Equation
(11) (on page 181) of that paper that
\begin{equation}
M_{xy}(a,b)=\sqrt{\mu _{x0}\mu _{0y}}\exp (\frac{\Phi _{xy}}{2}),
\label{MF:CS}
\end{equation}
in which the vectors of unassigned agents $(a,b)$ degenerates to $(\mu _{x0},\mu _{0y})$.
This matching function has several interesting properties: it is
homogeneous of degree 1, it has constant returns to scale, and the effect
of $\mu _{x0}$ and $\mu _{0y}$ is symmetric.\\

\noindent \textbf{b) Models with peer effects.}\\
\textcite{mourifie2021cobb} extend the \textcite{choo2006who} model
by introducing peer effects terms, $\psi _{x}\ln \mu _{xy}$ and
$\Psi _{y}\ln \mu _{xy}$ into the respective random utilities.
The utility of man $i$ of type $x,$ matched with a
type $y$ woman, is given by $\alpha _{xy}-\tau _{xy}+\psi _{x}\ln \mu
_{xy}+\epsilon _{iy}$,
and his utility from remaining unmatched is given by $\psi^0 _{x}\ln \mu
_{x0}+\epsilon _{i0}$.
The terms $\alpha _{xy},$ $\tau _{xy},$ and $\epsilon _{iy}$ are defined as in example a).
Similarly, the utility for woman $j$ of
type $y,$ matched with a man of type $x$ is, $\gamma _{xy}+\tau _{xy}+\Psi
_{y}\ln \mu _{xy}+\eta _{xj}$,
and her utility from being unmatched is given by $\Psi^0_{y} \ln \mu _{0y}+\eta _{0j}$.
Under logit distributional assumption, further computation yields
a Cobb-Douglas aggregate matching function,
\begin{equation}
M_{xy}(a,b)=\mu _{x0}^{k_{xy}} \mu _{0y}^{l_{xy}}\exp
\left( \frac{\Phi _{xy}}{2-\psi _{x}-\Psi _{y}}\right),
\label{MF:MS}
\end{equation}%
where $k_{xy}=\frac{1- \psi^0 _{x}}{2-\psi_x - \Psi_y}$ and $l_{xy}=\frac{1- \Psi^0 _{y}}{2-\psi_x - \Psi_y}$.

If we impose the restrictions $k_{xy}=k$, $l_{xy}=l$ and $k+l=1$, we recover the matching function of \textcite{ChiapporiSalanieWeiss2017}.
Note that in its most general formulation, this matching function does not
satisfy the homogeneity of degree 1 nor symmetry properties.\\

\noindent \textbf{c) Models with Non Transferable Utility (NTU).}\\
In the NTU framework, it is assumed that when a type $x$ man and type
$y$ woman match, they receive payoffs $\alpha _{xy}$ and $\gamma _{xy}$
respectively, and an idiosyncratic component which is assumed
to enter additively.
In this NTU setting,
\textcite{GalichonHsieh2017} shows that existence and computation of equilibria can be provided via an aggregate version of the \textcite{GaleShapley1962}
algorithm.
In the logit case, \textcite{GalichonKominersWeber2019} derive the matching function as
\begin{equation}
M_{xy}(a,b)=\min\{\mu_{x0} e^{\alpha_{xy}}, \mu_{0y} e^{\gamma_{xy}}\}.
\label{MF:Dag}
\end{equation}
This matching function is
homogeneous of degree 1 and features constant returns to scale.\footnote{Alternatively, \textcite{Dagsvik2000} and \textcite{menzel2015large}
provide a tractable expression for the aggregate matching function when the
number of market participants grows large. Equation (4.3) (on page 925) of \textcite{menzel2015large}  reads,
$
M_{xy}(\mu_{x0},\mu_{0y})=\mu _{x0}\mu _{0y}\exp (\alpha _{xy}+\gamma _{xy}),
$
which is homogeneous of degree 2 with increasing returns to scale.
}\\

\noindent \textbf{d) Models with Imperfectly Transferable Utility (ITU).}\\ Building on
this literature, \textcite{GalichonKominersWeber2019} (GKW hereafter) develops a general
framework for matching with ITU. This framework nests
both the TU and NTU settings\footnote{For a study of ITU settings using revealed preferences techniques instead of matching functions, see \textcite{CherchyeDemuynckDeRockEtAl2017}.}. The authors show that the equilibrium systematic utilities, $%
U_{xy}$ and $V_{xy}$, satisfy a key feasibility condition, $%
D_{xy}(U_{xy},V_{xy})=0$. The distance function $D_{xy}(\cdot)$ satisfies the property that
$D_{xy}(g+t,h+t)=t+D_{xy}(g,h)$.
Hence, when idiosyncratic taste shocks are
assumed to be logit \parencite[as in][]{choo2006who}, the feasibility condition
becomes $D_{xy}(\log \mu _{xy}-\log \mu _{x0},\log \mu _{xy}-\log \mu
_{0y})=0$, which combined with the property on $D_{xy}(\cdot)$, gives%
\begin{equation}
M_{xy}(a,b)=\exp \left( -D_{xy}(-\log \mu _{x0},-\log \mu
_{0y})\right).  \label{MF:GKW}
\end{equation}%
This matching function is also homogeneous of degree 1 but is not
necessarily symmetric in $\mu _{x0}$ and $\mu _{0y}$.\\

\noindent \textbf{e) Models with Exponentially Transferable Utility (ETU).}\\
\textcite{Schoen1981} introduces a matching function  based on the harmonic mean. Interestingly, the ITU-logit framework introduced in example d)
above allows us to recover a micro-founded version of that matching function. Assume that when a type $x$ man matches with a type $y$ woman, they bargain to split their income (which is normalized to $2$) into private consumption for the man and woman, denoted by $c_{xy}^a$ and  $c_{xy}^b$, respectively. Assume that the utility payoffs received by the man and woman are
$u_{xy}=\alpha_{xy}+\kappa_{xy}\log c_{xy}^a,$ and $v_{xy}=\gamma_{xy}+\kappa_{xy}\log c_{xy}^b$,
respectively. Assuming that the budget constraint is $c_{xy}^a + c_{xy}^b \leq 2$, one can verify that the distance function is given by
$D_{xy}(g,h)=\kappa_{xy}\log \left(\exp \left((g-\alpha_{xy})/ \kappa_{xy} \right) + \exp \left((h-\gamma_{xy}) / \kappa_{xy} \right)\right)/2$, 
hence by equation~\eqref{MF:GKW} we derive the following matching function,
\begin{equation}
M_{xy}(a,b)=\left[\frac{\exp \left(\frac{-\alpha_{xy}}{\kappa_{xy}}\right)\mu_{x0}^{-1/\kappa_{xy}}  +
\exp \left(\frac{-\gamma_{xy}}{\tau_{xy}}\right)\mu_{0y}^{-1/\kappa_{xy}}}{2} \right]^{-\kappa_{xy}}. \label{MF:Schoen}
\end{equation}%

We refer to the model that delivers the matching function in \eqref{MF:Schoen} as the Exponentially Transferable Utility (ETU) model.
Here, the parameter $\kappa_{xy}$, which is the elasticity of substitution between marital surplus and consumption,
captures the degree of utility transferability.
When $\kappa_{xy} \rightarrow 0$, we recover the NTU model with matching function given by \eqref{MF:Dag}.  As $\kappa_{xy} \rightarrow \infty$, we recover the TU model with matching function  given by \eqref{MF:CS}.
Moreover, when $\kappa_{xy} = 1$, we recover the harmonic mean matching function, as in Equation 1 (on page 281) in \textcite{Qian1998} (up to some multiplicative constants).
Hence, the ETU model is quite general and bridges the NTU and TU models.\\

\noindent \textbf{f) Dynamic Matching Models with TU.}\\
Extending the static TU CS model to a dynamic setting, \cite{choo2015dynamic} proposes an overlapping generations discrete time behavioural model of marriage. Single individuals can forgo matching today with the hope of finding a better match in the future.
It rationalizes a new marriage matching function where the marital surplus depends on the equilibrium numbers of unmatched individuals today and in the future.
The marital surplus parameter measure the discounted utility from being married today (till death of at least one spouse) relative to remaining single (for that same time span).
The number of unmatched individuals today and in the future captures the opportunity cost of committing to marriage and forgoing the opportunity to participate in the marriage market.
Under these settings, \cite{choo2015dynamic} derived a dynamic matching function, which takes the form,
\begin{align}
\label{eq: sec4_matchFunction_dynamicType2}
M_{xy}(\muxz,\muzy)=
\exp (\Pi_{xy}/2  )
\sqrt{n_xm_y}
 \prod_{k=0}^{T(x,y)-1}\bigg[\frac{\mu_{x+k0}}{n_{x+k}} \frac{\mu_{0y+k}}{m_{y+k}} \bigg]^{\frac{1}{2}[\beta(1-\delta)]^k},
\end{align}
where $\Pi_{xy}$ is the dynamic martial surplus, $T(x,y)$ is the marriage duration for a $(xy)$ match,
$\beta$ is the discount factor and $\delta$ is the divorce rate.
In this matching function, the equilibrium masses of matches depends on
the equilibrium masses of unmatched agents of all types.

\subsection{Existence and Uniqueness}

Under mild conditions on the matching function, there exists a unique equilibrium matching with partial assignment. These conditions are as follows:

\begin{assumption}
\label{ass: MFE} The aggregate matching function, $M_{xy}\left(\muxz,\muzy\right), $ satisfies the following five conditions$:$

$(i)$ $M_{xy}: \left(\muxz,\muzy\right) \mapsto M_{xy}\left(\muxz,\muzy\right) $ is continuous in all elements of $(\muxz,\muzy)$.

$(ii)$ $M_{xy}: \left(\muxz,\muzy\right) \mapsto M_{xy}\left(\muxz,\muzy\right) $ is differentiable.

$(iii)$ $M_{xy}: \left(\muxz,\muzy\right) \mapsto M_{xy}\left(\muxz,\muzy\right) $ is weakly increasing in $(\mu_{x0}, \mu_{0y})$ and weakly decreasing 
in $(\{\mu_{x^\prime 0}\}_{x^\prime \neq x}, \{\mu_{y^\prime 0}\}_{y^\prime \neq y})$.

$(iv)$ $ \vert \frac{\partial M_{xy}}{\partial \mu_{x0}} \vert \ge \vert \sum_{x^\prime \neq x} \frac{\partial M_{x^\prime y}}{\partial \mu_{x 0}} \vert$ 
and 
$ \vert \frac{\partial M_{xy}}{\partial \mu_{0y}} \vert \ge \vert \sum_{y^\prime \neq y} \frac{\partial M_{xy^\prime}}{\partial \mu_{0 y}} \vert$.

$(v)$ For all $\muxz>0$ and $\mu_{0y^\prime}>0$ with $y^\prime \neq y$ , $\lim_{\mu _{0y}\rightarrow 0^{+}}M_{xy}\left(
\muxz,\{\mu _{0y^\prime}\}_{y^\prime \neq y}, \mu_{0y}\right) =0;$ Similarly, for all $\muzy>0$
and $\mu_{0x^\prime}>0$ with $x^\prime \neq x$,
$\lim_{\mu _{x0}\rightarrow 0^{+}}M_{xy}\left( \mu _{x0}, \{\mu_{x^\prime 0}\}_{x^\prime \neq x}, \muzy \right) =0$.
\end{assumption}

Conditions $(i)$ and $(ii)$
are standard assumptions on continuity and differentiability of the matching function. Condition $(iii)$
states that the masses of matches between men of type $x$ and women of type $y$
are weakly increasing with the unassigned masses of own-type men and women 
and weakly decreasing with the unassigned masses of other-type men and women.
It implies that the matching function has the weak substitutes property.
Condition $(iv)$ restricts that
aggregate substitution effects of the unassigned masses of other-type agents
are smaller than the effects of the unassigned masses of own-type agents.
Condition $(v)$ is a standard limit assumption on the matching function.

In section \ref{sec: examples}, we provided several examples of
aggregate matching functions borrowed from the matching literature. It is
easy to show that conditions $(i)$-$(v)$ in assumption \ref{ass: MFE} are met in most examples. 
It turns out that these very mild conditions on $M_{xy}$ are sufficient to
prove the existence and uniqueness results in theorem \ref{thm:system}
below.
Our proof generalizes a result in GKW in two important aspects.
First,
we consider models with general matching functions which allow for the equilibrium masses of matched
agents to depend on the equilibrium masses of unassigned agents of all types,
while the result in GKW only applies to models with matching functions which restricts
the equilibrium masses of matched
agents to depend only on the equilibrium masses of unassigned own-type agents.
Second, our results extends GKW's results to matching functions that are not necessarily homogeneous of degree 1. 

\begin{theorem}
\label{thm:system} Under definition \ref{def: AMF} and assumption \ref{ass:
MFE}, there exists a unique equilibrium matching with partial assignment,
given by $\mu _{xy}^{\ast }=M_{xy}(\muxz^{\ast},\muzy^{\ast })$, for any $x\in \mathcal{X}$ and $y\in \mathcal{Y}$, where the pair of vectors,
$\muxz^{\ast}=\{\mu_{x0}^\ast\}_{x \in \mathcal{X}}$ and $\muzy^{\ast}=\{\mu_{0y}^\ast\}_{y \in \mathcal{Y}}$, is the unique solution to the system
\begin{equation}
\label{NL: theorem6}
\QATOPD\{ . {n_{x}=\mu _{x0}+\sum_{y\in \mathcal{Y}}M_{xy}(\muxz,\muzy)}{m_{y}=\mu_{0y}+\sum_{x\in \mathcal{X}}M_{xy}(\muxz,\muzy)}.
\end{equation}%
\begin{proof}
See Appendix \ref{app:proofs_thm1}.
\end{proof}
\end{theorem}

This result is closely related to \textcite{mourifie2019marriage} who proves existence and uniqueness of an equilibrium in an identical framework, using Brouwer’s fixed-point theorem to establish existence and \textcite{gale1965jacobian}'s results to establish uniqueness. Our proof of existence is constructive and relies on an iterative procedure
that we call {\it Generalized Iterative Projective Fitting Procedure} (Generalized IPFP). The IPFP algorithm has
been used in various fields under different names \parencite[see a survey in][]{Idel2016}, and more recently in economics \parencite[see][]{GalichonKominersWeber2015, GalichonKominersWeber2019}. GKW introduce the IPFP algorithm in their ITU-logit setting (see example d) in section \ref{sec: examples}).
In this paper, we propose a new algorithm, which generalizes the IPFP in GKW. It offers an efficient way of computing the unique matching equilibrium in definition \ref{def: AMF}. Because this algorithm
turns out to be very useful in practice (including estimation and counterfactuals), we provide a formal description of it below. Finally, our proof of
uniqueness relies on \textcite{berry2013connected}. 

\begin{Algorithm}
\label{IPFP Algorithm} The Generalized Iterative Projection Fitting
Procedure works as follows$,$
\begin{tabular}{r|p{5in}}
Step $0$ & Fix the value of $\mu _{x0}$ at  $\mu _{x0}^{0}=0$ for all $x \in \mathcal{X}$ and $\mu _{0y}$ at  $\mu _{0y}^{0}=0$ for all $y \in \mathcal{Y}$. \\
Step $1$ & Fix the value of $\mu _{x0}$ at  $\mu _{x0}^{1}=n_{x}$ for all $x \in \mathcal{X}$ and $\mu _{0y}$ at  $\mu _{0y}^{1}=m_{y}$ for all $y \in \mathcal{Y}$. \\
Step $2t$ & For each $x \in \mathcal{X}$, keep the values $\mu _{x^\prime 0}^{2t-2}$ for $x^\prime \neq x$, $\mu _{0y^\prime}^{2t-2}$ for all $y^\prime \neq y$ and $\mu _{0y}^{2t-1}$ for all $y \in \mathcal{Y}$ fixed,  solve for the value $\mu _{x0}^{2t}$, such
that the equality, $\mu^{2t}_{x0}+\sum_{y\in \mathcal{Y}}M_{xy}(\mu^{2t}_{x0},\{\mu_{x^\prime 0}^{2t-2}\}_{x^\prime \neq x},\mu^{2t-1}_{0y}, \{\mu_{0y^\prime}^{2t-2}\}_{y^\prime \neq y})=n_{x}$ holds$;$
For each $y \in \mathcal{Y}$, keep the values $\mu _{x^\prime 0}^{2t-2}$ for $x^\prime \neq x$,
$\mu _{x0}^{2t-1}$ for all $x \in \mathcal{X}$ and $\mu _{0y^\prime}^{2t-2}$ for all $y^\prime \neq y$ fixed, 
solve for the value $\mu _{0y}^{2t}$, such
that the equality, 
$\mu^{2t}_{0y}+\sum_{x\in \mathcal{X}}M_{xy}(\mu_{x0}^{2t-1},\{\mu_{x^\prime 0}^{2t-2}\}_{x^\prime \neq x}, \mu^{2t}_{0y},\{\mu_{0 y^\prime}^{2t-2}\}_{y^\prime \neq y})=m_{y}$ holds.\\
Step $2t+1$ & For each $x \in \mathcal{X}$, keep the values $\mu _{x^\prime 0}^{2t-1}$ for $x^\prime \neq x$, $\mu _{0y^\prime}^{2t-1}$ for all $y^\prime \neq y$ and $\mu _{0y}^{2t}$ for all $y \in \mathcal{Y}$ fixed,  solve for the value $\mu _{x0}^{2t+1}$, such
that the equality, $\mu^{2t+1}_{x0}+\sum_{y\in \mathcal{Y}}M_{xy}(\mu^{2t+1}_{x0},\{\mu_{x^\prime 0}^{2t-1}\}_{x^\prime \neq x},\mu^{2t}_{0y}, \{\mu_{0y^\prime}^{2t-1}\}_{y^\prime \neq y})=n_{x}$ holds$;$
For each $y \in \mathcal{Y}$, keep the values $\mu _{x^\prime 0}^{2t-1}$ for $x^\prime \neq x$,
$\mu _{x0}^{2t}$ for all $x \in \mathcal{X}$ and $\mu _{0y^\prime}^{2t-1}$ for all $y^\prime \neq y$ fixed, 
solve for the value $\mu _{0y}^{2t+1}$ such
that the equality, 
$\mu^{2t+1}_{0y}+\sum_{x\in \mathcal{X}}M_{xy}(\mu_{x0}^{2t},\{\mu_{x^\prime 0}^{2t-1}\}_{x^\prime \neq x}, \mu^{2t+1}_{0y},\{\mu_{0 y^\prime}^{2t-1}\}_{y^\prime \neq y})=m_{y}$ holds.
\end{tabular}
\newline
\newline
The algorithm terminates when,
$\sup \big\{
\sup_x \left| \mu^{2t+1}_{x0}-\mu^{2t-1}_{x0} \right|,
\sup_y \left| \mu^{2t+1}_{0y}-\mu^{2t-1}_{0y} \right|
\big \} < \epsilon$, where $\epsilon$ is a sufficiently small positive value.
\end{Algorithm}

According to the proof of theorem \ref{thm:system},  algorithm \ref{IPFP Algorithm} converges to a unique solution $(\muxz^*, \muzy^*)$ under assumption \ref{ass: MFE}.

% \begin{comment}

% \begin{Algorithm}
% \label{IPFP Algorithm extend} The extended Generalized Iterative Projection Fitting
% Procedure works as follow

% \begin{tabular}{r|p{5in}}
% Step $0$ & Fix the initial value of $\mu _{x0}$ and $\mu _{0y}$ at  $\mu _{x0}^{0}=n_{x}$ and $\mu _{0y}^{0}=m_{y}$ for all $x \in \mathcal{X}$
% and $y \in \mathcal{Y}$. \\
% Step $t+1$ & $(i)$ For each $x \in \mathcal{X}$, keep the values $\mu _{\acute{x}0}^{t}$ for any $\acute{x} \neq x$ and $\mu _{0y}^{t}$ for any $y \in \mathcal{Y}$ fixed, solve for the value, $\mu _{x0}^{t+1}$ of $\mu _{x0},$ such
% that the equality, $\mu_{x0}+\sum_{y\in \mathcal{Y}}M_{xy}(\mu _{x0},\{\mu_{\acute{x}0}^{t}\}_{\acute{x} \neq x},\{\mu_{0y}^{t}\}_{y \in \mathcal{Y}})=n_{x}$ holds. \\
% & $(ii)$ For each $y \in \mathcal{Y}$, keep the values $\mu _{0\acute{y}}^{t}$
% for any $\acute{y} \neq y$ and $\mu _{x0}^{t}$ for any $x \in \mathcal{X}$ fixed, solve for the value, $\mu _{0y}^{t+1}$ of $\mu _{0y},$ such
% that the equality, $\mu_{0y}+\sum_{x\in \mathcal{X}}M_{xy}(\{\mu_{x0}^{t}\}_{x \in \mathcal{X}}, \mu _{0y},\{\mu_{0\acute{y}}^{t}\}_{\acute{y} \neq y})=m_{y}$ holds.
% \end{tabular}
% \newline
% \newline
% The algorithm terminates when, $\sup\left|\{\mu _{x0}^{t+1}-\mu
% _{x0}^{t}\}_{x \in \mathcal{X}}, \{\mu _{0y}^{t+1}-\mu
% _{0y}^{t}\}_{y \in \mathcal{Y}} \right|<\epsilon$, where $\epsilon$ is a sufficiently small positive
% value.
% \end{Algorithm}

% \end{comment}

\section{Maximum Likelihood Estimation}\label{sec:MLE}

In this section, we show how to conduct parametric inference on the matching function equilibrium models with partial assignment, and how the
structural parameters can be estimated using maximum likelihood.
To reduce computational time, we provide an analytic expression for the gradient of the likelihood, as well as formulas to
compute confidence intervals when the matching function is homogeneous of degree 1.

Let us assume that $M_{xy}$ belongs
to a parametric family $M_{xy}^{\theta },$ where there is a unique value $\theta_0$ which rationalizes the observed data.\footnote{
This implicitly assumes that we consider the matching models which are parametrically identified.
All matching models listed in section \ref{sec: examples} are parametrically identified from the data on who marries whom.
}
Our goal is to estimate the parameters $\theta_0$ in $M_{xy}^{\theta }.$
Note that for given masses $n_{x}$
and $m_{y}$, and a given parameter $\theta$, we can obtain the (unique) equilibrium masses
of singles $(\muxz,\muzy)$
(using algorithm \ref{IPFP Algorithm}). We can then compute the predicted mass of
matching between type $x$ men and type $y$ women, $\mu _{xy}^{\theta
} $, using our aggregate matching function $M_{xy}^{\theta}$. These quantities are all we need to construct a likelihood.

\subsection{The Likelihood}

We first define the probability of forming a $(x,y)$ type
%$(x,y)\in \mathcal{XY}_{0}$-type of
household. 
Let $n=\sum_{x \in \mathcal{X}}n_x$ and $m=\sum_{y \in \mathcal{Y}}m_y$
denote the total masses of male and female.
Using the masses of singles $(\muxz,\muzy)$,
 we define the vector of model predicted matching
frequencies
$\Pi _{xy}(\theta ,n,m,\muxz,\muzy)$
as%
\begin{equation}
\Pi _{xy}(\theta ,n,m,\muxz,\muzy)=\frac{M_{xy}^{\theta }(a,b)}{%
N^\theta(n,m,\muxz,\muzy)}  \label{eq:PI}
\end{equation}%
for all $xy\in \mathcal{XY}_{0}$, where $N^\theta(n,m,\muxz,\muzy)=n+m-\sum_{xy \in \mathcal{XY}} M_{xy}^\theta(a, b)$
is the predicted total number of households.
If we observe the matching
$\hat{\boldsymbol{\mu}}=(\hat{\mu},\hat{\muxz},\hat{\muzy})$
from the data, then the log-likelihood is given by
\begin{equation}
l\left( \hat{\boldsymbol{\mu}}|\theta,n,m,\muxz,\muzy\right) =\sum_{xy\in
\mathcal{XY}_{0}}\hat{\mu}_{xy}\log \Pi _{xy}\left( \theta ,n,m,\muxz,\muzy\right).
\label{eq:loglikelihood}
\end{equation}

In practice, $n$ and $m$ are replaced by their efficient estimator $\hat{n},$
and $\hat{m}$ that can be computed from the observed matching,
$\hat{\boldsymbol{\mu}}$.
Hence, the maximum likelihood
estimator solves the following problem
\begin{equation}
\max_{\theta ,\muxz,\muzy} \:\:\: l\left( \hat{\boldsymbol{\mu}},\hat{n},\hat{m}|\theta
,\muxz,\muzy\right),  \label{eq:maxloglikelihood}
\end{equation}
\begin{equation*}
 \text{subject to }\qquad \QATOPD\{ . {\hat{n}_{x} =\mu _{x0}+\sum_{y\in \mathcal{Y}}M_{xy}^{\theta }(a,b)}{\hat{m}_{y} =\mu _{0y}+\sum_{x\in \mathcal{X}}M_{xy}^{\theta }(a,b)}.
\end{equation*}
Letting the constraints be rewritten as
$G(\theta, \muxz,\muzy)=0$,
we propose two computational
approaches to solve this estimation problem.

\subsection{Computation}
\subsubsection{The Nested approach}

The first approach is to get rid of the constraints,
$G(\theta ,\muxz,\muzy)=0,$
by solving
for the equilibrium
$(\muxz^{\theta},\muzy^{\theta })$
for any
value of $\theta $. From theorem \ref{thm:system} above, we know that such
an equilibrium always exists and is unique. From these unique values of
$\muxz^{\theta }$ and $\muzy^{\theta}$,
$\mu _{xy}^{\theta }$ is deduced
from $M_{xy}^{\theta }$ and the log-likelihood can be computed.

Estimation proceeds as follow: 
$(i)$ fix a value of $\theta $ ; 
$(ii)$ solve the system of equations (\ref{eq: nonlinear_system}) and obtain the unique
$\muxz^{\theta }$ and $\muzy^{\theta }$;
$(iii)$ deduce $\mu _{xy}^{\theta }$ from $M_{xy}^\theta(a^{\theta },b^{\theta })$ and
compute
$\Pi _{xy}(\theta , n, m,\muxz,\muzy)$
according to (\ref{eq:PI}) ; 
$(iv)$ compute the log-likelihood in equation (\ref{eq:loglikelihood}).

This approach has the advantage that by construction,
$\muxz^{\theta }$ and $\muzy^{\theta }$
solve
$G(\theta ,\muxz,\muzy)=0$
for any
value of $\theta $. Hence, we can apply the implicit function theorem to
compute the gradient of the unconstrained likelihood,
$l\left( \hat{\boldsymbol{\mu}},\hat{n},\hat{m}|\theta \right)$.
Most of the current available methods update
the parameters $\theta $ at each iteration using the gradient of the
objective function.
Numerical approximation of the gradient may be very time consuming  since evaluating the log-likelihood requires that we solve the system of equations (\ref{eq: nonlinear_system}).
In theorem \ref{thm:gradient} below, we provide an analytic expression of the gradient, which is particularly useful for applied work.

\begin{theorem}[Gradient of the log-likelihood]
\label{thm:gradient} Let
$N^{\theta}$
denote the abbreviation of $N^{\theta}(n,m,\muxz,\muzy)$.
Then, the derivative of the predicted frequency of a match pair $\left( x,y\right) \in
\mathcal{XY}_{0}$-type household with respect to $\theta ^{k}$ is given by,%
\begin{equation*}
\partial _{\theta ^{k}}\Pi _{xy}=\frac{\partial _{\theta ^{k}}\mu _{xy}}{%
N^{\theta}}+\frac{\mu _{xy}}{N^{\theta}\times N^{\theta}}\sum_{xy\in \mathcal{XY}
}\partial _{\theta ^{k}}\mu _{xy}
\end{equation*}%
where $\partial _{\theta ^{k}}\mu _{xy}= \sum_{x \in \mathcal{X}} \partial _{\mu _{x0}}M_{xy}^{\theta }\left(
a^{\theta },b^{\theta }\right) \partial _{\theta ^{k}}\mu
_{x0}+
\sum_{y \in \mathcal{Y}}
\partial _{\mu _{0y}}M_{xy}^{\theta }\left(a^{\theta },b^{\theta }\right) \partial _{\theta ^{k}}\mu _{0y}+\partial _{\theta
^{k}}M^\theta_{xy}\left( a^{\theta },b^{\theta }\right) $, whenever $xy\in \mathcal{XY},$ and
\begin{equation*}
\binom{\partial _{\theta ^{k}}a}
{\partial _{\theta ^{k}}b}%
=\Delta ^{-1}\binom{c^{k}}{d^{k}}
\end{equation*}%
where $\partial _{\theta ^{k}}a = 
\left(\partial _{\theta ^{k}}\mu_{x0} \right)^{\prime}_{x \in \mathcal{X}},$
$\partial _{\theta ^{k}}b = \left(\partial _{\theta ^{k}}\mu_{0y} \right)^{\prime}_{y \in \mathcal{Y}},$
$c^k=\left(c_x^k\right)^{\prime}_{x \in \mathcal{X}}$ with
$c_{x}^{k}=-\sum_{y\in \mathcal{Y}}\partial _{\theta
^{k}}M^\theta_{xy}\left( a^{\theta },b^{\theta }\right), $
and $d^k=\left(d_x^k\right)^{\prime}_{y \in \mathcal{Y}}$ with
$d_{y}^{k}=-\sum_{x\in \mathcal{X}}\partial _{\theta ^{k}}M^\theta_{xy}\left(a^{\theta },b^{\theta }\right) $, and $\Delta $ is expressed
blockwise by%
\begin{equation}
\Delta =%
\begin{pmatrix}
(1+\sum_{y\in \mathcal{Y}}\partial _{\mu _{10}}M_{1y}^{
}) 
& \sum_{y\in \mathcal{Y}} \partial_{\mu_{20}}M_{1y}
& \cdots 
& \sum_{y\in \mathcal{Y}} \partial_{\mu_{0|\mathcal{Y}|}}M_{1y}
\\
\sum_{y\in \mathcal{Y}} \partial_{\mu_{10}}M_{2y}
&(1+\sum_{y\in \mathcal{Y}}\partial _{\mu _{20}}M_{2y}^{
}) 
& \cdots 
& \sum_{y\in \mathcal{Y}} \partial_{\mu_{0|\mathcal{Y}|}}M_{2y}
\\
\vdots
&
\vdots
&
\vdots
&
\vdots 
\\
\sum_{x\in \mathcal{X}} \partial_{\mu_{10}}M_{x|\mathcal{Y}|}
& 
\sum_{x \in \mathcal{X}} \partial_{\mu_{20}}M_{x |\mathcal{Y}|}
& \cdots 
& (1+\sum_{x\in \mathcal{X}}\partial _{\mu _{0|\mathcal{Y}|}}M_{x|\mathcal{Y}|}^{
})
\end{pmatrix}.
\label{eq:delta}
\end{equation}%
\begin{proof}
See Appendix \ref{app:proofs_thm2}.
\end{proof}
\end{theorem}

\subsubsection{The MPEC approach}

In our second approach, we rewrite problem (\ref{eq:maxloglikelihood}) as
the Lagrangian%
\begin{equation*}
\min_{\lambda \in \mathbb{R}^{|\mathcal{X}|+|\mathcal{Y}|}}\max_{\theta \in \mathbb{R}^{d},\muxz\in \mathbb{R}^{|\mathcal{X}|},\muzy\in \mathbb{R}^{|\mathcal{Y}|}}l\left( \theta
,\muxz,\muzy\right) +\lambda G\left( \theta ,\muxz,\muzy\right)
\end{equation*}
where $\lambda \,$\ is the vector of Lagrange multipliers associated with the
constraint $G\left( \theta ,\muxz,\muzy\right) =0$.

The first order conditions are therefore,%
\begin{align*}
Z_{1}\left( \theta ,\muxz,\muzy,\lambda \right) & =0=\partial _{\theta }l\left(
\theta ,\muxz,\muzy\right) +\lambda \partial _{\theta }G\left( \theta ,\muxz,\muzy\right), \\
Z_{2}\left( \theta ,\muxz,\muzy,\lambda \right) & =0=\partial _{\muxz,\muzy}l\left( \theta
,\muxz,\muzy\right) +\lambda \partial _{\muxz,\muzy}G\left( \theta ,\muxz,\muzy\right), \mbox{ and }\\
Z_{3}\left( \theta ,\muxz,\muzy,\lambda \right) & =0=G\left( \theta ,\muxz,\muzy\right),
\end{align*}%
which defines a map $Z,$
\begin{equation*}
\begin{array}{ccc}
\mathbb{R}^{d}\times \mathbb{R}^{\left\vert \mathcal{X}\right\vert }\times
\mathbb{R}^{\left\vert \mathcal{Y}\right\vert }\times \mathbb{R}^{\left\vert
\mathcal{X}\right\vert +\left\vert \mathcal{Y}\right\vert } & \rightarrow &
\mathbb{R}^{d}\times \mathbb{R}^{\left\vert \mathcal{X}\right\vert }\times
\mathbb{R}^{\left\vert \mathcal{Y}\right\vert }\times \mathbb{R}^{\left\vert
\mathcal{X}\right\vert +\left\vert \mathcal{Y}\right\vert } \\
\left( \theta ,\muxz,\muzy,\lambda \right) & \rightarrow & Z=(Z_{1}\left( \theta
,\muxz,\muzy,\lambda \right) ,Z_{2}\left( \theta ,\muxz,\muzy,\lambda \right) ,Z_{3}\left(
\theta ,\muxz,\muzy,\lambda \right) ).%
\end{array}%
\end{equation*}%
Maximizing the likelihood is equivalent to finding the root of $Z$. In
general, numerical methods will require the knowledge of the Jacobian of $Z$, which is given by:
\begin{equation*}
JZ=%
\begin{pmatrix}
\partial _{\theta }^{2}l\left( \theta ,\muxz,\muzy\right) +\lambda \partial _{\theta
}^{2}G\left( \theta ,\muxz,\muzy\right) & \partial _{\theta }\partial _{\muxz,\muzy}l\left(
\theta ,\muxz,\muzy\right) +\lambda \partial _{\theta }\partial _{\muxz,\muzy}G\left( \theta
,\muxz,\muzy\right) & \partial _{\theta }G\left( \theta ,\muxz,\muzy\right) \\
\partial _{\theta }\partial _{\muxz,\muzy}l\left( \theta ,\muxz,\muzy\right) +\lambda
\partial _{\theta }\partial _{uv}G\left( \theta ,\muxz,\muzy\right) & \partial
_{\muxz,\muzy}^{2}l\left( \theta ,\muxz,\muzy\right) +\lambda \partial _{\muxz,\muzy}^{2}G\left(
\theta ,\muxz,\muzy\right) & \partial _{\muxz,\muzy}G\left( \theta ,\muxz,\muzy\right) \\
\partial _{\theta }G\left( \theta ,\muxz,\muzy\right) & \partial _{\muxz,\muzy}G\left( \theta
,\muxz,\muzy\right) & 0%
\end{pmatrix}
\label{eq:jacZ}
\end{equation*}%
We discuss the advantages and disadvantages of each approach in Online Appendix \ref{app:simulations}, and provide numerical experiments as well.

\subsection{Estimation in large markets}\label{sec:CI}

In this section, we shall discuss the difference between homogeneous and
non-homogeneous models in terms of estimation, and provide formulas to
compute the confidence intervals for the homogeneous case.

\subsubsection{Homogeneous and non-homogeneous matching functions}

Recall that in equilibrium, the scarcity constraints from equation \eqref{eq: nonlinear_system} are satisfied, that is $n_{x} =\mu_{x0}^{\ast }+\sum_{y\in \mathcal{Y}} M_{xy}\left(a^{\ast },b^{\ast }\right)$ and $m_{y} =\mu_{0y}^{\ast }+\sum_{x\in \mathcal{X}} M_{xy}\left(a^{\ast },b^{\ast }\right)$ for some matching function $M_{xy}$. Consider the equivalent system of equations, with rescaled quantities,
i.e.
\begin{equation*}
\frac{n_{x}}{K} =\frac{\mu _{x0}^{\ast }}{K}+\sum_{y\in \mathcal{Y}} \widetilde{M}_{xy}\left(\frac{a^{\ast }}{K},\frac{b^{\ast }}{K}\right)\text{ and }
\frac{m_{y}}{K} =\frac{\mu _{0y}^{\ast }}{K}+\sum_{x\in \mathcal{X}} \widetilde{M}_{xy}\left(\frac{a^{\ast }}{K},\frac{b^{\ast }}{K}\right)
\end{equation*}%
where we introduce the new matching function $\widetilde{M}_{xy}(a,b)\equiv\frac{1}{K}%
M_{xy}\left( Ka,Kb\right)$, for any given $K>0$.
When the matching function is homogeneous of degree $d$, $\widetilde{M}=K^{d-1} M$ for any $K>0$.
However, in the non-homogeneous case,
there is no guarantee that even when $K$ grows large, the matching function $%
\widetilde{M}$ will converge to a non trivial stable matching function $\bar{M}$. In addition, homogeneous models allow us to work with quantities that can be interpreted as frequencies instead of masses. Indeed, take
$K=N^\ast$,
the total number of household in equilibrium. Hence
\begin{equation}
\frac{n_{x}}{N^\ast} =\frac{\mu _{x0}^{\ast}}{N^\ast}+\sum_{y\in \mathcal{Y}} \widetilde{M}_{xy}\left(\frac{a^{\ast }}{N^\ast},\frac{b^{\ast }}{N^\ast}\right) \text{ and }
\frac{m_{y}}{N^\ast} =\frac{\mu _{0y}^{\ast }}{N^\ast}+\sum_{x\in \mathcal{X}} \widetilde{M}_{xy}\left(\frac{a^{\ast }}{N^\ast},\frac{b^{\ast }}{N^\ast}\right).
\end{equation}
Let $\zeta _{x}=\frac{n_{x}}{N^{\ast }}$, $\zeta _{y}=\frac{m_{y}}{
N^{\ast }}$, and $\zeta=\left( \{\zeta _{x}\}_{x \in \mathcal{X}}, \{\zeta _{y}\}_{y \in \mathcal{Y}} \right)$,
further denote $(\pi _{a}^{\ast },\pi _{b}^{\ast })=(\frac{a^{\ast }}{N^{\ast }},\frac{b^{\ast }}{N^{\ast }})$
and $\pi^{\ast} = \left( \pi^{\ast} _{a}, \pi^{\ast} _{b} \right)$.
 Then $\pi^{\ast}$ is the
unique solution to the system of equation $A\widetilde{M}(\pi)=\zeta$.\footnote{For convenience, we rewrite the system of equations in (\ref{eq: nonlinear_system}) as $A M(a,b) = \left(\begin{array}{c}
      n_x \\
      m_y
    \end{array}
  \right)$, where $A$ is an $\left(|\mathcal{X}|+|\mathcal{Y}|\right)\times\left(|\mathcal{X}||\mathcal{Y}| +|\mathcal{X}|+|\mathcal{Y}|\right)$ matrix.}
In this case, the key novelty is that the inputs of the aggregate matching
function are interpreted as the frequencies of single men of type $x$ and
single women of type $y$ in the population of households instead of masses,
while the output is interpreted as the frequency of married couples of type $(x,y)$ in the population of households instead of the mass. This is no longer
the case with non-homogeneous models, as there is no guarantee that the
output of the aggregate matching function can be interpreted as matching
frequencies when the inputs are also frequencies. Intuitively, this is also the reason why we can only compute confidence intervals for homogeneous models, which we provide in the next section.

In practice, homogeneous models imply that we observe the matching frequencies $\hat{%
\pi}$ from the data and estimate $\theta $ by maximum likelihood where the likelihood is given by
\begin{equation}
l\left( \hat{\pi}|\theta ,\zeta \right) =\sum_{xy\in \mathcal{XY}_{0}}\hat{%
\pi}_{xy}\log \Pi _{xy}\left( \theta ,\zeta \right),
\end{equation}%
and $\zeta $ will be replaced by its estimator
$A\widetilde{M}(\hat{\pi})$.
This does not
change the estimation, as $\Pi _{xy}\left( \theta ,\zeta \right) $ is
homogeneous of degree $0$ in $\zeta $ by the homogeneity property. It means
that in the homogeneous case, working with the observed $\hat{\pi}$ is
sufficient to carry on estimation, which is not the case in the
non-homogeneous setting, where we must work with $\hat{\mu}$.
Asymptotically, the noise added by $\hat{\pi}$ is normally distributed
around $\pi $, so that $N^{1/2}\left( \hat{\pi}-\pi \right) \sim \mathcal{N}%
\left( 0,V_{\pi }\right) $, where $V_{\pi }=diag\left( \pi \right) -\pi \pi
^{\prime }$. However, in the non-homogeneous case, we have instead, $%
N^{1/2}\left( \hat{\mu}-\mu \right) \sim \mathcal{N}\left( 0,V_{\mu }\right),
$ where $V_{\mu }=diag\left( N\mu \right) -\mu \mu ^{\prime }.$

\subsubsection{Confidence intervals for homogeneous models}

We now provide formulas to compute confidence intervals for estimates when the matching function is homogeneous of degree 1 and restricts the masses of matches to depend only on unassigned agents of own types. 
Note that these formulas are provided in
GKW
for imperfectly transferable utility matching models with logit heterogeneity in taste. In that paper, it is shown that such models lead to a matching function that is homogeneous of degree 1, as recalled in
section \ref{sec: examples}, example d).
Therefore, if we can show that any matching function that is homogeneous of degree 1 and restricts the masses of matches to depend only on unassigned agents of own types is in fact equivalent to a matching model with ITU and logit unobserved heterogeneity as introduced in GKW, then we can make use of their results. This is proven in the following theorem.

% \begin{comment}
% \subsubsection{ITU-logit matching and confidence intervals for homogeneous models}

% As mentioned in Section \ref{sec: examples}, the ITU-logit matching model introduced
% in GKW can be reformulated as a matching function equilibrium problem. In addition,
% the associated aggregate matching function is homogeneous of degree 1.
% Section \ref{sec:MLE} generalized the estimation strategy found in GKW to
% the case with matching functions that need not be homogeneous of degree 1.
% However, we are only able to compute confidence intervals in the homogeneous case,
% as the computation relies on the results from GKW. {\color{red} Am I right with this sentence? I am confused about the intuition that why we can only compute the confidence intervals in the homogeneous case. Is that because in order to compute the confidence intervals, we have to rely on the results from GKW?}

% In this section, we provide
% a formal proof of the result that the ITU-logit matching model introduced
% in GKW can be reformulated as a matching function model in which the associated aggregate matching function is homogeneous of degree 1.
% In addition, we show that conversely, any matching function
% equilibria satisfying homogeneity of degree 1 can be reformulated as an ITU matching model with logit
% heterogeneity. Theorem \ref{thm:equiv} below formally states this result.
% \end{comment}

\begin{theorem}\label{thm:equiv}
$(i)$ An ITU-logit model introduced in GKW generates an aggregate matching function satisfying homogeneous of degree 1 and depending only on the masses of unassigned agents of own types$;$
$(ii)$ A matching function equilibrium model defined in definition \ref{def: AMF} and with its aggregate matching function satisfying assumption \ref{ass: MFE} and homogeneity of degree $1$,
and depending only on the masses of unassigned agents of own types
 is equivalent to a matching model with ITU and logit unobserved heterogeneity as introduced in GKW.
\begin{proof}
See Appendix \ref{app:proofs_thm3}.
\end{proof}
\end{theorem}

The equivalence between this type of matching function equilibrium models and the
ITU-logit models in GKW suggests that we can simply make use of GKW's results for computing confidence intervals.
Note that in this case, $\zeta $ is
estimated by
%{\color{blue}
$A\widetilde{M}(\hat{\pi})$
%}
and thus, as noted earlier, doing so introduces additional
noise in the estimates of $\theta $, so that the computation of the
variance-covariance matrix of $\theta $ cannot rely on the standard
formulas. For the sake of clarity, we reproduce the results from GKW to compute
the variance-covariance matrix $V_{\theta }$ in closed-form, and give a detailed proof
of the result.

\begin{theorem}[Confidence Intervals]
\label{thm:confidence}
When $\theta $ is estimated by maximum likelihood as described
in section \ref{sec:MLE} and, 
the matching function satisfies
homogeneity of degree $1$ 
and depends only on the masses of unassigned agents of own types,
then
$V_{\theta }=\left( \mathcal{I}_{11}\right) ^{-1}+\mathcal{I}_{11}^{-1}%
\mathcal{I}_{12}AV_{\pi }A^{\prime }\mathcal{I}_{12}^{\prime }\mathcal{I}%
_{11}^{-1}$
where we denote $\mathcal{I}_{11} =-\left( D_{\theta }\log \Pi \right) ^{\prime }diag\left(
\pi \right) \left( D_{\theta }\log \Pi \right)$, and $\mathcal{I}_{12} =\left( D_{\theta }\log \Pi \right) ^{\prime }diag\left(
\pi \right) \left( D_{\zeta }\log \Pi \right)$ and $V_{\pi }=diag\left( \pi \right) -\pi \pi ^{\prime }$.
\begin{proof}
See appendix \ref{app:proofs_thm4}.
\end{proof}
\end{theorem}
Note that $D_{\theta }\log \Pi $ and $D_{\zeta }\log \Pi $ have analytic
expressions, which we provide in the proof.

\section{Estimation of counterfactuals}\label{sec:counterfactuals}

One key advantage of estimating a behavioural structural model is that it allows researchers
to conduct counterfactual experiments.
In the marriage market, researchers are often interested in how the marriage distribution or marital surplus is affected due to a demographic change in the number of available individuals, a policy change in birth control regulations, tax laws, divorce laws, or financial aid program, and so on.

In this section, we denote the parameterized matching function $M_{xy}$ by $M_{xy}(a, b; f_{xy}(\theta))$,
where $f_{xy}(\theta)$ is a function of parameterized matching surpluses.
The parameters $\theta$
are typically the parameters of interest that are to be estimated.
% \begin{comment}
% The extended notation $M_{xy}(\muv_{n}, \muv_{m}; f_{xy}(\theta))$ not only captures all matching functions in models (mostly static) listed in Section \ref{sec: examples},
% but also captures the matching functions of some dynamic matching models in the literature,
% such as the one derived by \cite{choo2015dynamic}.
% Extending the static transferable utility CS model to a stationary dynamic setting,
% \cite{choo2015dynamic} derives a dynamic matching function,
% \begin{align}
% \label{MF:Choo2015}
% M_{xy}(\boldsymbol{\mu}_{n}, \boldsymbol{\mu}_{m}; f_{xy}(\theta),\lambda)
% &=
% \exp \left( \frac{\Pi_{xy}}{2} \right)
% \sqrt{n_xm_y}
%  \prod_{k=0}^{T(x,y)-1}\bigg[\frac{\mu_{x+k0}}{n_{x+k}} \frac{\mu_{0y+k}}{m_{y+k}} \bigg]^{\frac{1}{2}[\beta(1-\delta)]^k}, \nonumber \\
%  &=
% f_{xy}(\theta)
% \sqrt{n_xm_y}
%  \prod_{k=0}^{T(x,y)-1}\bigg[\frac{\mu_{x+k0}}{n_{x+k}} \frac{\mu_{0y+k}}{m_{y+k}} \bigg]^{\frac{1}{2}[\beta(1-\delta)]^k},
%  \end{align}
% where $\Pi_{xy}$ is the dynamic martial surplus and $T(x,y)$ is the marriage duration for a $(xy)$ match,
% $\beta$ and $\delta$ are discount rate and divorce rate.
% The matching function \eqref{MF:Choo2015} relates matching numbers $M_{xy}$ to unmatched numbers of type $x$ men and type $y$
% women in all future periods during the marriage.
% \end{comment}
With some abuse of notation, let the observed equilibrium marriage distribution be denoted by the triple $(\mu^*, \muxz^*, \muzy^*)$ and the observed masses of available individuals are denoted by  $(n_x, m_y).$ %for all $x \in \mathcal{X}$ and $y \in \mathcal{Y}$.
Let the corresponding variables in the counterfactual scenario be denoted with superscript prime,
$(\mu^{* \prime}, \muxz^{* \prime}, \muzy^{* \prime}, n^\prime_x,m^\prime_y)$. Our main goal is to estimate the new equilibrium marriage distribution,
denoted by
$(\mu^{*\prime}, \muxz^{*\prime}, \muzy^{*\prime})$,
in the counterfactual scenario.

We propose two methods of conducting counterfactual experiments,
$i$) the parametric approach, which relies on estimating the structural parameters $\theta$, and
$ii$) our new {\emph{parameter-free approach}}, which does not require that we estimate $\theta$.
The latter works whenever the matching functions are  multiplicatively separable in parameters and take a Cobb-Douglas form.
The two approaches allow us to analyze three different types of counterfactuals scenario resulting from policy changes:
\begin{enumerate}
\item[(i)] changes in the available numbers of  individuals from  $(n_x,m_y)$ to $(n^\prime_x,m^\prime_y)$;
\item[(ii)] proportional changes in matching surpluses from  $f_{xy}(\theta)$ to $t_{xy}f_{xy}(\theta)$ for $t_{xy}>0$;
\item[(iii)] a combined simultaneous change in $(n_x,m_y)$ and $f_{xy}(\theta)$
stated in (i) and (ii).
\end{enumerate}
%Following the empirical industrial organization literature,
Both approaches assume that the matching preferences captured by $f(\theta)$ are unchanged by policy changes
in counterfactual scenarios.
To illustrate the two approaches, we hereafter focus on the most general type of counterfactuals experiment given by (iii), without any loss of generality.

\subsection{The parametric approach}\label{sec: para_app}
Consider a counterfactual change where the available numbers of individuals $(n_x,m_y)$ change to $(n^\prime_x,m^\prime_y)$
and matching surpluses $f_{xy}(\theta)$ to $t_{xy}f_{xy}(\theta)$.
The parametric approach follows a two-step procedure: (i) estimate the parameters, $\hat{\theta}$,
from the observed matches
$(\mu^*, \muxz^*, \muzy^*)$,
using the proposed nested or MPEC approaches discussed in section \ref{sec:MLE} ; (ii) compute the counterfactual equilibrium distribution of individuals who remain single,
$(\muxz^{* \prime}, \muzy^{* \prime})$ under $(n^\prime_x, m^\prime_y, t_{xy}f_{xy}(\hat{\theta}))$
by solving the system of equations,
\begin{equation}
\label{sec: count_nonlinear_system}
\QATOPD\{ . {n^\prime_{x}=\mu^{* \prime} _{x0}+\sum_{y\in \mathcal{Y}}
M_{xy}(a^{* \prime}, b^{* \prime};t_{xy} f_{xy}(\hat{\theta}))
}
{m^\prime_{y}=\mu^{* \prime} _{0y}+\sum_{x\in \mathcal{X}}
M_{xy}(a^{* \prime}, b^{* \prime};t_{xy} f_{xy}(\hat{\theta}))
}.
\end{equation}%
%\eqref{sec: count_nonlinear_system} below and
We then derive the counterfactual equilibrium distribution of matches using our matching function,
$\mu^{*\prime}_{xy}=M_{xy}(a^{* \prime}, b^{* \prime};t_{xy} f_{xy}(\hat{\theta}))$.

The existence and uniqueness results in section \ref{sec:MFE} ensure that there is a unique solution to the system of nonlinear equations \eqref{sec: count_nonlinear_system}. Hence, we have a unique prediction for the counterfactual matching distribution.
The main advantage of the parametric approach lies in its generality. It applies to all matching function equilibrium models that satisfy the conditions given in definition \ref{def: AMF}.
One shortcoming is that it is a computationally intensive procedure where the system of nonlinear equations \eqref{sec: count_nonlinear_system} is solved at each iteration of the estimation routine.
In the Online Appendix \ref{app:simulations}, we provide simulation results for the Nested and MPEC approaches. It is clear that the time required to solve the nonlinear system \eqref{sec: count_nonlinear_system} and the number of iterations required for estimation increase substantially with market size.

\subsection{The parameter-free approach}\label{sec:para_free_app}

In this section, we propose an alternative new approach, which we call the \emph{parameter-free approach}, to compute the  counterfactual matching equilibrium.
While the parameter-free approach applies only to a subset of matching function equilibrium models, it does not require estimating $\theta.$ Furthermore, it only requires that we solve the system of nonlinear equations \eqref{eq: nonlinear_system} once. This approach applies  only to matching functions that take the multiplicative homogeneous form. We begin by stating the two restrictions needed for this subset of matching functions:
\begin{assumption}
\label{ass: matchingFun_sep}
The matching function is multiplicatively separable in the parameters taking the form
$M_{xy}(a, b;f_{xy}(\theta))
=f_{xy}(\theta)g(a,b)$.
\end{assumption}
\noindent
\begin{assumption}
\label{ass: matchingFun_homo}
The function $g(a,b)$ takes the Cobb-Douglas form, $g(a,b)=\Pi_{x \in \mathcal{X}} \mu_{x0}^{k_x}
\Pi_{y \in \mathcal{Y}} \mu_{0y}^{\kappa_y}$
for any $\{k_x \in \mathbb{R}\}_{x\in\mathcal{X}}$ and $\{\kappa_y \in \mathbb{R}\}_{y\in\mathcal{Y}}$.
\end{assumption}

Assumption \ref{ass: matchingFun_sep} restricts the matching function to be multiplicatively separable in
$\theta$,
while assumption \ref{ass: matchingFun_homo} further restricts the matching function to be a Cobb-Douglas function.
These two assumptions may seem restrictive, but with the exception of the NTU and ETU models, all other matching function examples of section \ref{sec: examples} satisfy these two assumptions.

%{\color{blue}
%Comment: The generalization problem of assumption 2. Could a model without %Logit assumption satisfy this assumption?
%Looking for Literature.
%}

% \begin{comment}
% \begin{assumption}
% \label{ass: matchingFun_sep}
% The matching function is multiplicatively separable in parameters with the form
% $M^\theta_{xy}(\mu_{x0},\mu_{0y})=f(\theta)g(\mu_{x0},\mu_{0y})$.
% \end{assumption}
% \noindent
% \begin{assumption}
% \label{ass: matchingFun_homo}
% The matching function $M^{\theta}_{xy}(\mu_{x0}, \mu_{0y})$ is a homogeneous function in $(\mu_{x0}, \mu_{0y})$.
% \end{assumption}
% Assumption \ref{ass: matchingFun_sep} restricts the matching function to be multiplicatively separable in parameters,
% while assumption \ref{ass: matchingFun_homo} further restricts the matching function to be a homogeneous function.
% \end{comment}

Our parameter-free approach proceeds in two steps.
In the first step,
%by assumptions \ref{ass: matchingFun_sep} and \ref{ass: matchingFun_homo},
we show that the
ratio of the matches at the counterfactual equilibrium relative to those at the observed equilibrium is free of  $\theta$.
In step two, we substitute these ratios into the system of equations \eqref{eq: nonlinear_system}.
This generates a non-linear system of equations in terms of the ratios of the numbers of unmatched singles at the counterfactual equilibrium relative to those at the observed equilibrium.
This new system of equations in terms of these ratios of unmatched single individuals is also free of $\theta$.
We show that this new system has a unique solution in terms
of these ratios. We subsequently use this equilibrium ratio of unmatched single individuals to
calculate the changes in the matching distributions between the observed and counterfactual
equilibria. We now formally present these two steps.

Let us introduce the notation $\tilde{z} \equiv z^\prime/z$ which, for any variable $z$, denotes the ratio of the counterfactual equilibrium quantities to the observed quantities.
Consider again a counterfactual in which $(n_x,m_y)$ changes to $(n^\prime_x,m^\prime_y)$
and $f_{xy}(\theta)$ changes to $t_{xy}f_{xy}(\theta)$.
Taking the ratio of matching function $\mu_{xy}=M_{xy}(a, b;f_{xy}(\theta))$ evaluated at $(a^{*\prime}, b^{*\prime};t_{xy}f_xy(\theta))$ under the counterfactual equilibrium and $(a^{*}, b^{*};
f_{xy}(\theta))$ under the observed equilibrium yields
\begin{align}
\label{count: ratioMF}
\tilde{\mu}_{xy}=\frac{\mu^{*\prime}_{xy}}{\mu^{*}_{xy}}
=\frac{M_{xy}(a^{*\prime},b^{*\prime};t_{xy}f_{xy}(\theta))}{M_{xy}(a^{*},b^{*};f_{xy}(\theta))}
=\frac{t_{xy}g(a^{*\prime},b^{*\prime})}{g(a^{*},b^{*})}
=t_{xy} g(\tilde{a},\tilde{b}),
\end{align}
where the third equality is derived from assumption \ref{ass: matchingFun_sep} and the fourth equality is from the homogeneity assumption \ref{ass: matchingFun_homo}.

We next introduce $\tilde{\mu}_{xy}$ into the system \eqref{eq: nonlinear_system} by
expressing them in terms of ratios of counterfactual quantities relative to observed quantities,
$(\tilde{\mu}_{xy},\tilde{a},\tilde{b}, \tilde{n}_x, \tilde{m}_y)$.
Recall the system of equations  \eqref{eq: nonlinear_system}, i.e., $n_{x}=\mu _{x0}+\sum_{y\in \mathcal{Y}}M_{xy}(a,b;f_{xy}(\theta))$ and $m_{y}=\mu _{0y}+\sum_{x\in \mathcal{X}}M_{xy}(a,b;f_{xy}(\theta))$.
Evaluating the system at both the counterfactual and observed equilibria,  dividing them, and
substituting \eqref{count: ratioMF} into it,
we obtain
\begin{align}
\label{count: ratioSystem}
& \QATOPD\{. {\tilde{n}_x = p_{x0} \tilde{\mu}_{x0}+\sum_{y \in \mathcal{Y}} p_{xy} \cdot t_{xy} g(
\tilde{a},\tilde{b})}
{\tilde{m}_y = q_{0y} \tilde{\mu}_{0y}+\sum_{x \in \mathcal{X}} q_{xy} \cdot t_{xy} g(\tilde{a},\tilde{b})},
\end{align}
where the quantities $p_{x0}=\mu^{*}_{x0}/n_x$ and $p_{xy} =\mu^{*}_{xy}/n_x$ are the observed equilibrium probabilities that a type $x$ man remains single and marries a type $y$ woman, respectively. Similarly, $q_{0y}=\mu^{*}_{0y}/m_y$ and $q_{xy} =\mu^{*}_{xy}/m_y$ are the probabilities that a type $y$ woman remains single and marries a type $x$ man, respectively.
This new system of equations is free of the parameters $\theta$.
This system has $|\mathcal{X}|+|\mathcal{Y}|$  number of nonlinear equations with $|\mathcal{X}|+|\mathcal{Y}|$ unknowns $(\tilde{a}, \tilde{b})$.
%{\color{blue}
%for all $x \in \mathcal{X}$ and $y \in \mathcal{Y}$.
%}
If  a unique solution in $(\tilde{a}, \tilde{b})$ %to this new system \eqref{count: ratioSystem}
exists,
we need to solve the system for $(\tilde{a}, \tilde{b})$ only once.
Then, using the solved $(\tilde{a}, \tilde{b})$,
we can construct the new counterfactual equilibrium quantities, ($a^{*\prime}, b^{*\prime}, \mu^{*\prime}_{xy}$) using the definition $\tilde{z} \equiv z'/z$ and equation \eqref{count: ratioMF}.

The final issue left is to show
that the system \eqref{count: ratioSystem} has a unique solution in $(\tilde{a}, \tilde{b})$.
We show that this is indeed the case.
%under assumptions \ref{ass: MFE} , \ref{ass: matchingFun_sep}, and \ref{ass: %matchingFun_homo},
%there exist a unique solution in $(\tilde{\mu}_{x0}, \tilde{\mu}_{0y})$.
This result is formally stated in theorem \ref{thm: existence_change}.
\begin{theorem}
\label{thm: existence_change}
Under assumptions \ref{ass: MFE}, \ref{ass: matchingFun_sep}, and \ref{ass: matchingFun_homo}, there exists a unique solution of the system \eqref{count: ratioSystem}
with unknowns $(\tilde{a}, \tilde{b})$.

\begin{proof}
See Appendix \ref{app:proofs_thm5}.
\end{proof}
\end{theorem}
\noindent
The proof of existence relies on a revised algorithm based on algorithm \ref{IPFP Algorithm},
while the proof of uniqueness builds on \textcite{berry2013connected}.

The estimation procedure of our parameter-free approach works as follows: (i) compute
the ratios $(\tilde{n}_x,\tilde{m}_y)$,
and the probabilities $(p_{x0}, p_{xy})$ and $(q_{x0}, q_{xy})$, and then obtain $\tilde{\mu}_{xy}$ using
\eqref{count: ratioMF}; (ii) compute the changes in unmarrieds $(\tilde{a}, \tilde{b})$
 by solving \eqref{count: ratioSystem}, and then compute
 ($a^{*\prime}, b^{*\prime}, \mu^{*\prime}_{xy}$) accordingly.
 In the Online Appendix \ref{app:example_choo2015},
 we also provide  details on how
 %to derive equation \eqref{count: ratioMF} and the system \eqref{count: ratioSystem} by using
 this approach also applies to the dynamic matching function \eqref{eq: sec4_matchFunction_dynamicType2} in \textcite{choo2015dynamic}.

Our approach highlights a previously undocumented new property in the matching literature,
that ratios in the equilibrium numbers of unmarried men and women can be inferred from the
observed matching equilibrium by using the system \eqref{count: ratioSystem},
which themselves are free of the model parameters for some class of matching models.
The first paper that proposed this `hat' approach is \textcite{eaton2007unbalanced}.
While this approach is quite commonly used in the trade literature, it does
not typically generate a transformed model that is free of structural parameters.
Unlike our case here, parameters such as the elasticity of substitution typically remain in the `hat'
model and these parameters need to be calibrated or estimated.
In our context, the parameter-free approach has an advantage of not requiring the estimation of model parameters. Moreover, we only need to solve the system of non-linear equations \eqref{count: ratioSystem} once.

%%%%%%%%%

\section{Empirical Application - The 1982 Elimination of the Social Security Student Benefit Program}
\label{sec:application}
The goal in this section is to investigate how the elimination of the Social Security Student Benefit Program in $1982$ has affected the 1987/88 age-education marriage distributions in the United States.
Using our proposed empirical approach,  we estimate candidate models from the ETU family of matching models to conduct this counterfactual analysis.
As discussed in example e) of section \ref{sec: examples}, the ETU family encompasses the TU CS model, the NTU model of \textcite{GalichonHsieh2017},
and the Harmonic Mean Matching function of \textcite{Schoen1981}.
The ETU models  are also a subset of ITU models.  To pick the best fitting model,  we perform model selection using information based criteria.

\subsection{The background}
Under the $1939$ Amendments to the Social Security Act, the children of deceased, disabled, and retired Social Security beneficiaries
could receive Social Security payments until they reach the age of eighteen.
In $1965$, these payments were extended to persons up to twenty two years of age still enrolled as full-time college students.
% in college because the Social Security Administration recognized the fact that children who are full-time students after age $18$ are often still dependent on their parents for their support.
 The social security student benefits were paid to eligible college students as monthly lump sums. %, similar to the G.I. Bill benefits paid to college-attending veterans. %, based on the earnings history of their parents.
 The benefits were extremely generous, especially considering the cost of public four-year colleges and universities at that time.\footnote{The average annual benefit in $1980$ paid to a child of deceased parent was $\$6,700$ while the average tuition and fees for public four-year colleges and universities was $\$1,900$.}
In the peak year of $1977$, there were about $900,000$ benefit recipients. In the peak pay-out year of $1981$, about $\$2.4$ billion were paid as student benefits.\footnote{These statistics are obtained from ``Research Note $\#11$: The History of Social Security Student Benefits published by the Historian's Office''.}

\begin{figure}[t!]
    \centering
    \begin{subfigure}[t]{0.5\textwidth}
        \centering
        \includegraphics[height=2.2in]{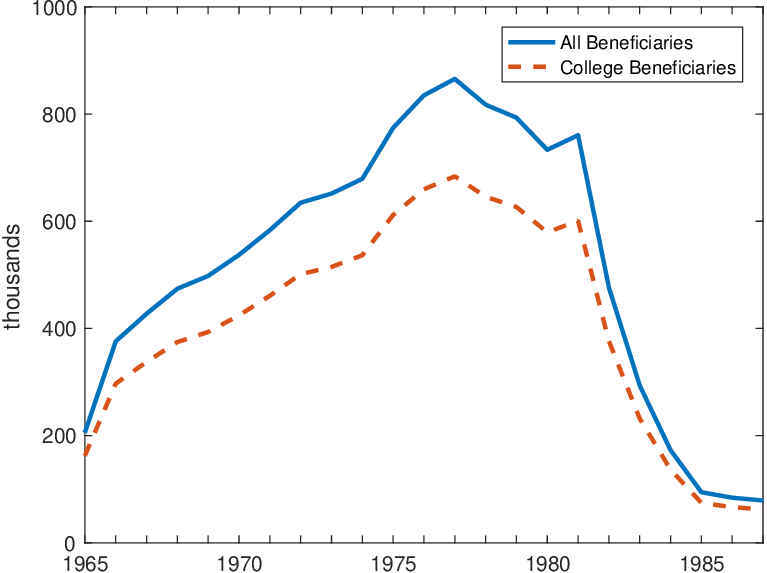}
        \caption{}
         \label{figure1:numbers}
    \end{subfigure}%
    ~
    \begin{subfigure}[t]{0.5\textwidth}
        \centering
        \includegraphics[height=2.2in]{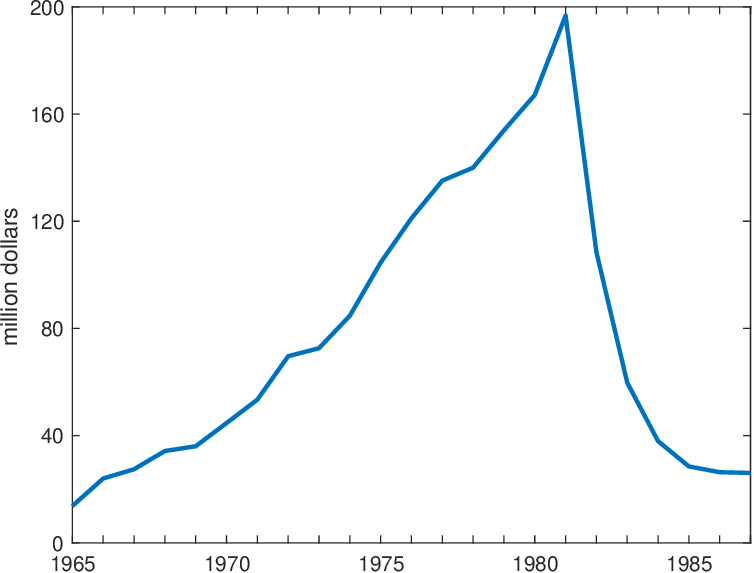}
        \caption{}
             \label{figure1:benefits}
    \end{subfigure}

    \caption{Number of beneficiaries and average monthly benefits payment}
    \floatfoot{Panel (a): Number of social security beneficiaries and college student beneficiaries;
    Panel (b): Average monthly benefit payment. Source: Social Security Administration (SSA) Research Note $\#11$.}
\end{figure}

In $1981$, Congress voted to eliminate the Social Security Student Benefits Program from $1982$ onwards.\footnote{According to the `Social Security Administration Research Note $\#11$', ``Benefits paid to post-secondary students ages 18-21 are to be phased-out; The phase-out is to be completed by April 1985; Benefits to elementary and/or secondary school students older than 18, are to end in August 1982.''} Since then, the number of student benefit recipients and the program spending dropped dramatically.
As shown in figure \ref{figure1:numbers},
the number of student beneficiaries dropped from around $760,000$ in $1981$ to $84,000$ recipients in $1986.$ The number of college student beneficiaries is estimated to have dropped from about $600,000$ to $66,000$  recipients in $1986$.\footnote{We do not have access to annual share of college student beneficiaries. According to a $1977$ Social Security Administration (SSA) survey, about $79\%$ student beneficiaries are in post-secondary institutions.}
The amount paid to eligible students was reduced immediately after the elimination of the program in 1981 (see figure \ref{figure1:benefits}).
The average monthly payment fell substantially from about $\$196$ million in $1981$ to $\$26$ million in $1986$.

Except for the introduction of the Pell Grant program in the early $1970$'s and the various G.I. Bills, the elimination of the Social Security Student Benefit program is the largest policy change in financial aid for college students.
 The impacts of various financial aid programs on students' college attendance and completion has been well studied in the literature.\footnote{
See e.g. \textcite{manski1983college} and \textcite{kane1994college} on Pell grant introduced in 1973, \textcite{reyes1997educational} on the Middle Income
Student Assistance Act, which eliminated the income cutoff for the Guaranteed Student Loan, and  \textcite{angrist1993effect} on War II G.I. Bills, which provided a generous monthly stipend to veterans in college.}
For the Social Security Student Benefit program,
\textcite{dynarski2003} found that the elimination of the program has had a large significant causal effect on students' college attendance and completion.
It is well known that education is a primary attribute in the marriage market.
%{\color{blue}
Moreover, marriage matching has important implications on fertility and population growth, labor-force
participation, income inequality, etc \parencite[see e.g.][]{grossbard1988women, Becker1991, doepke2019bargaining, eika2019educational}.
%}
Our goal is to understand how the 1982 elimination of the Social Security Student Benefit program affects the marriage matching distribution through its impact on students' college attendance and completion.

\subsection{The data}
We attempt to answer this question by estimating the counterfactual marriage distribution in 1987/88 had the Social Security Student Benefit program not been eliminated and compare it with the observed 1987/88 marriage matching distribution using
the ETU model. Since the ETU model
is static, we need to choose a specific year for our analysis. The year 1987/88 is sufficiently far along after the policy change to allow those most affected by the policy to reach a marriageable age. The elimination of aid in 1982 would most likely affect the then and soon-to-be high school seniors. It will also affect those individuals attending college and those   who were considering going to  college in the near future.\footnote{The year 1988 also happens to be the last year for which we have access to educational attainment of newly weds. We also wanted to minimize the effect of other educational policy that came into effect towards the end of the 1980s that would have confounded our results, such as `The Emergency Immigration Education Act' of 1984. }

%Our approach provides a convenient way of constructing the counterfactual distribution without estimating the model primitives.
% We illustrate our approach by estimating the changes of marriage matching distributions by ages and educations in
% $1987/88$ due to the elimination of the Social Security Student Benefit program using equations \eqref{matching_changes_CS}, \eqref{singleMen_changes2_CS} and \eqref{singleWomen_changes2_CS} derived from the CS model.
For this exercise, we require three data inputs:
\begin{itemize}
\item[i)] the observed number of available single men and women  by age and education as a result of the elimination of the financial aid program, $(n_x,m_y)$,  for all $x \in  \mathcal{X}$ and $y \in \mathcal{Y}$,
\item[ii)] the counterfactual number of available single men and women  by age and education had the financial aid program not been eliminated, $(n^{\prime}_x,m^{\prime}_y)$, for all $x \in  \mathcal{X}$ and $y \in \mathcal{Y}$,
\item[iii)] the observed flow of new marriages by age and education as a result of the elimination of the financial aid program, $\mu$.
\end{itemize}
Using these inputs, our empirical framework allows us to construct the counterfactual marriage distribution, $\mu^{\prime},$ i.e. the marriage distribution by age and education in 1987/88 had the financial aid program not been eliminated in 1981.
We are assuming that this policy change only affects the number of available individuals $(n_x,m_y)$, and not  the matching surplus, $f_{xy}(\theta)$.\footnote{
This is a common assumption in the literature as we expect marital preferences to remain unchanged in the short-run.
}

%Note that in this application, the unobserved counterfactual marriage distribution before {\color{red} XXX(Simon): without?} policy change is denoted by $\mu,$
%while the observed marriage distribution after policy is denoted by $\mu^\prime$.
%{\color{red} XXX(Simon): If I get it correctly, the notation is inconsistent with the previous section. Maybe we should choose new notations altogether and be consistent throughout the paper.}

Data on the flow of new marriages  as a result of the financial aid program elimination (that is, item iii)) is constructed using the $1987/88$ Vital Statistics marriage records obtained from the National Bureau of Economic Research data website \parencite{VitalStatistics2022}.\footnote{Like in CS, the flow of new marriages in constructed by taking a two year average of the new marriages in 1987 and 1988. This helps reduce the number of marriage pairs with zero new marriages.} Before 1989, the Vital Statistics tracks new marriages by educational attainment for 22 reporting states.\footnote{These $22$ states include California, Connecticut, Hawaii, Illinois, Kansas, Kentucky, Louisiana, Maine, Mississippi, Missouri, Montana, Nebraska, New Hampshire, New York, North Carolina, Rhode Island, Tennessee, Utah, Vermont, Virginia, Wisconsin and Wyoming.} However, from 1989 onwards, information on educational attainment of newly weds are no longer recorded. Our analysis will focus on these 22 reporting states. We construct item i), the observed number of available single men and women as a result of the financial aid program elimination, using the Integrated Public-Use Microdata
(IPUMS hereafter) files of the 1986 U.S. Current Population Survey (CPS) for these 22 reporting states \parencite{CPS2022}. More details about the data construction can be found in Online Appendix \ref{app:data}.

\begin{table}[ht!]
\centering
  \caption
  [Available Numbers of Male and Female \newline in millions from the 1986 CPS for the 22 reporting states]
{\tabular[t]{@{}c@{}}Numbers of available single male and female \\ in millions from the 1986 CPS for the 22 reporting states\endtabular}
\begin{tabular}{ l c c}
\hline\hline
%\multicolumn{3}{c}{U.S. CPS data in $1986$} \\
&  Male & Female\\\hline
 High school or less (HS)    &   8.79   &    10.41  \\
                         &   (63.3\%) &   (65.3\%)  \\
College (Col)                 &   4.24   &    4.72   \\
                         &  (30.5\%)  &  (29.6\%)   \\
Graduate school (GS)          &   0.86   &    0.80   \\
                         &   (6.2\%)   &  (5.1\%)  \\
Total  & 13.88&15.94  \\
\hline \hline
\multicolumn{3}{c}{\footnotesize Percentage of total in parenthesis} \\
\end{tabular}
\label{table:AvailableMarriages}
\end{table}

Individuals are differentiated by their age and educational attainment. We divide educational attainment into three levels - high school diploma or less, some years of college or college degree, and graduate school.
%{\color{red} XXX(Simon): I have two remarks here: (i) is there any particular reason to choose this categorization? It seems to me that high school (sometimes split into below HS and HS degree), some college, and college (4 years and +) is more natural and widely used in the literature. (ii) are there any thin cell problems in the application? Could we increase the number of education categories?}
Where convenient, we will refer to these three groups with the abbreviation HS, Col and GS, respectively. Table \ref{table:AvailableMarriages} provides count statistics for our sample of single individuals by education in $1986$. These are constructed by taking the average of unmarried individuals from the twelve CPS monthly surveys in $1986$. There are around 16 million single women and around 14 million single men between the ages of 16 and 75 in our sample from the 22 reporting states.
There are dramatically fewer single adults with Col and GS educational attainment compared to those with HS education.
Around 63\% of single men and 65\% of single women has qualification up to a high school diploma. Only around 30\% of single men and women have some years in a college or a college degree and only 6\% of single men and 5\%  of single women have post-college qualification.

\begin{table}[ht!]
\centering
  \caption
  [Number of marriages by education in thousands \newline from the 1987/88 Vital Statistics for the 22 reporting states]
{\tabular[t]{@{}c@{}}Number of marriages by education in thousands  \\from the 1987/88 Vital Statistics for the 22 reporting states\endtabular}
\begin{tabular}{ c l c c c}
\hline\hline
%\multicolumn{5}{c}{Vital Statistics data in $1987/88$} \\
  & &  \multicolumn{3}{c}{Female} \\
  & & High School  & College  & Graduate School  \\\hline
\multirow{3}{*}{Male} &High School  &573.96 & 167.71 &11.35 \\
& College  & 153.47& 303.81 & 34.10 \\
 & Graduate School & 14.40& 53.21 & 40.39\\
\hline \hline
\end{tabular}
\label{table: numberMarriages}
\end{table}
Table \ref{table: numberMarriages}  tabulates the 1987/88 marriage distribution by education groups for the 22 reporting states. There is an average of 1.36 million marriages over this two year period. As evident from the table, there is strong assortative matching by education groups. With the exception of men with graduate school qualification, each of the remaining 5 groups of single men and women are most likely to marry a spouse with the same educational attainment. Men with graduate school qualification are more likely to marry college educated than graduate school educated women. This pattern does not hold for women with graduate school qualification.\footnote{
With age ranging from  $16$ to $75$ years and the 3 education groups, we have $180$ types (or age-education combination) of both men and women.
Since individuals who are younger than $23$ years of age rarely completed graduate school education, we exclude individuals younger than $23$ with graduate school qualification. This reduces the number of types from $180$ to $173$ for both men and women.}

Our methodology also requires data on $(n^{\prime}_x, m^{\prime}_y)$, the supply of single men and women had the student benefit program not been eliminated (that is, item ii) above). Since this counterfactual is unobserved to the econometrician, we use the estimates from \textcite{dynarski2003} to construct the counterfactual changes in available single men and women.
Using data from the CPS, \textcite{dynarski2003} proxies the benefits eligibility by the death of a parent during the individual's childhood.
The author employs the difference-in-differences framework to analyze the impact of the elimination of the Social Security Student Benefit program
on the probability of college attendance and completion for students who were eligible to the program.
Let the causal effect on college attendance and completion be denoted by $\gamma$ and $\delta$, respectively. \textcite{dynarski2003} finds that
eliminating the financial aid program on average reduces the probability of attending college by about $\hat{\gamma} = 24.3\%$ and the probability of completing any year of college by $\hat{\delta} = 16.1\%$ for the eligible students.\footnote{
Since the aid program was only imposed on students younger than 23 years old,
its effects on graduate study were negligible and ignored by \textcite{dynarski2003}.
We follow \textcite{dynarski2003} in  assuming that the aid program has no effect on the number of people who achieved graduate education.}

\begin{table}[ht!]
\caption{Effect on single pool had the student benefit program not been eliminated.}
\centering
\begin{tabular}{ c c l}
\hline\hline
Age in 82 & Age in 86 & Aid Effects \\
\hline
14  &   18 &    $\rho_{18}\cdot \gamma$ \\
15  &   19 &     $\rho_{18}\cdot \gamma+\rho_{19}\cdot \delta$  \\
16  &   20 &    $\rho_{18}\cdot \gamma+\rho_{19}\cdot \delta+\rho_{20}\cdot \delta$ \\
17  &   21 &    $\rho_{18}\cdot \gamma+\rho_{19}\cdot \delta+\rho_{20}\cdot \delta+\rho_{21}\cdot \delta$ \\
18  &   22 &    $\rho_{18}\cdot \gamma+\rho_{19}\cdot \delta+\rho_{20}\cdot \delta+\rho_{21}\cdot \delta$ \\
19  &   23 &    $ \rho_{19}\cdot \delta+\rho_{20}\cdot \delta+\rho_{21}\cdot \delta$ \\
20  &   24 &    $\rho_{20}\cdot \delta+\rho_{21}\cdot \delta$ \\
21  &   25 &    $\rho_{21}\cdot \delta$ \\
\hline \hline
\end{tabular}
\label{table:aidEffects}
\end{table}

Since these estimates are for eligible students and we do not observe actual benefit recipients in our data, we need to also compute the fraction of the population who are eligible for these benefits.
The program did not differentiate between male or female recipients.
Hence, we construct the counterfactual numbers of available singles without distinguishing their gender.
Using the similar approach as in \textcite{dynarski2003}, we proxy the proportion of age $i$  benefit eligible individuals (males or females), $\rho_i,$ by the fraction of age $i$ individuals  whose father are deceased, retired or disable in their cohort from the $1980$ U.S. census. We assume that individuals (males or females) are in high school till they are 18 and that a college degree takes 4 years.% The policy change affects the number of senior high school students attending colleges from $1982$ and $1986$.
%Thus it affects the available numbers of men and women with high school or less and some years college or college degree from ages $18$ to $22$ in $1986$.

Consider now a counterfactual setting where the financial aid program had not been eliminated. Eighteen  years old high school seniors  in 1986 would have been fourteen years old  high schoolers in 1982 when the program was eliminated. Our estimates suggest that an additional $\rho_{18}\gamma$ proportion of high school graduates would have attended college had the program not been eliminated. As for nineteen year olds in 1986, a $\rho_{18}\gamma$ fraction of these individuals would have attended college when they were eighteen years old and $\rho_{19}\delta$ of them would not have dropped out of college that year. Hence, an additional $(\rho_{18}\gamma+\rho_{19}\delta)$ of the nineteen years old high school graduates would have attended college in 1986. We repeat this calculation for individuals aged between eighteen and twenty five years old in our 1986 supply of single men and women. Table \ref{table:aidEffects} tabulates the calculations of changes to the population of single high school graduates by age in the counterfactual setting.

\begin{table}[ht!]
\caption{Available number of single men and women  between ages 18 and 22 (millions)}
\centering
\begin{tabular}{ l c c c}
\hline\hline
&\multicolumn{3}{c}{Male} \\
&  CPS in $1986$ &  Counterfactual Policy & \% Change\\
High school or less                    &  2.52  & 2.36  & -6.12\% \\
College                                &  1.23  & 1.39  & 12.50\%  \\
\hline
&\multicolumn{3}{c}{Female} \\
&  CPS in $1986$ &  Counterfactual Policy & \% Change\\
High school or less                    &  2.05  & 1.93 &  -5.90\% \\
College                                &  1.28  & 1.40 &  9.47\%  \\
\hline \hline
\end{tabular}
\label{table:Available1822}
\end{table}

%\begin{figure}[ht!]
%\includegraphics[scale=0.7,trim = 0 0 100 400, width=\textwidth]{FFG_MFSupp.pdf}
%\caption{Changes in available HS and Col single men and women}
%\label{fig:CompareS}
%\end{figure}

Table \ref{table:Available1822} compares the observed  number of single men and women between the ages of 18 and 22 with the counterfactuals computed using the procedure just outlined. While the estimated causal effect in \textcite{dynarski2003} was statistically significant on those eligible for the benefits, the overall effect of the program elimination on the number of single men and women remains modest due to the small fraction of eligible individuals in the population. Our calculation suggests that the number of college graduated men and women between the ages of 18 and 22 would have increased by approximately 136,000 and 121,000 respectively. This represents an increase of around 12.5\% and 9.47\%  more college educated men and women aged between the ages of 18 and 22, respectively.
Figure \ref{fig:CompareS} in Online Appendix \ref{app:figures} shows the observed and counterfactual available single men and women by age between 16 and 22 in 1986.

\subsection{Implementations and specifications}

We employ the nested maximum likelihood approach proposed in section \ref{sec:MLE} to estimate seven candidate parameterizations  of  the ETU model in example e) of section \ref{sec: examples}.\footnote{We could also estimate the models using moment matching techniques.
For a discussion on the use of maximum likelihood and moment matching techniques in matching models, see \textcite{galichon2012}.}
These different parameterizations allow us to consider different candidate models from  our general class of matching
functions that best fit out data. Using information criteria, we then select the  best fitting model. With the matching surpluses estimated, we  solve for the counterfactual number of single individuals $(\mu^{\prime}_{x0}, \mu^{\prime}_{0y}),$ by substituting the estimated matching surpluses
into the system \eqref{sec: count_nonlinear_system}, and calculating the counterfactual matches $\mu^{\prime}_{xy}$ from the matching function accordingly.

Recall that the ETU model delivers the following harmonic mean matching function,
\begin{equation}
M_{xy}(\mu _{x0},\mu _{0y})=\left[\frac{\exp \left(\frac{-\alpha_{xy}}{\kappa_{xy}}\right)\mu_{x0}^{-1/\kappa_{xy}}  +
\exp \left(\frac{-\gamma_{xy}}{\kappa_{xy}}\right)\mu_{0y}^{-1/\kappa_{xy}}}{2} \right]^{-\kappa_{xy}}, \label{MF:Schoen2}
\end{equation}
where $\kappa_{xy}$ captures the degree of utility transferability.\footnote{Note that we recover the matching function in the NTU model of \textcite{GalichonHsieh2017} as $\kappa_{xy} \rightarrow 0$ and that in the TU model of CS as $\kappa_{xy} \rightarrow \infty$.}
We estimate $\alpha_{xy}$ and $\gamma_{xy}$ in \eqref{MF:Schoen2}
given six different values of $\kappa_{xy}$:
$\big\{\kappa_{xy} \rightarrow 0, \kappa_{xy}=\{0.1,1,10,100\}, \kappa_{xy} \rightarrow +\infty \big\}$ for all $x \in \mathcal{X}$ and $y \in \mathcal{Y}$.
Moreover, we also estimate $\alpha_{xy}$ and $\gamma_{xy}$ in the case where we treat $\kappa_{xy}$ as an unknown parameter to be estimated (assuming that $\kappa_{xy}=\kappa$, $\forall\,\, x y\in\mathcal{XY}$).
In total, we estimate seven candidate parameterizations  followed by model selection.

To parameterize $\alpha_{xy}$ and $\gamma_{xy},$ we need to discretize the male and female types as captured by age and education. We denote the types of men and women by $x=(x_a, x_e),$ and $y=(y_a, y_e),$ respectively, where age is captured by $(x_a, y_a)$ and education by  $(x_e, y_e).$ The age of individuals goes from 16 to 75 in increments of one. We label this by $(x_a, y_a) \in  \{1, \cdots, 60\}.$ Educational attainment are divided into three categories, high school diploma or less, some years of college or college degree, and graduate school, which we label as $(x_e, y_e) \in  \{1, 2, 3\}$, respectively.
% \begin{comment}
% We parameterize $\alpha_{xy}$ and $\gamma_{xy}$ as follows.
% Let the type of a man be defined as $x=(x_a, x_e)$, where $x_a \in  \{16, \cdots, 75\}$ (we relabel it as $x_a \in  \{1, \cdots, 60\}$)
% denotes the man's age and $x_e \in  \{hs, col, grad\}$ (we relabel it as $x_e \in  \{1, 2, 3\}$) denotes the man's education level.
% Similarly, we define the type of a woman as $y=(y_a, y_e)$ with $y_a$ the age of the woman relabelled as $y_a \in  \{1, \cdots, 60\}$
% and $y_e$ her education relabelled as $y_e \in  \{1, 2, 3\}$.
% \end{comment}
Our parametric specifications for $\alpha_{xy}$ and $\gamma_{xy}$ are given by
 \begin{align}
 \label{eq: specific_surplus_men}
\alpha_{xy}({\Lambdav})&=\Lambda_0+ \sum^{60}_{i=2} \Lambda^{ma}_{i} \mathbbm{1}\{x_a=i\}+\sum^{3}_{j=2} \Lambda^{me}_j \mathbbm{1}\{x_e=j\}
+ \sum^{60}_{i=2} \Lambda^{wa}_i \mathbbm{1}\{y_a=i\}+\sum^{3}_{j=2} \Lambda^{we}_j \mathbbm{1}\{y_e=j\} \nonumber \\
&+\sum^{59}_{i=1} \Lambda^{mwa}_i\mathbbm{1}\{|x_a-y_a|=i\} +\sum^{2}_{j=1} \Lambda^{mwe}_j\mathbbm{1}\{|x_e-y_e|=j\}, \\
 \label{eq: specific_surplus_women}
\gamma_{xy}({\Gammav})&=\Gamma_0+ \sum^{60}_{i=2} \Gamma^{ma}_{i} \mathbbm{1}\{x_a=i\}+\sum^{3}_{j=2} \Gamma^{me}_j \mathbbm{1}\{x_e=j\}
+ \sum^{60}_{i=2} \Gamma^{wa}_i \mathbbm{1}\{y_a=i\}+\sum^{3}_{j=2} \Gamma^{we}_j \mathbbm{1}\{y_e=j\} \nonumber \\
&+\sum^{59}_{i=1} \Gamma^{mwa}_i\mathbbm{1}\{|x_a-y_a|=i\} +\sum^{2}_{j=1} \Gamma^{mwe}_j\mathbbm{1}\{|x_e-y_e|=j\}.
\end{align}
The second and third terms in each equation capture the surplus from male traits,
the fourth and fifth terms capture the surplus from female traits,
and the sixth and seventh terms capture the surplus from the interactions of male and female traits.\footnote
{
This specification is quite flexible as it includes a number of parameters.
Although the interaction terms enter  the specification in absolute values,
this specification is able to partially capture asymmetric effects.
The men's preferences for marrying someone one level more educated and one level less educated are different, holding all else equal.
For instance, a man's surplus for a match with male traits $x=(x_a=6, x_e=2)$ and female trait $y=(y_a=8, y_e=3)$
is given by $\alpha^1_{xy}(\Lambdav)=\Lambda_0+\Lambda^{ma}_{6}+\Lambda^{me}_{2}+\Lambda^{wa}_{8}+\Lambda^{we}_{3}+\Lambda^{mwa}_{2}+\Lambda^{mwe}_{1}$,
while a man's surplus for a match with man's traits $x=(x_a=6, x_e=2)$ and woman's trait $y=(y_a=8, y_e=1)$
is given by $\alpha^2_{xy}(\Lambdav)=\Lambda_0+\Lambda^{ma}_{6}+\Lambda^{me}_{2}+\Lambda^{wa}_{8}+\Lambda^{we}_{1}+\Lambda^{mwa}_{2}+\Lambda^{mwe}_{1}$.
$\alpha^1_{xy}(\Lambdav)$ and $\alpha^2_{xy}(\Lambdav)$ have different fixed effects terms for women's education.
We also attempted to estimate the models using a specification with asymmetric interaction terms
and the estimated joint systematic surpluses have similar qualitative features.
See the Online Appendix \ref{app:asy_specification} for the asymmetric specification and estimated joint systematic surpluses using it.
}
The specification requires us to estimate  368 parameters. That is, the vectors $\Lambdav$ and $\Gammav$ each consist of 184 parameters, and the vector of parameters to be estimated is $\thetav = \{\Lambdav, \Gammav \}$.\footnote{
It is well known that only the joint surpluses $\alpha_{xy}+\gamma_{xy}$ are identified in the TU model of CS, in which case we only have 184 parameters to estimate.}

\subsection{Model selection}

\begin{figure}[t!]
        \centering
        \includegraphics[height=3in]{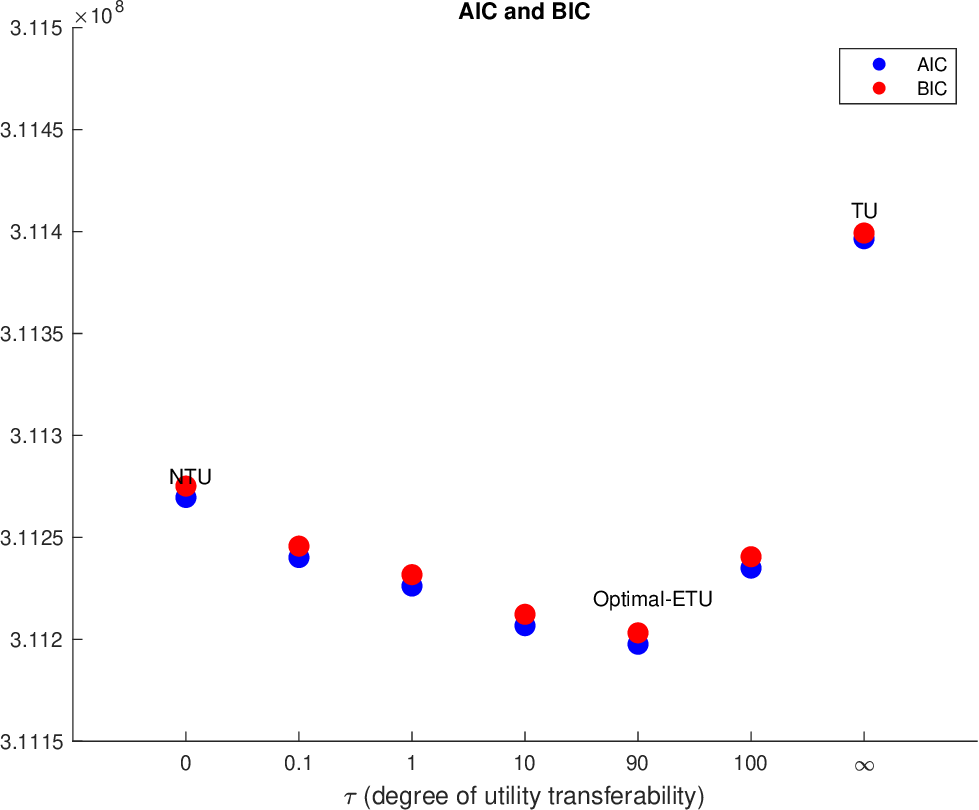}
        \caption{AIC and BIC}
         \label{figure2: AIC}
\end{figure}

After estimating these parameters,
we conduct our model selection procedure across our seven candidate models using the Akaike Information Criterion (AIC) and Bayesian Information Criterion (BIC).\footnote{$AIC=2k-2\ln(\hat{L})$ and  $BIC=k \ln(n) - 2\ln(\hat{L})$,
where $k$ is the number of parameters, $\hat{L}$ is the estimated maximum
value of the likelihood, and $n$ is the sample size for the model.
}
Figure \ref{figure2: AIC} reports the AIC and BIC for the seven candidate models.
The model where $\kappa_{xy}$ has  to be estimated, yields the largest log-likelihood and is our preferred  model.
This ITU model, which we label as the `optimal-ETU' model, generates the lowest AIC and BIC relative to those in the NTU and TU models.
The `optimal-ETU' model has
$\widehat{\kappa}_{xy} \approx 90.$
 Since $\kappa_{xy}$ bridges the NTU model at one end ($\kappa_{xy} \rightarrow 0$), and the TU model at the other ($\kappa_{xy} \rightarrow +\infty$),
our model selection procedure is hence selecting between three models: the TU, optimal-ETU, and NTU models.
The TU model generates the largest AIC and BIC.
We highlight the importance of our model selection procedure
%{\color{blue}
in the following sections, where we compare the differences in the  estimated joint surpluses with the  in-sample predicted matching distributions, and
their corresponding counterfactual matching distributions.
%}

\subsection{The estimated joint systematic surpluses}

\begin{figure}[!htb]
\minipage{0.32\textwidth}
  \includegraphics[width=\linewidth]{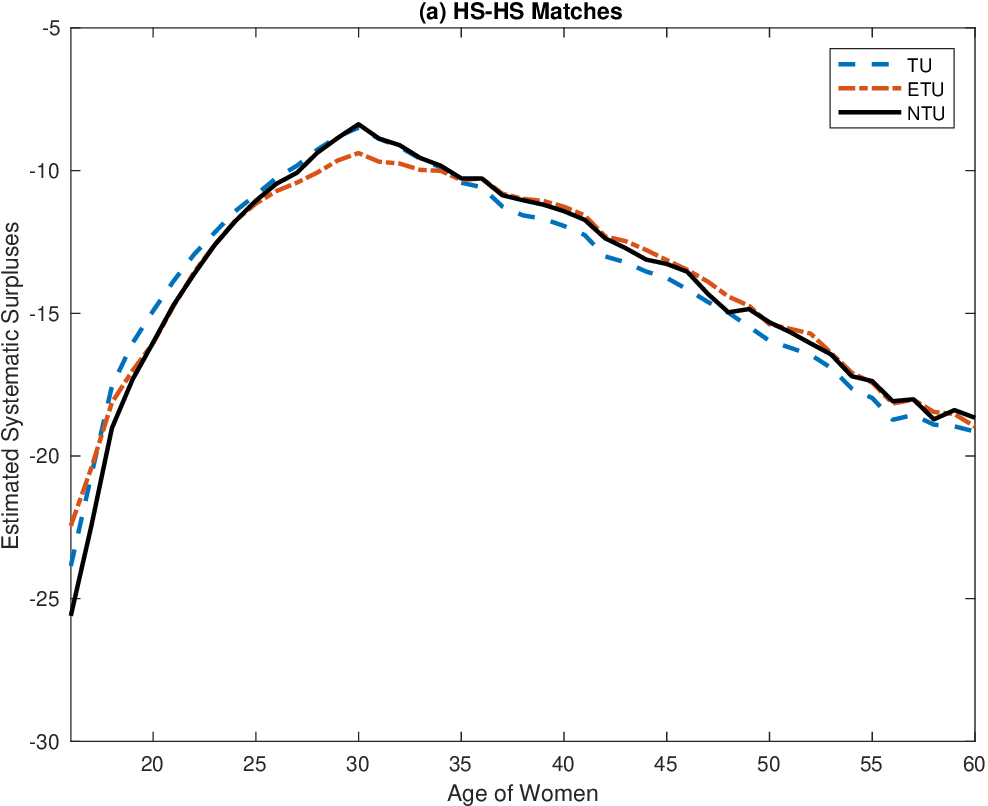}
%  \caption{A really Awesome Image}\label{fig:awesome_image1}
\endminipage\hfill
\minipage{0.31\textwidth}
  \includegraphics[width=\linewidth]{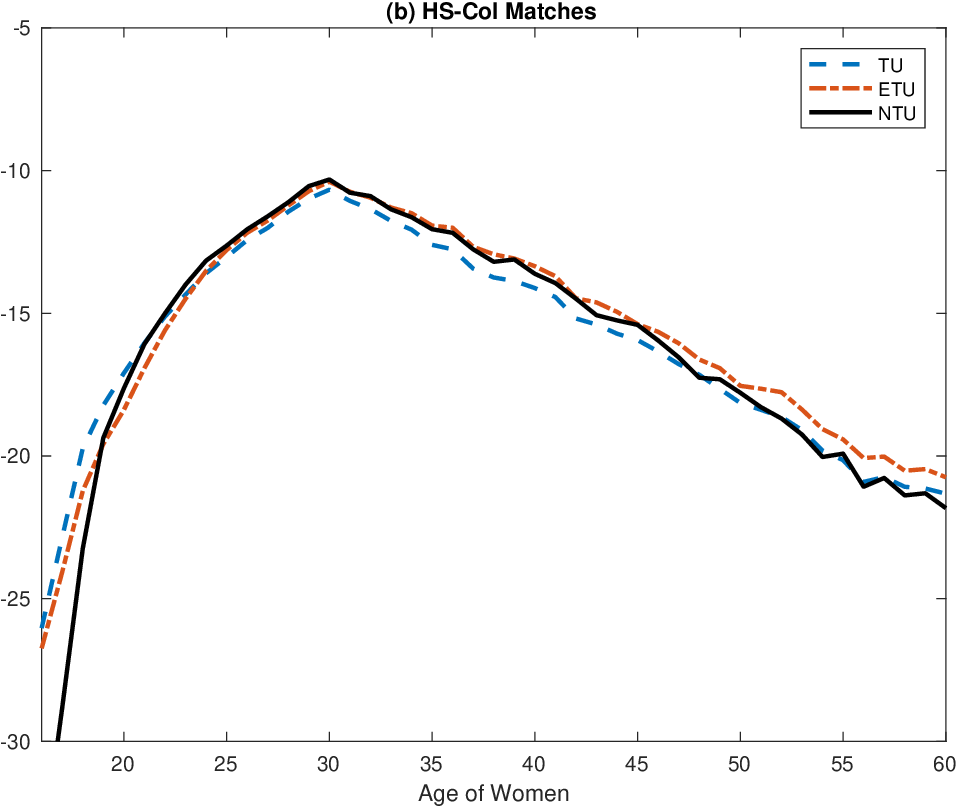}
%  \caption{A really Awesome Image}\label{fig:awesome_image2}
\endminipage\hfill
\minipage{0.31\textwidth}%
  \includegraphics[width=\linewidth]{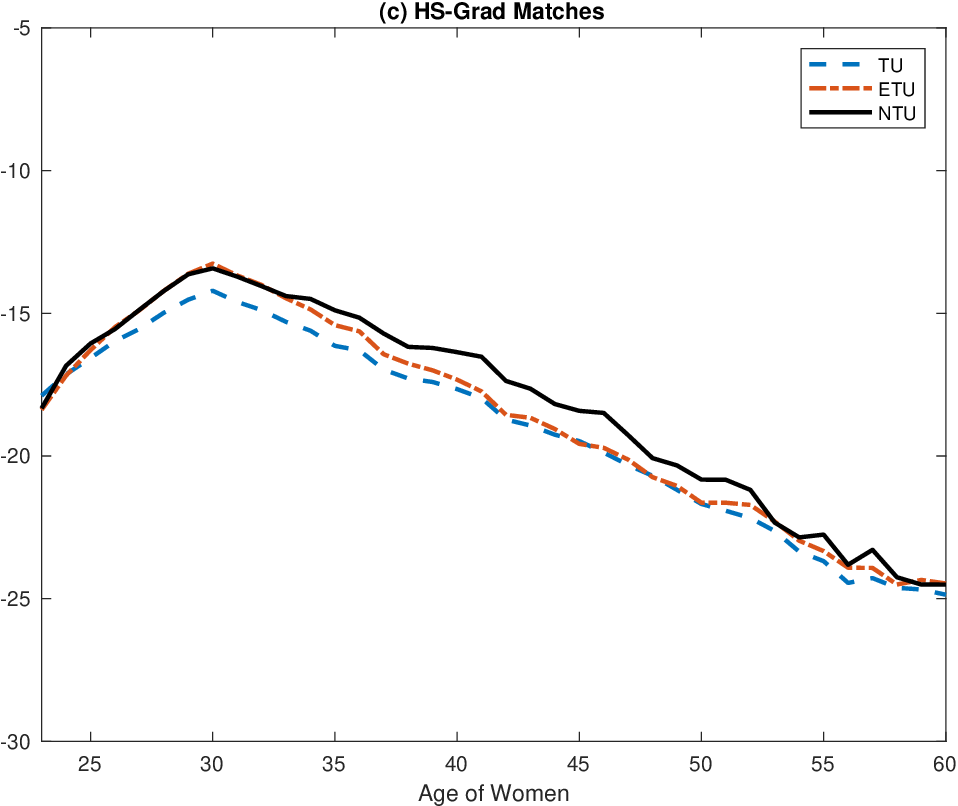}
%  \caption{A really Awesome Image}\label{fig:awesome_image3}
\endminipage
\centering
\caption{Estimated $U_{xy}(\hat{\Lambdav})+V_{xy}(\hat{\Gammav})$ for high school men of 30 years old}
\label{fig: joint_surplus_hs}
\end{figure}

We first compare the  estimated joint systematic surpluses for the TU, optimal-ETU and NTU models.
In figure \ref{fig: joint_surplus_hs}, we graph the joint systematic utilities estimates, $U_{xy}(\hat{\Lambdav})+V_{xy}(\hat{\Gammav})$, for a 30 years old, high school educated male matched with a partner of different ages and educational attainment.\footnote{The qualitative features of the joint systematic utilities estimates for college and above college educated male are similar, and can be found in figures \ref{fig: joint_surplus_col} and \ref{fig: joint_surplus_grad} of the Online Appendix \ref{app:figures}.}
These figures highlight two main features:
(1) the estimates for all three models exhibit strong assortativeness in education
and age. That is, couples with similar education and age generally obtain the highest joint systematic utilities;
(2) the estimates from the three models track each other very closely across age and education, with the estimates from the optimal-ETU estimates generally lying in between the estimates from the NTU and TU model.

\begin{figure}[!htb]
\minipage{0.32\textwidth}
  \includegraphics[width=\linewidth]{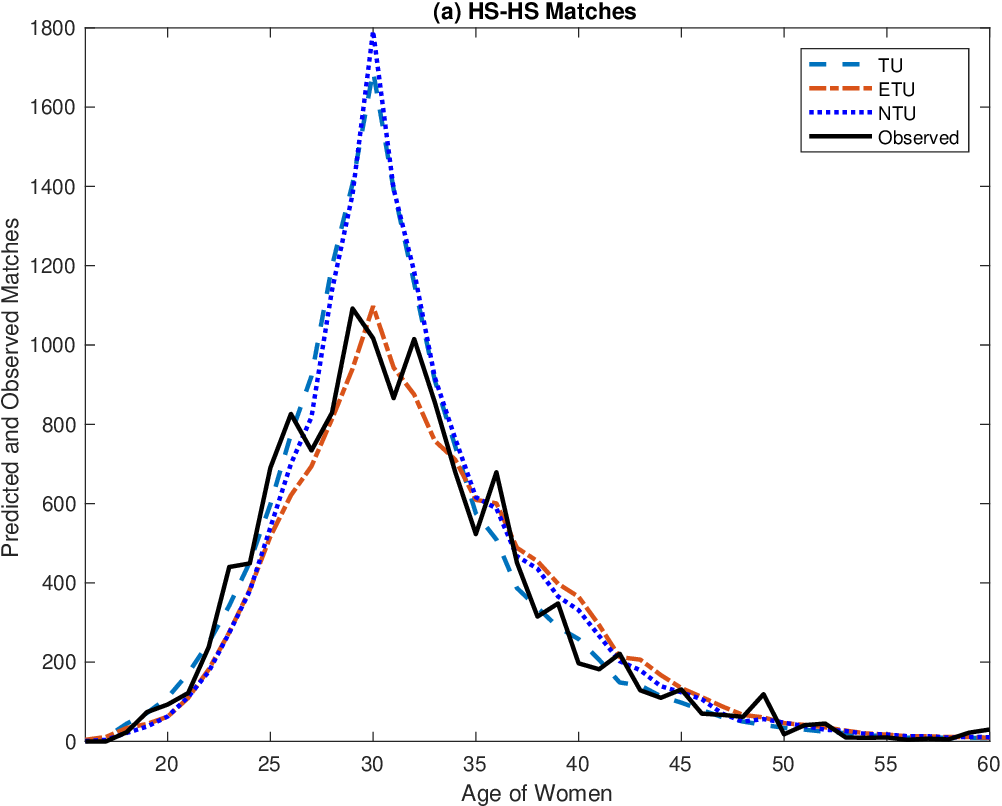}
%  \caption{A really Awesome Image}\label{fig:awesome_image1}
\endminipage\hfill
\minipage{0.31\textwidth}
  \includegraphics[width=\linewidth]{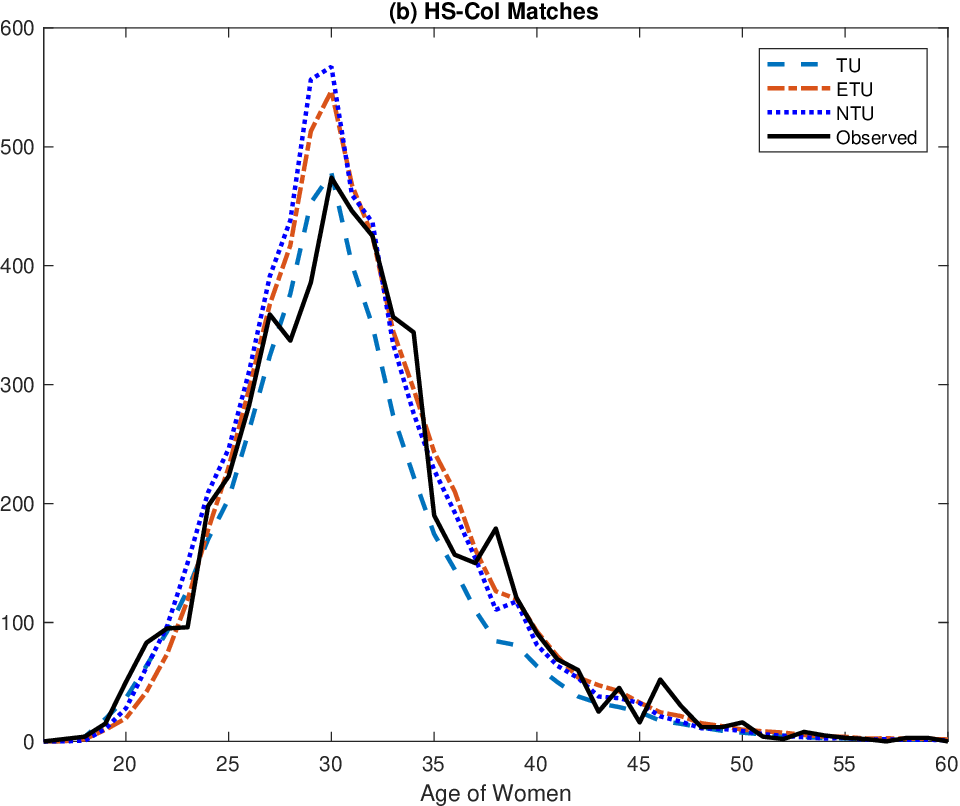}
%  \caption{A really Awesome Image}\label{fig:awesome_image2}
\endminipage\hfill
\minipage{0.31\textwidth}%
  \includegraphics[width=\linewidth]{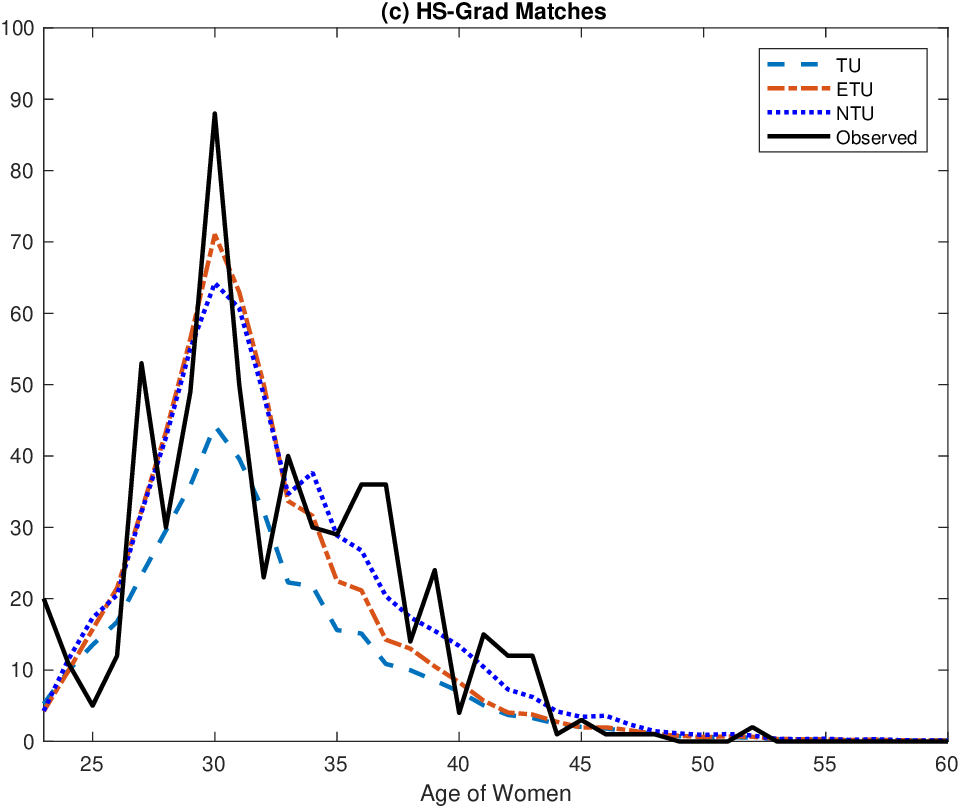}
%  \caption{A really Awesome Image}\label{fig:awesome_image3}
\endminipage
\centering
\caption{Predicted and observed matches for high school men of 30 years old}
\label{fig: matches_hs}
\end{figure}

% \begin{comment}
% (2) the joint surpluses estimated from the parametric specification are smoother than those estimated from nonparametric specification, especially for marriages involving old GS educated women.
% This is due to the fact that we observe very thin cells for the marriages between
% high school men with old or GS educated women,\footnote{There are about 63\% zero cells for the number of marriages between high school men at 25,30, or 35 years old and women older than 55 years old, and for the marriages between high school men of the three ages with graduate school women, there are about 49.7\% zero cells in the observed data.}
% which gives us imprecise nonparametric estimates for the joint surpluses.
% \end{comment}

\subsection{Model fitting}
In this section, we compare  the goodness of fit for the TU, optimal-ETU and NTU models.
We first predict the matches under the observed supply of men and women using the estimated parameters from the three models.
We compare these with the observed matches in the data.
In figure \ref{fig: matches_hs}, we graph the observed and predicted number of matches  for 30 years old, high school educated men, matched with a partner of different ages and educational attainment.
These graphs highlight a salient feature: the numbers of predicted matches from all three models exhibit a pattern similar to the observed matches, with the optimal-ETU model prediction being the closest.
Figures \ref{fig: matches_col} and \ref{fig: matches_grad} provided in the Online Appendix \ref{app:figures}
suggest similar qualitative features.

\subsection{The changes in marriage distributions due to the policy change}

We now consider how the elimination of the Social Security Benefit Program in 1982 has affected the marriage distributions in the United States.
To do so,
we first estimate the counterfactual marriage distributions for the 22 reporting states using our three models.
We then compare them with the predicted actual marriage distributions from the three models, to obtain the changes in the marriage distributions brought about by the elimination of the Social Security Benefit Program in 1982.\footnote{
The changes in matches between the counterfactual and observed marriage distributions
compound the effects of the program elimination on population supply and the model restrictions.}
 While the elimination of the aid program would have affected the marriage distribution of the whole United States, our analysis is unfortunately confined to the 22 reporting states for which we have data on.

\begin{figure}[!htb]
\minipage{0.32\textwidth}
  \includegraphics[width=\linewidth]{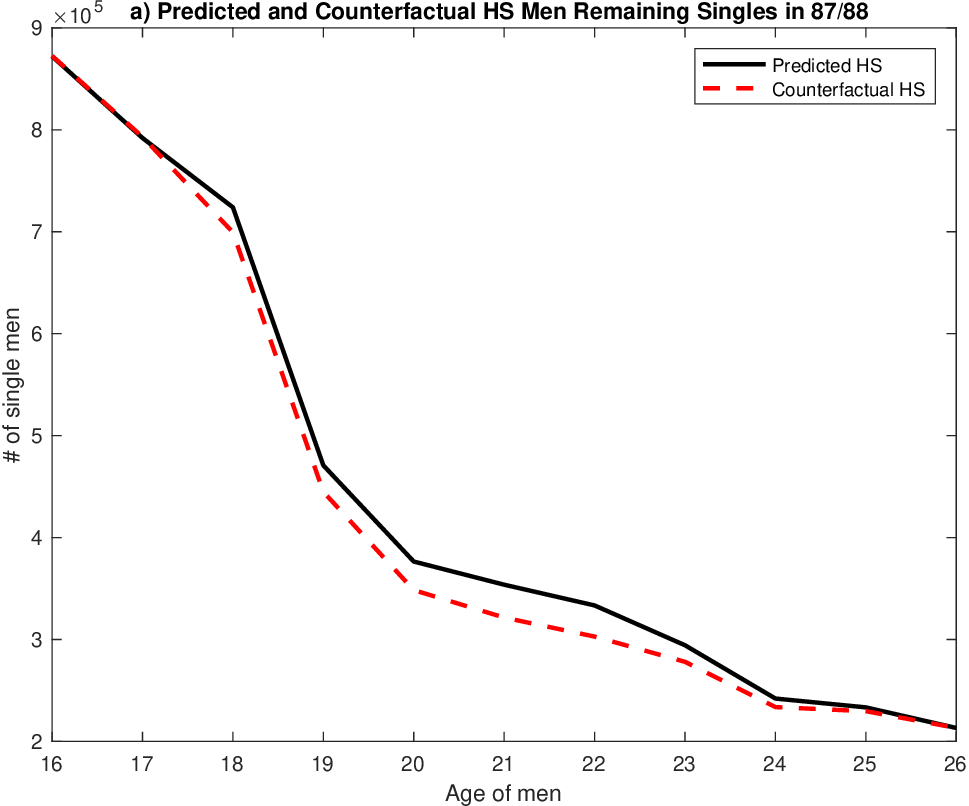}
%  \caption{A really Awesome Image}\label{fig:awesome_image1}
\endminipage\hfill
\minipage{0.32\textwidth}
  \includegraphics[width=\linewidth]{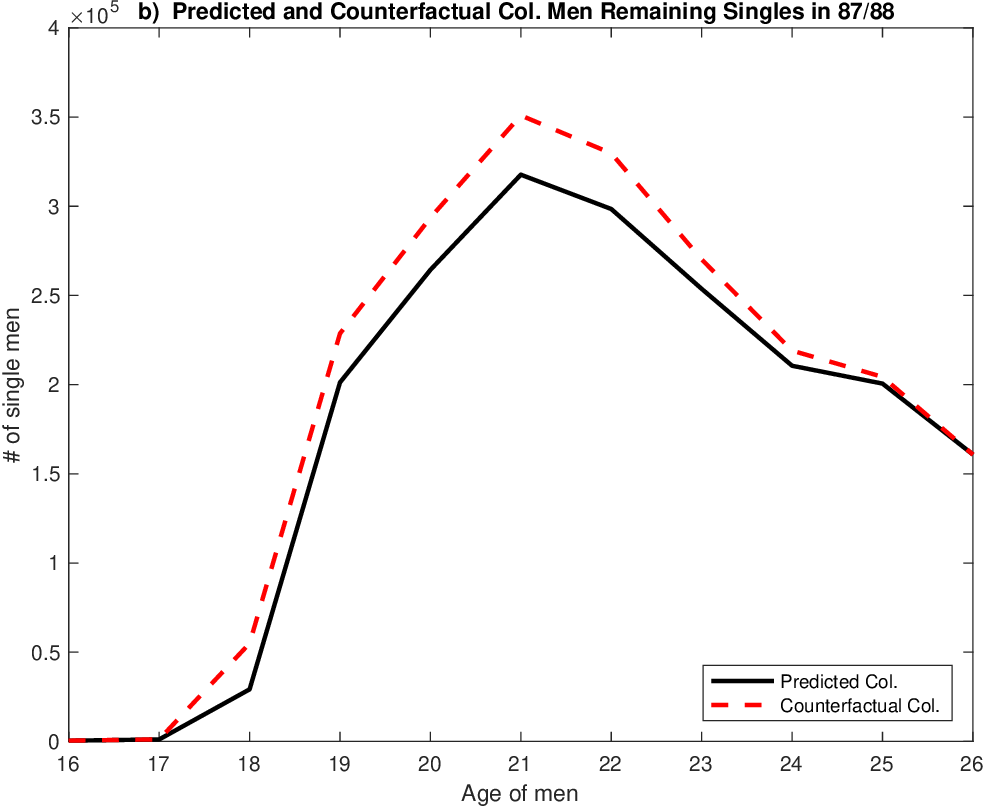}
%  \caption{A really Awesome Image}\label{fig:awesome_image2}
\endminipage\hfill
\minipage{0.32\textwidth}%
  \includegraphics[width=\linewidth]{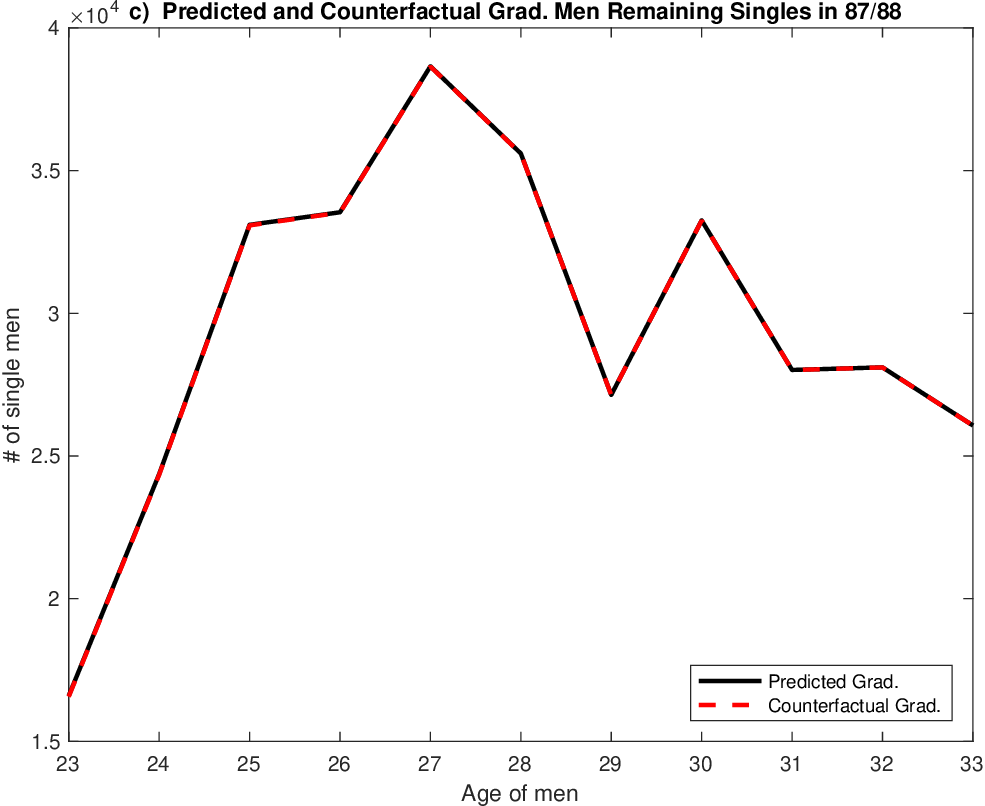}
%  \caption{A really Awesome Image}\label{fig:awesome_image3}
\endminipage\hfill
\minipage{0.32\textwidth}
  \includegraphics[width=\linewidth]{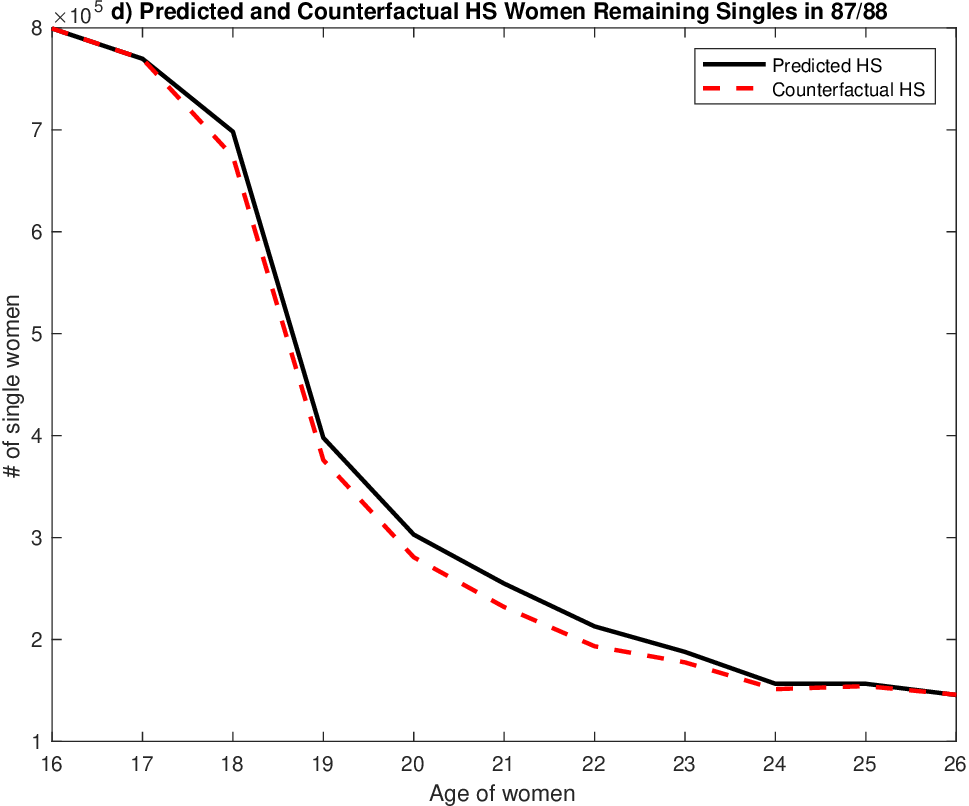}
%  \caption{A really Awesome Image}\label{fig:awesome_image1}
\endminipage\hfill
\minipage{0.32\textwidth}
  \includegraphics[width=\linewidth]{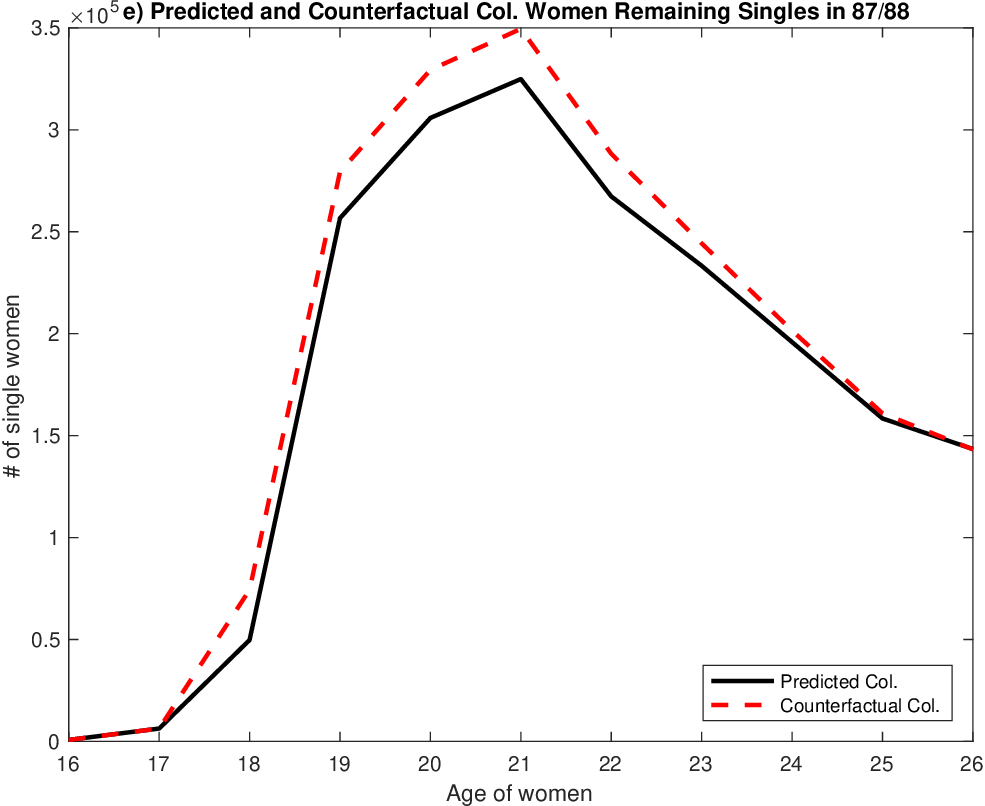}
%  \caption{A really Awesome Image}\label{fig:awesome_image2}
\endminipage\hfill
\minipage{0.32\textwidth}%
  \includegraphics[width=\linewidth]{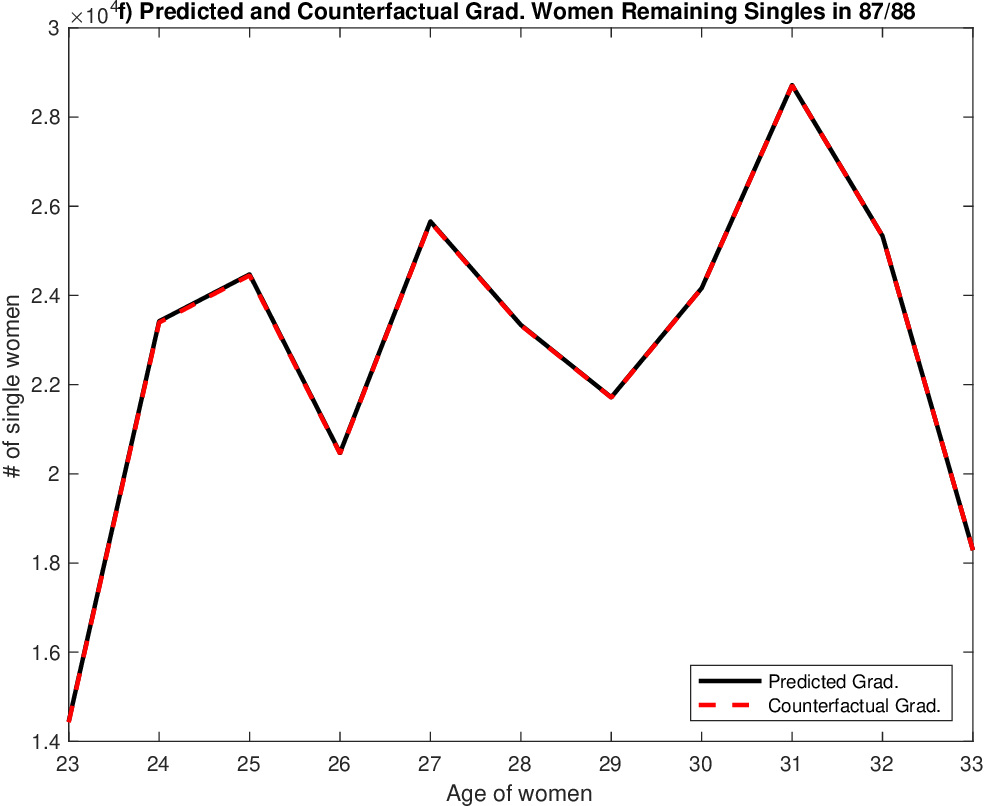}
%  \caption{A really Awesome Image}\label{fig:awesome_image3}
\endminipage
\centering
\caption{Predicted and counterfactual numbers of men and women remaining singles \\
\centering (ETU-Optimal Model)}
\label{fig: pred_count_singles_ETU}
\end{figure}

Figure \ref{fig: pred_count_singles_ETU} displays the equilibrium numbers of individuals remaining singles by age and education
in both the predicted and counterfactual marriage distributions estimated from the ETU-Optimal model.\footnote{
The corresponding marriage distributions estimated from the TU and NTU models are provided in figures \ref{fig: pred_count_singles_TU} and
\ref{fig: pred_count_singles_NTU}, which exhibit similar patterns.}
Our estimates suggest that had the financial aid program not been eliminated, there would  have been less high-school educated adults remaining singles, and more college educated adults remaining singles. There would be very little change to the equilibrium numbers of individuals with above college educational attainment remaining singles.
This is consistent with the impacts of the financial aid program on the available numbers of single individuals,
shown in figure \ref{fig:CompareS} in Online Appendix \ref{app:figures}.

\begin{table}[ht!]
\caption{Changes in Number of New Marriages by Education}
\centering
\begin{tabular}{ c l c c c}
\hline\hline
%\multicolumn{5}{c}{Change in Marriages in $1987/88$ } \\
   \multicolumn{5}{c}{Optimal-ETU Model} \\
\hline
  & &  \multicolumn{3}{c}{Female} \\
  & & High School  & College  & Grad. School  \\\hline
 & High School     &  -17150.8 (-3.00\%)  &  -291.6 (-0.17\%) &  -66.3 (-0.59\%)\\
Male & College     &  1527.6 (0.98\%)     &  8154.8 (2.68\%) &  213.3(0.07\%)\\
 & Grad. School    &  -53.0 (-0.36\%)      & 189.8 (0.37\%)    & -11.0 (-0.03\%)\\
\hline \hline

   \multicolumn{5}{c}{TU Model} \\
\hline
  & &  \multicolumn{3}{c}{Female} \\
  & & High School  & College  & Grad. School  \\\hline
 & High School     &  -14572.8 (-2.55\%)  &  -356.9 (-0.21\%) &  -114.5 (-1.13\%)\\
Male & College     &  2191.9 (1.42\%)     &  11801.9 (3.87\%) &  535.6 (0.18\%)\\
 & Grad. School    &  -74.3 (-0.48\%)      & 246.0 (0.49\%)    & -23.8 (-0.06\%)\\
\hline \hline
%\multicolumn{5}{c}{Change in Marriages in $1987/88$ } \\
 \multicolumn{5}{c}{NTU Model} \\
\hline
  & &  \multicolumn{3}{c}{Female} \\
  & & High School  & College  & Grad. School  \\\hline
 & High School     &  -13895.4 (-2.43\%)  &     -981.6 (-0.58\%) &  -110.1 (0.95\%)\\
Male & College     &  385.7 (0.25\%)     &  10555.2(3.46\%) &  333.7 (0.11\%)\\
 & Grad. School    &  102.3 (-0.73\%)      & 527.3(1.03\%)    & -34.1 (-0.08\%)\\
\hline \hline

\end{tabular}
\label{table: changeMarriages}
\end{table}

Table \ref{table: changeMarriages} tabulates the changes in the marriage distributions by education, between the predicted and the counterfactual
distribution as estimated from the three models.
%\footnote{
%Recall that the nonparametric specification and parameter-free approach provide identical estimates of the counterfactual marriage distribution.
%We thus omit the results from the parameter-free approach in Table \ref{table: changeMarriages}.}
%The continuation of this aid program would have increased the number of available college educated single men and women and decreased the number of available high school educated single men and women.
We  use  HS-Col to refer to marriages between a high school educated male and a college educated female.\footnote{The types of marriages HS-HS, HS-Grad, Col-HS, Col-Col, Col-Grad, Grad-HS, Grad-Col and Grad-Grad follow the same convention.}
Since the optimal-ETU model provides the best fit of the data, we will focus our discussion on its estimates in the top panel of table \ref{table: changeMarriages}. The high degree of positive assortative matching by age and education implies that we expect to see the biggest changes
among couples with the same educational attainment, that is in HS-HS, Col-Col and Grad-Grad matches.
Had the Social Security Benefit Program not been eliminated in 1982,
the marriage distribution in 1987/88 would have seen a decrease in  HS-HS marriages of around 17,150 (3.00\%) matches,  an increase in Col-Col marriages of around 8,154 (2.68\%), and an decrease in Grad-Grad marriages of 11 (0.03\%).
A large change is also seen among Col-HS marriages, which would increase by around 1,527  (0.98\%)  compared to the small decrease of around 292 (0.17\%) marriages among HS-Col marriages. This is probably a reflection of social norms of men preferring spouses who are not more educated than themselves, which is embedded in the preference parameters.

The middle and bottom panels of table \ref{table: changeMarriages} provide the estimates using the TU and NTU models.
The qualitative result for couples with the same educational attainment is similar to the estimates from the optimal ETU model. We observe the biggest difference in `off-diagonal' marriage involving individuals with above college educational attainment, that is in HS-GS, Col-GS, GS-HS and GS-Col marriages.
This is a reflection of large differences in estimates across the three models in the preference parameters for GS educated individuals.

\section{Conclusion}\label{sec:con}

In the context of matching models, researchers are often interested in how the equilibrium matching distribution
would change in response to a change in the structure of the market (e.g. for marriage markets, changes in the number of available men or women). We study this question in matching function equilibrium models, a class of matching models that's characterized by a matching function and a system of demographic constraints. We point out that a surprisingly large number of models in the matching literature belong to this class. In this paper, we focus on the partial assignment case. We show how one can parametrically estimate the matching functions of these
models by maximum likelihood; we provide efficient computing techniques, an analytic expression for the gradient of the log-likelihood, and formulas to compute confidence intervals.

We study counterfactuals arising from policy changes that affect the number of available men and women on the market but  leave the matching surplus parameters unchanged.  We show how to compute the counterfactual equilibrium matching distribution when the structural parameters of the matching function have been previously estimated. In addition, for a certain subclass of matching function equilibrium models, we propose a new parameter-free approach of identifying the counterfactual matching equilibrium. We illustrate our framework by
analyzing the impact of the elimination of the Social Security Student Benefit Program in
1982 on the marriage market.
We show that, had the Social Security Benefit Program not been eliminated in 1982, there would have been around
%{\color{blue}
17,150 (3.00\%)
%}
fewer marriages among high school educated individuals,
%{\color{blue}
8,154 (2.68\%),
%}
more marriages among college educated individuals, and
%{\color{blue}
a negligible 11 (0.03\%) less
%}
marriages among individuals with above college educational attainment in 1987/88 in the 22 reporting U.S. states for which we have data.
\newpage
\printbibliography

\newpage
\appendix

\section{Proofs}\label{app:proofs}
\subsection{Proof of theorem \protect\ref{thm:system} \label{app:proofs_thm1}}

\begin{proof}
\underline{Part $(i)$. Proof of Existence}.
The proof of existence is similar as that in the proof of theorem 1.
It relies on a generalized IPFP algorithm, algorithm \ref{IPFP Algorithm}, provided in main text.
We show that $(\muxz^\ast,\muzy^\ast)$ defined by system \eqref{NL: theorem6} converges to a solution under assumption \ref{ass: MFE}.
The proof consists of two steps.

First, we show that the construction of $\mu^{2t}_{x0}$ for all $x \in \mathcal{X}$,
$\mu^{2t}_{0y}$ for all $y \in \mathcal{Y}$, 
$\mu^{2t+1}_{x0}$ for all $x \in \mathcal{X}$,
and 
$\mu^{2t+1}_{0y}$ for all $y \in \mathcal{Y}$,
is well defined at each step. 
Consider step $2t$ for $t \ge 1$.
For each $x \in \mathcal{X}$, recall that the equation to solve for $\mu^{2t}_{x0}$ given
$\mu^{2t-2}_{x^\prime 0}$ for all $x^\prime  \neq x$,
$\mu^{2t-1}_{0y}$ and $\mu^{2t-1}_{0y^\prime }$ for all $y^\prime \neq y$,
\begin{align}
\label{balance_n_extended}
\textstyle
\mu^{2t}_{x0}+\sum_{y\in \mathcal{Y}}M_{xy}(\mu^{2t}_{x0},\{\mu_{x^\prime 0}^{2t-2}\}_{x^\prime \neq x},\mu^{2t-1}_{0y}, \{\mu_{0y^\prime}^{2t-2}\}_{y^\prime \neq y})=n_{x}
\end{align}
Under assumption \ref{ass: MFE},
the left-hand side of above equation is a continuous and increasing function of $\mu^{2t}_{x0}$.
It tends to $0$ when $\mu^{2t}_{x0} \rightarrow 0$,
and tends to $+\infty$ when $\mu^{2t}_{x0} \rightarrow +\infty$.
Hence $\mu^{2t}_{x0}$ is well defined and belongs in $(0, +\infty)$ for all $x \in \mathcal{X}$.
Similarly, it can be shown that $\mu^{2t+1}_{x0}$ is also well defined and belongs in $(0, +\infty)$ for all $x \in \mathcal{X}$,
 $\mu^{2t}_{0y}$ and  
 $\mu^{2t+1}_{0y}$
 are also well defined and belong in $(0, +\infty)$ for all $y \in \mathcal{Y}$.
 
Second, we show the mapping from $(\muxz^{2t-1}, \muzy^{2t-1}, \muxz^{2t},\muzy^{2t})$ to $(\muxz^{2t+1}, \muzy^{2t+1})$ are isotone in $(\muxz^{2t-1}, \muzy^{2t-1})$ and antitone in $(\muxz^{2t},\muzy^{2t})$.
Denote the mapping $\mu_{x0}^{2t+1} \equiv  \mathcal{H}_x(\muzy^{2t})$ for all $x \in \mathcal{X}$
and $\mu_{0y}^{2t+1} \equiv  \mathcal{H}_y(\muxz^{2t})$ for all $y \in \mathcal{Y}$.
Note that the mapping $\mathcal{H}$ is antitone, meaning that
$\muxz^{2t} \le \tilde{\muxz}^{2t}$ and $\muzy^{2t} \le \tilde{\muzy}^{2t}$ implies $\mathcal{H}_x(\muzy^{2t}) \ge \mathcal{H}_x(\tilde{\muzy}^{2t})$ 
for all $x \in \mathcal{X}$ and 
$\mathcal{H}_y(\muxz^{2t}) \ge \mathcal{H}_y(\tilde{\muxz}^{2t})$ for all $y \in \mathcal{Y}$.
By the same token, we define the mapping
 $\mu_{x0}^{2t+1} \equiv  \mathcal{G}_x(\muxz^{2t-1},\muzy^{2t-1})$ for all $x \in \mathcal{X}$ and $\mu_{0y}^{2t+1} \equiv  \mathcal{G}_y(\muxz^{2t-1},\muzy^{2t-1})$ for all $y \in \mathcal{Y}$,
 when we can verify that $\mathcal{G}$ is also isotone.
 As a result, the mapping from $(\muxz^{2t-1}, \muzy^{2t-1}, \muxz^{2t},\muzy^{2t})$ to $(\muxz^{2t+1}, \muzy^{2t+1})$ are isotone in $(\muxz^{2t-1}, \muzy^{2t-1})$ and antitone in $(\muxz^{2t},\muzy^{2t})$.
Similarly,
we can show the mapping from $(\muxz^{2t-2}, \muzy^{2t-2}, \muxz^{2t-1},\muzy^{2t-1})$ to $(\muxz^{2t}, \muzy^{2t})$ are isotone in $(\muxz^{2t-2}, \muzy^{2t-2})$ and antitone in 
$(\muxz^{2t-1},\muzy^{2t-1})$.
We also know that $\mu^0_{x0}\le\mu^2_{x0}$ and $\mu^1_{x0}\ge\mu^3_{x0}$ for all $x \in \mathcal{X}$,  and $\mu^0_{0y} \le \mu^2_{0y}$
and  $\mu^1_{0y} \ge \mu^3_{0y}$  for all $y \in \mathcal{Y}$.
Therefore, $(\muxz^{2t-1}, \muzy^{2t-1}) \le (\muxz^{2t+1}, \muzy^{2t+1})$ for all $t$.
Hence the sequence $((\muxz^1,\muzy^1), (\muxz^3,\muzy^3), (\muxz^5,\muzy^5), ..., (\muxz^{2t+1},\muzy^{2t+1}), ...)$ is a decreasing sequence bounded from below by $0$.
As a result, $(\muxz^{2t+1},\muzy^{2t+1})$ converges.
Letting $(\bar{\muxz},\bar{\muzy})$ be its limit,
it is easy to see that $(\bar{\muxz},\bar{\muzy})$ is the solution of system 
\eqref{NL: theorem6}.

\underline{Part $(ii)$. Proof of Uniqueness}.
The proof of uniqueness also relies on the main result in theorem 1 of
\citet{berry2013connected}. First, introduce the quantities $p_{x}=\mu _{x0}$ and $p_{y}=-\mu _{0y}$
and thus denote $p=(\muxz,-\muzy)$.
Solving the system (\ref{NL: theorem6}) is equivalent to finding the root of
the following system
\begin{equation}
\label{NL: theorem6_p_appendix}
\QATOPD\{ . {\sigma _{x}(p) = p_{x}+\sum_{y}M_{xy}(p_x,\{p_{x^\prime}\}_{x^\prime \neq x},-p_y,\{-p_{y^\prime}\}_{y^\prime \neq y})-n_{x}}
{\sigma _{y}(p) = p_{y}-\sum_{x}M_{xy}(p_x,\{p_{x^\prime}\}_{x^\prime \neq x},-p_y,\{-p_{y^\prime}\}_{y^\prime \neq y})+m_{y}}.
\end{equation}
We then introduce $\sigma
_{0}(p)=1-\sum_{x\in \mathcal{X}}\sigma _{x}(p)-\sum_{y\in \mathcal{Y}%
}\sigma _{y}(p)$.

Assumption 1 in \citet{berry2013connected} is satisfied as $\sigma $ is
defined over the Cartesian product of intervals $\Pi _{x\in \mathcal{X}%
}[0,n_{x}]\Pi _{y\in \mathcal{Y}}[-m_{y},0]$.
Note that $\forall z^{\prime }\neq z$, $\sigma _{z}(p)$ is weakly decreasing
in $p_{z^{\prime }}$ from assumption \ref{ass: MFE} on $M$.
To see this, first consider $z=x$ and $z^\prime=x^\prime$ for any $x^\prime \neq x$.
Taking partial derivative of $\sigma _{x}(p)$ with respect to $p_{x^\prime}$ yields,
\begin{align*}
\sum_{y \in \mathcal{Y}} \frac{\partial M_{xy}}{\partial p_{x^\prime}}  \le 0,
\end{align*}
where the inequality is from assumption \ref{ass: MFE}-$(iii)$ stating $M_{xy}$ is weakly decreasing in $\mu_{x^\prime 0}$ for all $x^\prime \neq x$.
Second, consider $z=x$ and $z^\prime=y$ for any $y \in \mathcal{Y}$.
Taking partial derivative of $\sigma _{x}(p)$ with respect to $p_{y}$ yields,
\begin{align*}
-\frac{\partial M_{xy}}{\partial (-p_{y})} - \sum_{y^\prime \neq y} \frac{\partial M_{xy^\prime}}{\partial (-p_{y})} \le 0,
\end{align*}
where the inequality is from assumption \ref{ass: MFE}-$(iii)$ stating that $M_{xy}$ is weakly increasing in $\mu_{0 y}$ and that $M_{xy}$ is weakly decreasing in $\mu_{0 y^\prime}$ for all $y^\prime \neq y$ and
assumption \ref{ass: MFE}-$(iv)$ stating $\vert \frac{\partial M_{xy}}{\partial \mu_{0y}} \vert \ge \vert \sum_{y^\prime \neq y} \frac{\partial M_{xy^\prime}}{\partial \mu_{0 y}} \vert$.
Similarly, we can also show that $\forall z^{\prime }\neq z$, $\sigma _{z}(p)$ is weakly decreasing
in $p_{z^{\prime }}$ holds for $z=y$ for all $y \in \mathcal{Y}$ and $z^\prime \neq y$ from assumption \ref{ass: MFE} on $M$.
Thus, assumption 2 from \citet{berry2013connected} is satisfied.

Moreover, let us rewrite $\sigma_{0}(p)$ as
\begin{equation*}
\textstyle
\sigma _{0}(p)=1+\sum_{x}n_{x}-\sum_{y}m_{y}-\sum_{x}p_{x}-\sum_{y}p_{y},
\end{equation*}
which is strictly decreasing in any element of $\muxz$ and $\muzy$
and then $\sigma_0(p)$ is strictly decreasing in $p_z$, $\forall z \in \mathcal{X} \cup \mathcal{Y}$.
Thus, assumption 3 in \citet{berry2013connected} is also satisfied by applying Lemma 1 in that paper.
Hence, $\sigma $ is inverse
isotone by theorem 1 in \citet{berry2013connected}, which provides uniqueness. Indeed, assume that $\sigma (p)=\sigma
(p^{\prime })$, so that $\sigma (p)\leq \sigma (p^{\prime })$ and $\sigma
(p)\geq \sigma (p^{\prime })$ which implies by inverse isotonicity that $p \leq
p^{\prime }$ and $p\geq p^{\prime }$, hence $p=p^{\prime }$.
Therefore,
there is a unique root $p^{\ast }$ to the system of equation \eqref{NL: theorem6_p_appendix}.
Hence, there is a unique solution to system \eqref{NL: theorem6}, with $\mu _{x0}^{\ast }=p_{x}^{\ast }$ and $\mu _{0y}^{\ast
}=-p_{y}^{\ast }$. QED.
\end{proof}

%%%%%%%%%%%%%%%%%%%%%%%%%
\subsection{Proof of theorem \protect\ref{thm:gradient} \label{app:proofs_thm2}}

\begin{proof}
The expression for $\partial _{\theta ^{k}}\Pi _{xy}$ follows immediately
from the fact that
$\Pi _{xy}=\frac{M _{xy}}{N^\theta}$ and
that
$N^\theta=n+m-\sum_{xy \in \mathcal{XY}} M^\theta(a, b)$.

By the implicit function theorem in (\ref{eq: nonlinear_system}), one has%
\begin{eqnarray*}
\partial _{\theta ^{k}}\mu _{x0}+\sum_{y\in \mathcal{Y}}(
\sum_{x \in \mathcal{X}}
\partial _{\theta
^{k}}\mu _{x0}\partial _{\mu _{x0}}M_{xy}
+
\sum_{y \in \mathcal{Y}}
\partial _{\theta ^{k}}\mu
_{0y}\partial _{\mu _{0y}}M_{xy}) &=&-\sum_{y}\partial _{\theta ^{k}}M_{xy},
\end{eqnarray*}
for all $x \in \mathcal{X}$,
and 
\begin{eqnarray*}
\partial _{\theta ^{k}}\mu _{0y}+\sum_{x\in \mathcal{X}}(
\sum_{x \in \mathcal{X}}
\partial _{\theta
^{k}}\mu _{x0}\partial _{\mu _{x0}}M_{xy}
+
\sum_{y \in \mathcal{Y}}
\partial _{\theta ^{k}}\mu
_{0y}\partial _{\mu _{0y}}M_{xy}) &=&-\sum_{x}\partial _{\theta ^{k}}M_{xy},
\end{eqnarray*}%
for all $y \in \mathcal{Y}$,
which can be rewritten using the expression of $\Delta $ (\ref{eq:delta}) as%
\begin{equation*}
\Delta=
 \begin{pmatrix}
(1+\sum_{y\in \mathcal{Y}}\partial _{\mu _{10}}M_{1y}^{
}) 
& \sum_{y\in \mathcal{Y}} \partial_{\mu_{20}}M_{1y}
& \cdots 
& \sum_{y\in \mathcal{Y}} \partial_{\mu_{0|\mathcal{Y}|}}M_{1y}
\\
\sum_{y\in \mathcal{Y}} \partial_{\mu_{10}}M_{2y}
&(1+\sum_{y\in \mathcal{Y}}\partial _{\mu _{20}}M_{2y}^{
}) 
& \cdots 
& \sum_{y\in \mathcal{Y}} \partial_{\mu_{0|\mathcal{Y}|}}M_{2y}
\\
\vdots
&
\vdots
&
\vdots
&
\vdots 
\\
\sum_{x\in \mathcal{X}} \partial_{\mu_{10}}M_{x|\mathcal{Y}|}
& 
\sum_{x \in \mathcal{X}} \partial_{\mu_{20}}M_{x |\mathcal{Y}|}
& \cdots 
& (1+\sum_{x\in \mathcal{X}}\partial _{\mu _{0|\mathcal{Y}|}}M_{x|\mathcal{Y}|}^{
})
\end{pmatrix}
\end{equation*}%
and $\Delta $ being a strictly diagonally dominant matrix, is invertible,
QED.
\end{proof}

%%%%%%%%%%%%%%%%%%%%%%%%%
\subsection{Proof of theorem \protect\ref{thm:equiv} \label{app:proofs_thm3}}

\begin{proof}
\underline{Proof of Part $(i)$.}
Recall that in the ITU-logit model, the aggregate matching function is given by $M_{xy}(\mu_{x0},\mu_{0y}) = \exp\left(-D_{xy}(-\log(\mu_{x0}),-\log(\mu_{0y})\right)$ where the distance function $D_{xy}$ is defined by
$D_{xy}(g_x,h_y) = \min\{z\in\mathbb{R}:(g_x-z, h_y-z)\in \mathcal{F}_{xy}\}$
and where $\mathcal{F}_{xy}$ is the bargaining set for the $xy$ pair. By definition, $D_{xy}(g_x+t,h_y+t) = D_{xy}(g_x,h_y)+t$, which
implies $M_{xy}(\lambda\mu_{x0},\lambda\mu_{0y}) = \lambda  M_{xy}(\mu_{x0},\mu_{0y})$.
This is a homogenous of degree 1 matching function that depends only on the masses
of unassigned agents of own types.

\underline{Proof of Part $(ii)$.} Denote $M_{xy}(\mu_{x0},\mu_{0y})$ as a matching function in definition \ref{def: AMF} satisfying assumption \ref{def: AMF} and homogeneous of degree $1$, and 
depending only on the masses of unassigned masses of own types.
Then
introduce the mapping $D_{xy}(g,h) = -\log M_{xy}\left(e^{-g}, e^{-h}\right)$
(in the following, we will drop the indices for convenience).
We will show that $D$ is the distance function associated with some proper bargaining set introduced in GKW and that $M$ is the associated aggregate matching function in ITU-logit model of GKW.

\underline{Step 1.} We begin by constructing the bargaining set $\mathcal{F}$ as follows
\[
\mathcal{F} = {(g,h)\in \mathbb{R}^2: D(g,h) \leq 0}.
\]

\underline{Step 2.} Let us show that the set $\mathcal{F}$ is a proper bargaining set. First, note that assumption \ref{ass: MFE} does not ensure that $\mathcal{F}$ is non-empty. However, this is not much of a concern in our setting: there will simply be no match between the two corresponding individuals in equilibrium (only a mild additional assumption is required to obtain non-emptiness: ${M}_{xy}(\mu_{x0},\mu_{0y})$ is bounded below by $1$ as $\mu_{x0}$ ($\mu_{0y}$) approaches infinity while $\mu_{0y}$ ($\mu_{x0}$) is bounded below by 0 ; as a matter of fact, it is satisfied on all of our introductory examples). Closedness follows from the continuity of $M$ by assumption \ref{ass: MFE} -(i). From assumption \ref{ass: MFE} -(ii), we can deduce that $\mathcal{F}$ is lower comprehensive. Indeed, assume that $(g,h)\in\mathcal{F}$. By construction, $D(g,h)\leq0$. Take $(g^\prime,h^\prime)$ with $g^\prime\leq g$ and $h^\prime\leq h$. By weak isotonicity of $M$, we have $D(g^\prime,h^\prime)\leq D(g,h) \leq 0$, hence $(g^\prime, h^\prime)\in \mathcal{F}$. Finally, we can show that $\mathcal{F}$ is bounded above. Indeed, assume $g_n\rightarrow +\infty$ and $h_n$ bounded below, then for $n$ large enough, $M(g_n,h_n)<1$ by assumption \ref{ass: MFE}-(iii), so that $D(g_n,h_n)>0$, that is $(g_n,h_n)\notin \mathcal{F}$ (the same reasoning applied with $h_n\rightarrow +\infty$ and $h_n$ bounded below).

\underline{Step 3.} Let us now show that $D$ is the distance function associated with the bargaining set $\mathcal{F}$. The distance from the point $(g,h)$ to the frontier is the value $z$ such that $(g-z,h-z)$ belongs to the frontier of the bargaining set. By construction, $D(g-z,h-z)=0$ but homogeneity of degree one implies that $D(g+t,h+t)=D(g,h)+t$, therefore $z=D(g,h)$.

\underline{Step 4.} GKW showed that in equilibrium, individuals receive a payoff that is the sum of two components: a systematic component that depends on the observable characteristics of the partners, denoted respectively for men and women, $U_{xy}$ and $V_{xy}$ ; and an idiosyncratic component $\epsilon_{iy}$ and $\eta_{xj}$. Assume that when a man of type $x$ meet with a woman of type $y$, they decide upon a utility wedge $w$ and receive $U_{xy} = -D_{xy}(0,-w)$ and $V_{xy} = -D_{xy}(w,0)$. Note that the functions $U$ and $V$, as defined here, explicitly represent the bargaining frontier. Assuming logit heterogeneities, the systematic component of utility can be recovered from the marriage patterns by the usual formulas $U_{xy} = \log \frac{\mu_{xy}}{\mu_{x0}}$ and $V_{xy} = \log \frac{\mu_{xy}}{\mu_{0y}}$. The equilibrium condition in GKW is $D_{xy}(U_{xy},V_{xy})=0$, that is $D_{xy}(\log \mu_{xy} - \log \mu_{x0},\log \mu_{xy}- \log \mu_{0y})=0$
which yields to the aggregate matching function
\begin{equation*}
M_{xy}(\mu_{x0},\mu_{0y}) = \exp\left(-D_{xy}\left(-\log \mu_{x0}, -\log \mu_{0y}\right)\right) = \mu_{xy}
\end{equation*}
This concludes the proof.
\end{proof}

%%%%%%%%%%%%%%%%%%%%%%%%%
\subsection{Proof of theorem \ref{thm:confidence} \label{app:proofs_thm4}}
\label{app:CI}
In the following, $%
\Pi\left( \theta ,\zeta \right) $ denotes the predicted frequencies given $%
\theta $ and the frequencies $\zeta $. We also introduce the rescaling
operator $\Pi^{\theta }(\pi _{0})=\frac{M^{\theta }(\pi _{0})}{1^{\prime
}M^{\theta }(\pi _{0})}$, and note that%
$\Pi\left( \theta ,\zeta \right) =\Pi ^{\theta }\circ \left( AM^{\theta
}\right) ^{-1}\left( \zeta \right)$.

\underline{Part (i). A first expression}. In the maximization of the
log-likelihood, the first order conditions with respect to $\theta $ are $\hat{\pi}^{\prime }\partial _{\theta }\ln \Pi\left( \theta ,A\hat{\pi}\right) =0$ that we denote $F\left( \hat{\theta},\hat{\pi}\right) =0$. From a serie of Taylor expansions around the true value of $\theta $ and $\pi $, we can then deduce
that $\left( \hat{\theta}-\theta \right) =-\left( D_{\theta }F\right) ^{-1}\left(
D_{\pi }F\right) \left( \hat{\pi}-\pi \right)$ where we use the notation $D$ for the Jacobian matrix. Note that
asymptotically, $N^{1/2}\left( \hat{\pi}-\pi \right) \sim \mathcal{N}\left(
0,V_{\pi }\right) $ where $V_{\pi }=diag\left( \pi \right) -\pi \pi ^{\prime
}.$ Hence, it follows that
\begin{equation*}
N^{1/2}\left( \hat{\theta}-\theta \right) \Rightarrow \mathcal{N}\left(
0,V_{\theta }\right)
\end{equation*}
where $V_{\theta }=\left( D_{\theta }F\right) ^{-1}\left( D_{\pi }F\right)
V_{\pi }\left( D_{\pi }F\right) ^{\prime }\left( \left( D_{\theta }F\right)
^{\prime }\right) ^{-1}$.

\underline{Part (ii). Analytic expressions for each component}. Let us begin
with $D_{\theta }F$. Note that by definition, we have $\Pi ^{\prime }1=1$.
Hence, $\Pi ^{\prime }\partial _{\theta ^{i}}\log \Pi =0$ and $\Pi ^{\prime
}\partial _{\theta ^{i}\theta ^{j}}^{2}\log \Pi +\partial _{\theta ^{j}}\Pi
^{\prime }\partial _{\theta ^{i}}\log \Pi =0$. Finally, we get%
$\pi ^{\prime }\partial _{\theta ^{i}\theta ^{j}}^{2}\log \Pi =-\pi ^{\prime
}\partial _{\theta ^{i}}\log \Pi ^{\prime }\partial _{\theta ^{j}}\log \Pi$
so
\begin{equation*}
D_{\theta }F=-D_{\theta }\log \Pi ^{\prime }diag(\pi )D_{\theta }\log \Pi
\end{equation*}%
and $D_{\theta }\log \Pi $ can be obtained from our results on the gradient
of the log-likelihood.

We may now turn to $D_{\pi }F$. We have

\begin{equation*}
D_{\pi }F:=\left( D_{\theta }\log \Pi \right) ^{\prime }-\left( D_{\theta
}\log \Pi \right) ^{\prime }\left( D_{\zeta }\Pi \right) A
\end{equation*}%
where, as before, $D_{\theta }\log \Pi $ appears in theorem \ref%
{thm:gradient}. We obtain $D_{\zeta }\Pi $ as%
\begin{equation*}
D_{\zeta }\Pi =\left( D_{\pi _{0}}\Pi ^{\theta }\right) \left( \left( D_{\pi
_{0}}AM^{\theta }\right) ^{-1}\right)
\end{equation*}
The expressions above allow us to prove the announced formula for $V_{\theta }$. We have%
\begin{eqnarray*}
V_{\theta } &=&\left( D_{\theta }F\right) ^{-1}\left( D_{\pi }F\right)
V_{\pi }\left( D_{\pi }F\right) ^{\prime }\left( \left( D_{\theta }F\right)
^{\prime }\right) ^{-1} \\
&=&T^{-1}\left( D_{\pi }F\right) V_{\pi }\left( D_{\pi }F\right) ^{\prime
}\left( T^{\prime }\right) ^{-1} \\
&=&T^{-1}\left( \left( D_{\theta }\log \Pi \right) ^{\prime }-\left(
D_{\theta }\log \Pi \right) ^{\prime }\left( D_{\zeta }\Pi \right) A\right)
V_{\pi }\left( \left( D_{\theta }\log \Pi \right) ^{\prime }-\left(
D_{\theta }\log \Pi \right) ^{\prime }\left( D_{\zeta }\Pi \right) A\right)
^{\prime }\left( T^{\prime }\right) ^{-1} \\
&=&T^{-1}\left( -\left( D_{\theta }\log \Pi \right) ^{\prime }+\left(
D_{\theta }\log \Pi \right) ^{\prime }\left( D_{\zeta }\Pi \right) A\right)
V_{\pi }\left( A^{\prime }\left( D_{\zeta }\Pi \right) ^{\prime }\left(
D_{\theta }\log \Pi \right) -\left( D_{\theta }\log \Pi \right) \right)
\left( T^{\prime }\right) ^{-1} \\
&=&T^{-1}\left( D_{\theta }\log \Pi \right) ^{\prime }\left( D_{\zeta }\Pi
\right) AV_{\pi }A^{\prime }\left( D_{\zeta }\Pi \right) ^{\prime }\left(
D_{\theta }\log \Pi \right) \left( T^{\prime }\right) ^{-1}+H
\end{eqnarray*}%
where $H=G+G^{\prime }+T^{-1}\left( D_{\theta }\log \Pi \right) ^{\prime }V_{\pi
}\left( D_{\theta }\log \Pi \right) \left( T^{\prime }\right) ^{-1}$
and is composed of three terms, and two of them are symmetric. Let us start
with these symmetric terms:%
\begin{equation*}
G=T^{-1}\left( D_{\theta }\log \Pi \right) ^{\prime }V_{\pi }A^{\prime
}\left( D_{\zeta }\Pi \right) ^{\prime }\left( D_{\theta }\log \Pi \right)
\left( T^{\prime }\right) ^{-1}
\end{equation*}%
this is
\[
T^{-1}\left( D_{\theta }\log \Pi \right) ^{\prime }A^{\prime
}diag(\pi )\left( D_{\zeta }\Pi \right) ^{\prime }\left( D_{\theta }\log \Pi
\right) \left( T^{\prime }\right) ^{-1}-T^{-1}\left( D_{\theta }\log \Pi
\right) ^{\prime }\pi \pi ^{\prime }A^{\prime }\left( D_{\zeta }\Pi \right)
^{\prime }\left( D_{\theta }\log \Pi \right) \left( T^{\prime }\right)
^{-1}.
\]

The first term in this sum is $0$ because $\left( D_{\theta }\log
\Pi \right) ^{\prime }A^{\prime }=0$. The second term is
\begin{eqnarray*}
&&T^{-1}\left( D_{\theta }\Pi \right) ^{\prime }diag(\mu )^{-1}\mu \zeta
^{\prime }\left( D_{\zeta }\Pi \right) ^{\prime }\left( D_{\theta }\log \Pi
\right) \left( T^{\prime }\right) ^{-1} \\
&=&T^{-1}\left( D_{\theta }\Pi \right) ^{\prime }1\zeta ^{\prime }\left(
D_{\zeta }\Pi \right) ^{\prime }\left( D_{\theta }\log \Pi \right) \left(
T^{\prime }\right) ^{-1}
\end{eqnarray*}%
where $\left( D_{\theta }\Pi \right) ^{\prime }1=0$. Hence,%
\begin{eqnarray*}
H &=&T^{-1}\left( D_{\theta }\log \Pi \right) ^{\prime }V_{\pi }\left(
D_{\theta }\log \Pi \right) \left( T^{\prime }\right) ^{-1} \\
&=&\left( T^{\prime }\right) ^{-1}
\end{eqnarray*}%
We obtain
\begin{equation*}
V_{\theta }=\left( \mathcal{I}_{11}\right) ^{-1}+\mathcal{I}_{11}^{-1}%
\mathcal{I}_{12}AV_{\pi }A^{\prime }\mathcal{I}_{12}^{\prime }\mathcal{I}%
_{11}^{-1}
\end{equation*}
where $\mathcal{I}_{11} =-\left( D_{\theta }\log \Pi \right) ^{\prime }diag\left(
\pi \right) \left( D_{\theta }\log \Pi \right)$ and $\mathcal{I}_{12} =\left( D_{\theta }\log \Pi \right) ^{\prime }diag\left(\pi \right) \left( D_{\zeta }\log \Pi \right)$. This concludes the proof.

%%%%%%%%%%%%%%%%%%%%%%%%%

\subsection{Proof of theorem \ref{thm: existence_change} \label{app:proofs_thm5}}
\label{app: Change}

\begin{proof}
\underline{Part (i). Proof of Existence}. The proof of existence relies on a revised procedure based on
algorithm \ref{IPFP Algorithm}, which is stated below.

\begin{Algorithm}
\label{IPFP Algorithm revised} The revised algorithm works as follows$,$

\begin{tabular}{r|p{5in}}
Step $0$ & Fix the value of $\tilde{\mu} _{x0}$ at  $\tilde{\mu} _{x0}^{0}=0$ for all $x \in \mathcal{X}$ and $\tilde{\mu} _{0y}$ at  $\tilde{\mu} _{0y}^{0}=0$ for all $y \in \mathcal{Y}$. \\
Step $1$ & Fix the value of $\tilde{\mu} _{x0}$ at  $\tilde{\mu} _{x0}^{1}=\tilde{n}_{x}$ for all $x \in \mathcal{X}$ and $\tilde{\mu} _{0y}$ at  $\tilde{\mu} _{0y}^{1}=\tilde{m}_{y}$ for all $y \in \mathcal{Y}$. \\
Step $2t$ & For each $x \in \mathcal{X}$, keep the values $\tilde{\mu} _{x^\prime 0}^{2t-2}$ for 
$x^\prime \neq x$, $\tilde{\mu} _{0y^\prime}^{2t-2}$ for all $y^\prime \neq y$ and $\tilde{\mu} _{0y}^{2t-1}$ for all $y \in \mathcal{Y}$ fixed,  solve for the value $\tilde{\mu} _{x0}^{2t}$, such
that the equality, $\tilde{\mu}^{2t}_{x0}+\sum_{y\in \mathcal{Y}}M_{xy}(\tilde{\mu}^{2t}_{x0},\{\tilde{\mu}_{x^\prime 0}^{2t-2}\}_{x^\prime \neq x},\tilde{\mu}^{2t-1}_{0y}, \{\tilde{\mu}_{0y^\prime}^{2t-2}\}_{y^\prime \neq y})=\tilde{n}_{x}$ holds$;$
For each $y \in \mathcal{Y}$, keep the values $\tilde{\mu} _{x^\prime 0}^{2t-2}$ for $x^\prime \neq x$,
$\tilde{\mu}_{x0}^{2t-1}$ for all $x \in \mathcal{X}$ and $\tilde{\mu} _{0y^\prime}^{2t-2}$ for all $y^\prime \neq y$ fixed, 
solve for the value $\tilde{\mu} _{0y}^{2t}$, such
that the equality, 
$\tilde{\mu}^{2t}_{0y}+\sum_{x\in \mathcal{X}}M_{xy}(\tilde{\mu}_{x0}^{2t-1},\{\tilde{\mu}_{x^\prime 0}^{2t-2}\}_{x^\prime \neq x}, \tilde{\mu}^{2t}_{0y},\{\tilde{\mu}_{0 y^\prime}^{2t-2}\}_{y^\prime \neq y})=\tilde{m}_{y}$ holds.\\
Step $2t+1$ & For each $x \in \mathcal{X}$, keep the values $\tilde{\mu} _{x^\prime 0}^{2t-1}$ for $x^\prime \neq x$, $\tilde{\mu} _{0y^\prime}^{2t-1}$ for all $y^\prime \neq y$ and $\tilde{\mu} _{0y}^{2t}$ for all $y \in \mathcal{Y}$ fixed,  solve for the value $\tilde{\mu} _{x0}^{2t+1}$, such
that the equality, $\tilde{\mu}^{2t+1}_{x0}+\sum_{y\in \mathcal{Y}}M_{xy}(\tilde{\mu}^{2t+1}_{x0},\{\tilde{\mu}_{x^\prime 0}^{2t-1}\}_{x^\prime \neq x},\tilde{\mu}^{2t}_{0y}, \{\tilde{\mu}_{0y^\prime}^{2t-1}\}_{y^\prime \neq y})=\tilde{n}_{x}$ holds$;$
For each $y \in \mathcal{Y}$, keep the values $\tilde{\mu} _{x^\prime 0}^{2t-1}$ for $x^\prime \neq x$,
$\tilde{\mu} _{x0}^{2t}$ for all $x \in \mathcal{X}$ and $\tilde{\mu} _{0y^\prime}^{2t-1}$ for all $y^\prime \neq y$ fixed, 
solve for the value $\tilde{\mu} _{0y}^{2t+1}$ such
that the equality, 
$\tilde{\mu}^{2t+1}_{0y}+\sum_{x\in \mathcal{X}}M_{xy}(\tilde{\mu}_{x0}^{2t},\{\tilde{\mu}_{x^\prime 0}^{2t-1}\}_{x^\prime \neq x}, \tilde{\mu}^{2t+1}_{0y},\{\tilde{\mu}_{0 y^\prime}^{2t-1}\}_{y^\prime \neq y})=\tilde{m}_{y}$ holds.
\end{tabular}
\newline
\newline
The algorithm terminates when,
$\sup \big\{
\sup_x \left| \tilde{\mu}^{2t+1}_{x0}-\tilde{\mu}^{2t-1}_{x0} \right|,
\sup_y \left| \tilde{\mu}^{2t+1}_{0y}-\tilde{\mu}^{2t-1}_{0y} \right|
\big \} < \epsilon$, where $\epsilon$ is a sufficiently small positive value.
\end{Algorithm}

\underline{Part (ii). Proof of Uniqueness}. The proof of uniqueness relies
on the main result in \textcite{berry2013connected} similarly as it in the proof of theorem \ref{thm:system}.
\end{proof}

%%%%%%%%%%%%%%%%%%%%%%%%%%%%%%%%%%%%%%%%%%
%%%%%%%%%%%%%%%%%%%%%
\clearpage
\newpage
\section{Online Appendix}

\subsection{Parameter-free approach for the model in \textcite{choo2015dynamic} \label{app: dynamic_matching} \label{app:example_choo2015}}
In \textcite{choo2015dynamic}, the observed equilibrium marriage distribution
$(\mu^*, \muxz^*, \muzy^*)$
must satisfy the matching function
\eqref{eq: sec4_matchFunction_dynamicType2},
 \begin{align}
 \label{app: MF_choo2015_observed}
M_{xy}(\muxz^*, \muzy^*; f_{xy}(\theta))=
f_{xy}(\theta)
\sqrt{n_xm_y}
 \prod_{k=0}^{T(x,y)-1}\bigg[\frac{\mu^*_{x+k0}}{n_{x+k}} \frac{\mu^*_{0y+k}}{m_{y+k}} \bigg]^{\frac{1}{2}[\beta(1-\delta)]^k},
 \end{align}
 and also satisfy the nonlinear equation system,
 \begin{equation}
\label{NL: app_observed}
\QATOPD\{ . {n_{x}=\mu^*_{x0}+\sum_{y\in \mathcal{Y}}
M_{xy}(\muxz^*, \muzy^*; f_{xy}(\theta))
}
{m_{y}=\mu^*_{0y}+\sum_{x\in \mathcal{X}}
M_{xy}(\muxz^*, \muzy^*; f_{xy}(\theta))
}.
\end{equation}%

Consider a counterfactual change in which the number of available individuals $(n_x,m_y)$ changes to $(n^\prime_x,m^\prime_y)$
and matching surplus $f_{xy}(\theta)$ changes to $t_{xy}f_{xy}(\theta)$.
In this counterfactual scenario, the dynamic matching function is given by
 \begin{align}
 \label{app: MF_choo2015_counterfactual}
M_{xy}(\muxz^{* \prime}, \muzy^{* \prime}; f_{xy}(\theta))=
t_{xy} f_{xy}(\theta)
\sqrt{n^\prime_x m^\prime_y}
 \prod_{k=0}^{T(x,y)-1}\bigg[\frac{\mu^{* \prime}_{x+k0}}{n^\prime_{x+k}} \frac{\mu^{* \prime}_{0y+k}}{m^\prime_{y+k}} \bigg]
^{\frac{1}{2}[\beta(1-\delta)]^k},
\end{align}
which also satisfy the nonlinear equation system,
 \begin{equation}
\label{NL: app_counterfactual}
\QATOPD\{ . {n^\prime_{x}=\mu^{* \prime}_{x0}+\sum_{y\in \mathcal{Y}}
M_{xy}(\muxz^{* \prime}, \muzy^{* \prime}; f_{xy}(\theta))
}
{m^\prime_{y}=\mu^{* \prime} _{0y}+\sum_{x\in \mathcal{X}}
M_{xy}(\muxz^{* \prime}, \muzy^{* \prime}; f_{xy}(\theta))
}.
\end{equation}%

Taking the ratio between \eqref{app: MF_choo2015_observed} and \eqref{app: MF_choo2015_counterfactual} yields the ratio of matches between
observed and counterfactual equilibria,
\begin{align}
\label{app: ratio_matching}
\frac{\mu^\prime_{xy}}{\mu_{xy}}=\frac{
M_{xy}(\muxz^{* \prime}, \muzy^{* \prime}; f_{xy}(\theta))
}
{
M_{xy}(\muxz^{*}, \muzy^{*}; f_{xy}(\theta))
}
=t_{xy}
\sqrt{\tilde{n}_x \tilde{m}_y}
 \prod_{k=0}^{T(x,y)-1}\bigg[\frac{\tilde{\mu}_{x+k0}}{\tilde{n}_{x+k}} \frac{\tilde{\mu}_{0y+k}}{\tilde{m}_{y+k}} \bigg]
 ^{\frac{1}{2}[\beta(1-\delta)]^k}.
\end{align}

Taking the ratio between \eqref{NL: app_observed} and \eqref{NL: app_counterfactual} and substituting
\eqref{app: ratio_matching}
gives us,
 \begin{equation}
\label{NL: app_ratio}
\QATOPD\{ . {\tilde{n}_{x}=p_{x0}\tilde{\mu}_{x0}+\sum_{y\in \mathcal{Y}}p_{xy}
t_{xy}\sqrt{\tilde{n}_x \tilde{m}_y}
 \prod_{k=0}^{T(x,y)-1}\bigg[\frac{\tilde{\mu}_{x+k0}}{\tilde{n}_{x+k}} \frac{\tilde{\mu}_{0y+k}}{\tilde{m}_{y+k}} \bigg]
 ^{\frac{1}{2}[\beta(1-\delta)]^k} }
{\tilde{m}_{y}=q_{0y} \tilde{\mu} _{0y}+\sum_{x\in \mathcal{X}} q_{xy}
t_{xy}\sqrt{\tilde{n}_x \tilde{m}_y}
 \prod_{k=0}^{T(x,y)-1}\bigg[\frac{\tilde{\mu}_{x+k0}}{\tilde{n}_{x+k}} \frac{\tilde{\mu}_{0y+k}}{\tilde{m}_{y+k}} \bigg]
 ^{\frac{1}{2}[\beta(1-\delta)]^k}
}
\end{equation}%
which contain $|\mathcal{X}|+|\mathcal{Y}|$ nonlinear equations for solving $|\mathcal{X}|+|\mathcal{Y}|$ unknowns of
$(\tilde{\mu}_{x0}, \tilde{\mu}_{0y})$ for all $x \in \mathcal{X}$ and $y \in \mathcal{Y}$.

\subsection{Simulations \label{app:simulations}}

In this section, we conduct simulations to investigate the numerical performance of the nested and MPEC approaches for maximum likelihood estimation.
In the nested approach,
it is crucial (i)
to be able to solve system (\ref{eq: nonlinear_system}) and (ii) to do so in an
efficient way, for the sake of minimizing computation time.
Therefore, we first investigate the performance of the IPFP algorithm and Newton descent method for solving the system (\ref{eq: nonlinear_system})
before turning  our attention to the nested and MPEC approaches.

\subsubsection{Solving system (\ref{eq: nonlinear_system}) for 
$(a^{\protect\theta },b^{\protect\theta })$}

Theorem \ref{thm:system}
and algorithm \ref{IPFP Algorithm} address both capability and efficiency concerns of solving system (\ref{eq: nonlinear_system}).
However in
practice, algorithm \ref{IPFP Algorithm} is not necessarily the most
efficient way to solve for $(a^{\theta },b^{\theta })$. When
the Jacobian of system (\ref{eq: nonlinear_system}) is known, it can be solved very
efficiently using Newton descent methods, which we recall below.

\begin{Algorithm}
\label{Newton Descent Algorithm} Rewrite the system of nonlinear equations 
$(\ref{eq: nonlinear_system})$ as $\sigma (\muxz,\muzy)=0$. The Newton's descent method works as follows

\begin{tabular}{r|p{5in}}
Step $0$ & Fix the initial value of $(\muxz,\muzy)$ at $(\muxz^0,\muzy^0)=(n^{\prime
},m^{\prime })^{\prime }$. \\
Step $t$ & Given $(\muxz^{t-1},\muzy^{t-1})$, solve $J\sigma (\muxz^{t-1},\muzy^{t-1})\delta
=-\sigma (\muxz^{t-1},\muzy^{t-1})$, where $J\sigma (\muxz^{t-1},\muzy^{t-1})$ is the Jacobian
matrix at $(\muxz^{t-1},\muzy^{t-1})$. Update $(\muxz^{t},\muzy^{t})=(\muxz^{t-1}+\delta,\muzy^{t-1}+\delta)$.%
\end{tabular}
\newline
\newline
The algorithm terminates when $\sup_{y}|(\muxz^{t+1},\muzy^{t+1})-(\muxz^{t},\muzy^{t})|<\epsilon $.
\end{Algorithm}

To benchmark these different methods, we consider the Exponentially
Transferable Utility model in GKW. The
aggregate matching function is given by
\begin{align*}
M_{xy}(\mu _{x0},\mu _{0y})=\exp
(-D_{xy}(-\log \mu _{x0},-\log \mu _{0y}))
\end{align*}
 where $D_{xy}(g,h)=\kappa _{xy}\log\left((
\exp((g-\alpha _{xy})/\kappa _{xy}) + \exp((h-\alpha _{xy})/\kappa _{xy}))/2\right)$.
We draw the types $x$ and $y$ from two uniform distributions, assume that $%
\alpha _{xy}=xy$ and $\gamma _{xy}=xy$, and fix $\kappa _{xy}=\kappa =1$,$%
\forall xy\in \mathcal{XY}$. In the experiment, we vary $|\mathcal{X}|$, the
number of types on the men side of the market, and fix $|\mathcal{Y}|=1.5|%
\mathcal{X}|$. Finally, we assume that $n_{x}=m_{y}=1$, $\forall x\in
\mathcal{X}$, $\forall y\in \mathcal{Y}$.

Table \ref{table:system} below summarizes the number of iteration and
computation time averaged over $50$ replications of the experiment for three
numerical methods: the IPFP algorithm described in algorithm \ref{IPFP
Algorithm}, its parallelized version, and the Newton's method described in
algorithm \ref{Newton Descent Algorithm}.

\begin{table}[th]
\caption{IPFP and Newton method}
\label{table:system}\centering
%\begin{tabular}{ccccccccc}
%\hline\hline
%& \multicolumn{2}{c}{IPFP} &  & \multicolumn{2}{c}{par. IPFP} &  &
%\multicolumn{2}{c}{Newton} \\ \hline
%Mkt. Size & Iter. & Time &  & Iter. & Time &  & Iter. & Time \\ \hline
%10 & 17.78 & 0.58 &  & 17.78 & 1.78 &  & 26.76 & 0.02 \\
%50 & 11.98 & 1.41 &  & 11.98 & 1.51 &  & 37.2 & 0.15 \\
%100 & 10.14 & 2.32 &  & 10.14 & 1.67 &  & 37.12 & 0.49 \\
%200 & 8.48 & 4.46 &  & 8.48 & 2.88 &  & 41.72 & 1.64 \\
%300 & 7.64 & 7.17 &  & 7.64 & 4.27 &  & 49.46 & 4.5 \\
%500 & 7.16 & 14.43 &  & 7.16 & 7 &  & 38.64 & 13.14 \\ \hline
%\end{tabular}%
%\end{table}
%
%\begin{table}[th]
%\caption{IPFP and Newton method (second test)}\centering%
\begin{tabular}{lllllllll}
\hline\hline
& \multicolumn{2}{c}{IPFP} &  & \multicolumn{2}{c}{par. IPFP} &  &
\multicolumn{2}{c}{Newton} \\ \hline
Mkt. Size & Iter. & Time &   & Iter. & Time &   & Iter. & Time \\
  \hline
10 & 14.64 & 0.36 &  & 14.64 & 1.45 &  & 26.42 & 0.01 \\
  50 & 9.2 & 1.17 &  & 9.2 & 1.12 &  & 27.08 & 0.09 \\
  100 & 8 & 2.16 &  & 8 & 1.33 &  & 23.22 & 0.29 \\
  200 & 7 & 4.35 &  & 7 & 2.43 &  & 23.62 & 1.15 \\
  300 & 7 & 7.21 &  & 7 & 3.99 &  & 25.9 & 3.08 \\
  500 & 6 & 13.54 &  & 6 & 7.4 &  & 31.38 & 12.12 \\
  1000 & 6 & 40.92 &  & 6 & 18.32 &  & 37.56 & 81.53 \\
  2000 & 5 & 106.93 &  & 5 & 47.77 &  & 24.02 & 680.16 \\
  5000 & 5 & 497.02 &  & 5 & 199.14 &  & 26 & 7934.81 \\
   \hline
\end{tabular}
\end{table}

This table raises three comments. First, to improve the computational efficiency of the
Newton's descent method, we provide the analytic expression of the Jacobian matrix of
system (\ref{eq: nonlinear_system}). Such analytic expression is not always available,
in which case the Jacobian must be approximated numerically, which will greatly
increase computation time for this method (at least for large markets).
Second, there is no guarantee of convergence when using the Newton's descent
algorithm. We notice no such issues when performing this simple experiment, but
nonconvergence may well be an issue with more complex models. For these two reasons,
and since Newton's method  performs only better for smaller market, this method is not our preferred algorithm.
Furthermore, it should be added that the IPFP
algorithm is very suitable for parallel computing. The gains are negative
for small markets, but as market size grows, we manage to reduce
computation time by a factor of two. The parallel IPFP runs on four
processors (which is what is currently available on most high-end personal
computers). This suggests that performance could be further improved when
running on computing clusters.

\subsubsection{Estimation}

We test the numerical performance of our maximum likelihood estimator, using
the nested and MPEC approaches. The setup of the
experiments remain the same as before, but we assume $\alpha _{xy}=\alpha \times
x\times y$ and $\gamma _{xy}=\gamma \times x\times y$, where $\alpha $ and $%
\gamma $ are arbitrarily chosen. Given $\theta _{0}=(\alpha ,\gamma )$, we
compute the equilibrium matching $\mu ^{\theta _{0}}$ using the IPFP
algorithm and set $\hat{\mu}=\mu ^{\theta _{0}}$. Then, we test if we are
able to recover $\theta _{0}$ from the observed $\hat{\mu}$ using our
maximum likelihood estimator in this correctly specified case. The results
are reported in table \ref{table:estim} below.
%\begin{center}
%TCIMACRO{\TeXButton{B}{\begin{table}[tbp] \centering}}%
%BeginExpansion
\begin{table}[h] \centering%
%EndExpansion
\caption{Estimation}%
\begin{tabular}{cccccccc}
\hline\hline
& \multicolumn{3}{c}{Nested Approach} &  & \multicolumn{3}{c}{MPEC} \\ \hline
Mkt. Size & Iter. & Time & \% Failure &  & Iter. & Time & \% Failure \\
\hline
10 & 18.64 & 5.4 & 6 &  & 15.91 & 5.1 & 8 \\
50 & 21.6 & 21.12 & 0 &  & 16.2 & 8.01 & 20 \\
100 & 23.34 & 42.05 & 0 &  & 18.05 & 17.58 & 22 \\
200 & 26.98 & 101.67 & 0 &  & 36.1 & 102.56 & 16 \\
300 & 26.8 & 179.25 & 0 &  & 20.64 & 121.44 & 16 \\
500 & 25.67 & 343.16 & 4 &  & 27.27 & 549.76 & 26 \\ \hline
\end{tabular}%
\label{table:estim}%
\end{table}%
%\end{center}

First, a word of caution, the nested approach
we implement here relies on a simple version of the IPFP algorithm, so its
performance can be further improved using the parallel version. It is
difficult to interpret the results in table \ref{table:estim}. The MPEC algorithm seems to perform better for small
market sizes as it converges faster to the correct value of $\theta .$ For
larger markets, however, the IPFP approach does better in some cases, for
example with $100$ or $500$ men. Note that the number of iterations is
relatively similar across methods, but performing one iteration can be
computationally burdensome in the MPEC case. Indeed, the nested approach only
requires  solving for the equilibrium matching using the IPFP algorithm and
computing the gradient as in theorem \ref{thm:gradient}. The MPEC approach,
on the other hand, requires the computation of the Jacobian matrix in
equation \eqref{eq:jacZ}, which involves second order and cross partial
derivatives. Although we do have analytic expressions for these components,
it can still be cumbersome to compute due to the size of these objects. For
example, in the case with $500$ men, $750$ women and two parameters, the
Jacobian matrix is a $2502\times 2502$ matrix. Finally, table \ref{table:estim} illustrates a common issue with Newton-like methods, that is, non-convergence.

\subsection{Data Construction} \label{app:data}

As discussed in the main text, we require three data inputs to implement our approach:
\begin{enumerate}
\item[i)] the number of available single men and women  by age and education had the financial aid program not been eliminated, $(n_x,m_y)$, for all $x \in \mathcal{X}$ and $y \in \mathcal{Y}$,
\item[ii)]  the number of available single men and women  by age and education as a result of the elimination of the financial aid program, $(n_x,m_y)$, for all $x \in \mathcal{X}$ and $y \in \mathcal{Y}$,  and
\item[iii)] the flow of new marriages by age and education as a result of the elimination of the financial aid program, $\mu^\prime$.
\end{enumerate}

The flow of new marriages as a result of the policy change $\mu'$ (item iii) above) is collected from the Vital Statistics in $1987/88$ obtained from the National Bureau of Economic Research data website.
Vital Statistics recorded the education and age of married couples until $1988$ for 22 reporting states.
The $22$ reporting states  are
California, Connecticut, Hawaii, Illinois, Kansas, Kentucky, Louisiana, Maine, Mississippi, Missouri, Montana, Nebraska, New Hampshire, New York, North Carolina, Rhode Island, Tennessee, Utah, Vermont, Virginia, Wisconsin, Wyoming.

The number of available single men and women  by age and education after the policy change, $(n_x,m_y)$ (item ii) above
are constructed from the IPUMS files of the U.S. CPS data. The sample used in this study is the monthly data in $1986$. In order for the CPS data to match the marriage data from the Vital Statistics, our sample comprises  only of individuals  from the  $22$
states reporting states.
We take an average of 12 monthly CPS surveys for the 22 matching states in $1986$ to construct the yearly available population vectors.

The age range studied is between $16$ and $75$ years old.
The education information is obtained from the variable ``{\tt EDUC}'' in the US CPS data.
The education attainment is divided into three groups: high school graduate or less, some years of college or college graduate, more than college.
There are $180$ possible age-education combinations from the $60$ age groups and $3$ education levels. We exclude $5$ groups - these are individuals who are less than 23 years of age with more than college education. This leaves us with $173$ types of men and women.
The variable ``{\tt marst}'' in IPUMS CPS data provides us with marital status information. It distinguishes an individual  either married, separated, divorced, widowed or never married/single.
We consider separated, divorced, widowed, never married/single individuals as available single individuals in the marriage market and calculated the number of available men and women for each type by adding the weight from each sample in the dataset.

\subsection{Asymmetric Specification} \label{app:asy_specification}

The specification that captures the asymmetric effects of interaction is given by
 \begin{align*}
% \label{eq: asy_specific_surplus_men}
\alpha_{xy}({\Lambdav})&=\Lambda_0+ \sum^{60}_{i=2} \Lambda^{ma}_{i} \mathbbm{1}\{x_a=i\}+\sum^{3}_{j=2} \Lambda^{me}_j \mathbbm{1}\{x_e=j\}
+ \sum^{60}_{i=2} \Lambda^{wa}_i \mathbbm{1}\{y_a=i\}+\sum^{3}_{j=2} \Lambda^{we}_j \mathbbm{1}\{y_e=j\} \nonumber \\
&+\sum^{59}_{i=1} \Lambda^{mwa1}_i\mathbbm{1}\{x_a-y_a=i\} \mathbbm{1}\{x_a>y_a\}+
\sum^{59}_{i=1} \Lambda^{mwa2}_i\mathbbm{1}\{y_a-x_a=i\} \mathbbm{1}\{x_a<y_a\} \nonumber \\
&+\sum^{2}_{j=1} \Lambda^{mwe1}_j\mathbbm{1}\{x_e-y_e=j\}\mathbbm{1}\{x_e>y_e\}+
\sum^{2}_{j=1} \Lambda^{mwe2}_j\mathbbm{1}\{y_e-x_e=j\}\mathbbm{1}\{x_e<y_e\}, \\
% \label{eq: asy_specific_surplus_women}
 \alpha_{xy}({\Gammav})&=\Gamma_0+ \sum^{60}_{i=2} \Gamma^{ma}_{i} \mathbbm{1}\{x_a=i\}+\sum^{3}_{j=2} \Gamma^{me}_j \mathbbm{1}\{x_e=j\}
+ \sum^{60}_{i=2} \Gamma^{wa}_i \mathbbm{1}\{y_a=i\}+\sum^{3}_{j=2} \Gamma^{we}_j \mathbbm{1}\{y_e=j\} \nonumber \\
&+\sum^{59}_{i=1} \Gamma^{mwa1}_i\mathbbm{1}\{x_a-y_a=i\} \mathbbm{1}\{x_a>y_a\}+
\sum^{59}_{i=1} \Gamma^{mwa2}_i\mathbbm{1}\{y_a-x_a=i\} \mathbbm{1}\{x_a<y_a\} \nonumber \\
&+\sum^{2}_{j=1} \Gamma^{mwe1}_j\mathbbm{1}\{x_e-y_e=j\}\mathbbm{1}\{x_e>y_e\}+
\sum^{2}_{j=1} \Gamma^{mwe2}_j\mathbbm{1}\{y_e-x_e=j\}\mathbbm{1}\{x_e<y_e\}.
\end{align*}
The estimated joint systematic utilities using the above asymmetric specification are provided below.

\begin{figure}[!htb]
\minipage{0.32\textwidth}
  \includegraphics[width=\linewidth]{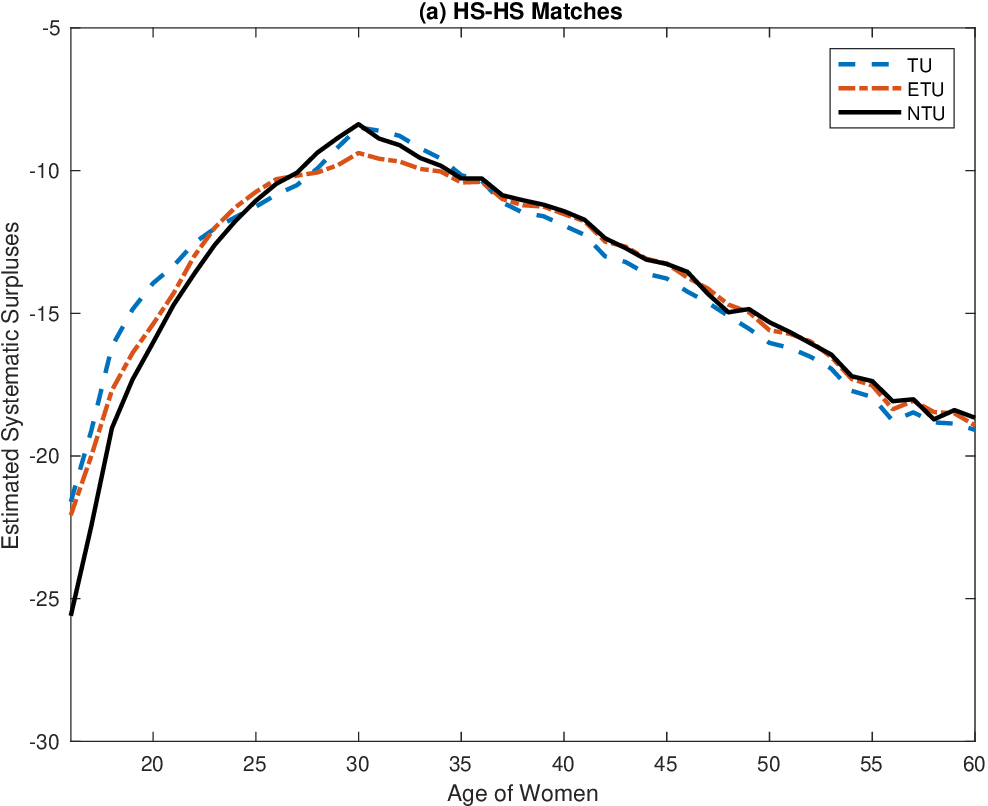}
%  \caption{A really Awesome Image}\label{fig:awesome_image1}
\endminipage\hfill
\minipage{0.31\textwidth}
  \includegraphics[width=\linewidth]{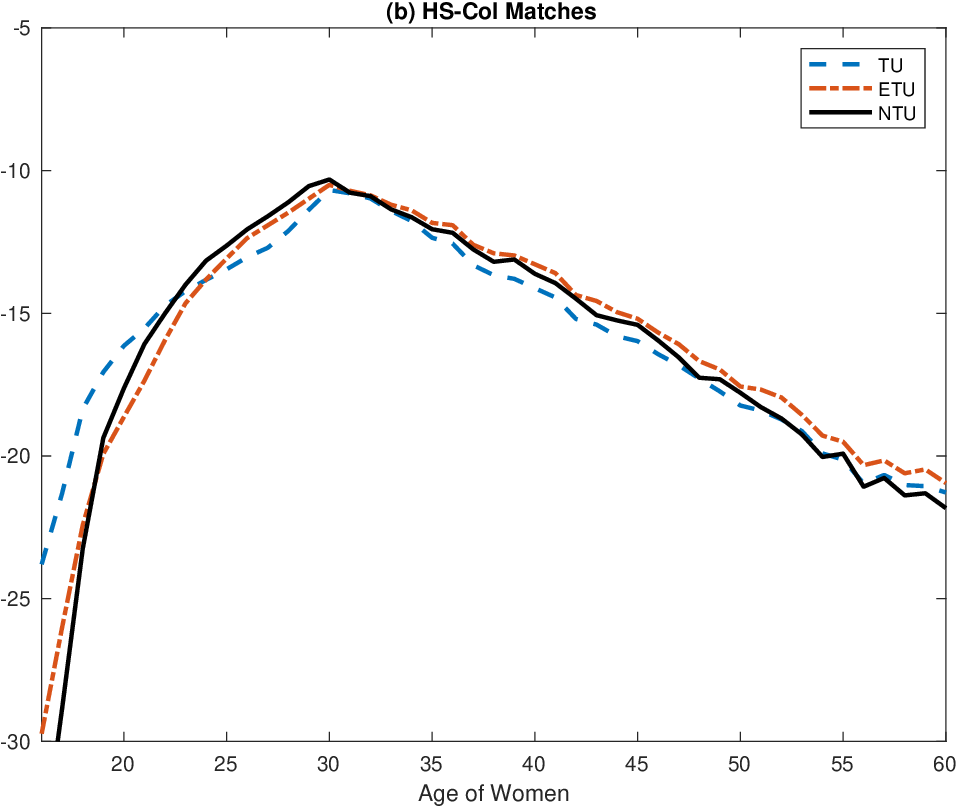}
%  \caption{A really Awesome Image}\label{fig:awesome_image2}
\endminipage\hfill
\minipage{0.31\textwidth}%
  \includegraphics[width=\linewidth]{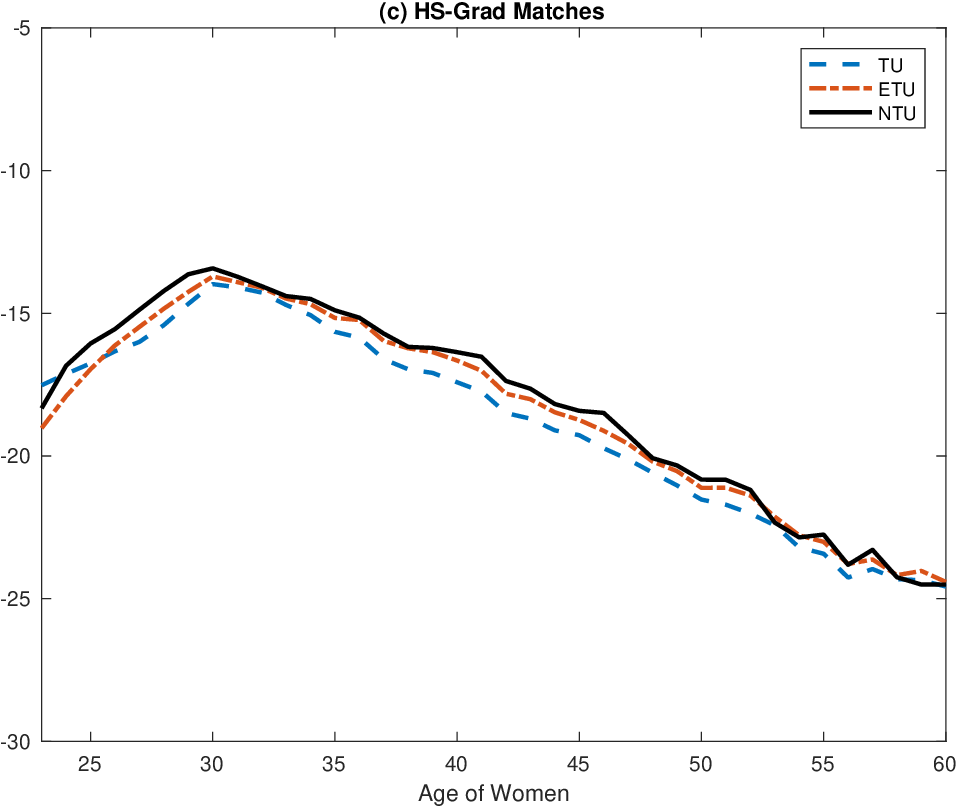}
%  \caption{A really Awesome Image}\label{fig:awesome_image3}
\endminipage
\centering
\caption{\centering Estimated $U_{xy}(\hat{\Lambdav})+V_{xy}(\hat{\Gammav})$ for high school men of 30 years old
 (Asymmetric Specification)}
\label{fig: joint_surplus_hs_asy}
\end{figure}

\begin{figure}[!htbp]
\minipage{0.32\textwidth}
  \includegraphics[width=\linewidth]{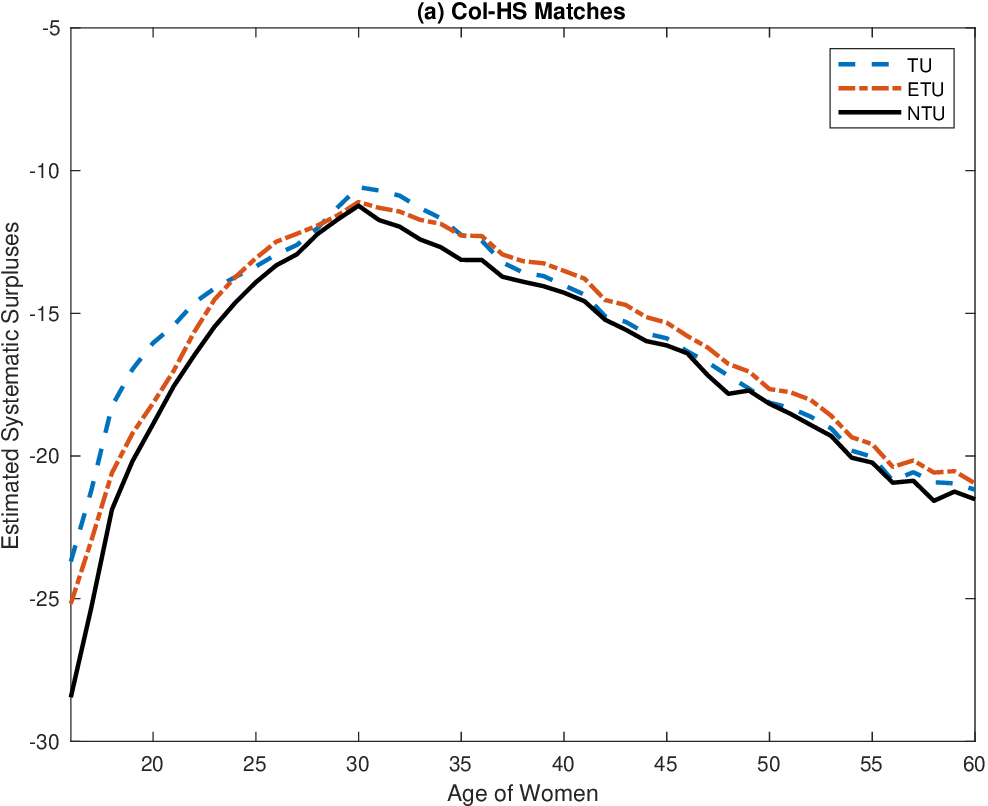}
%  \caption{A really Awesome Image}\label{fig:awesome_image1}
\endminipage\hfill
\minipage{0.31\textwidth}
  \includegraphics[width=\linewidth]{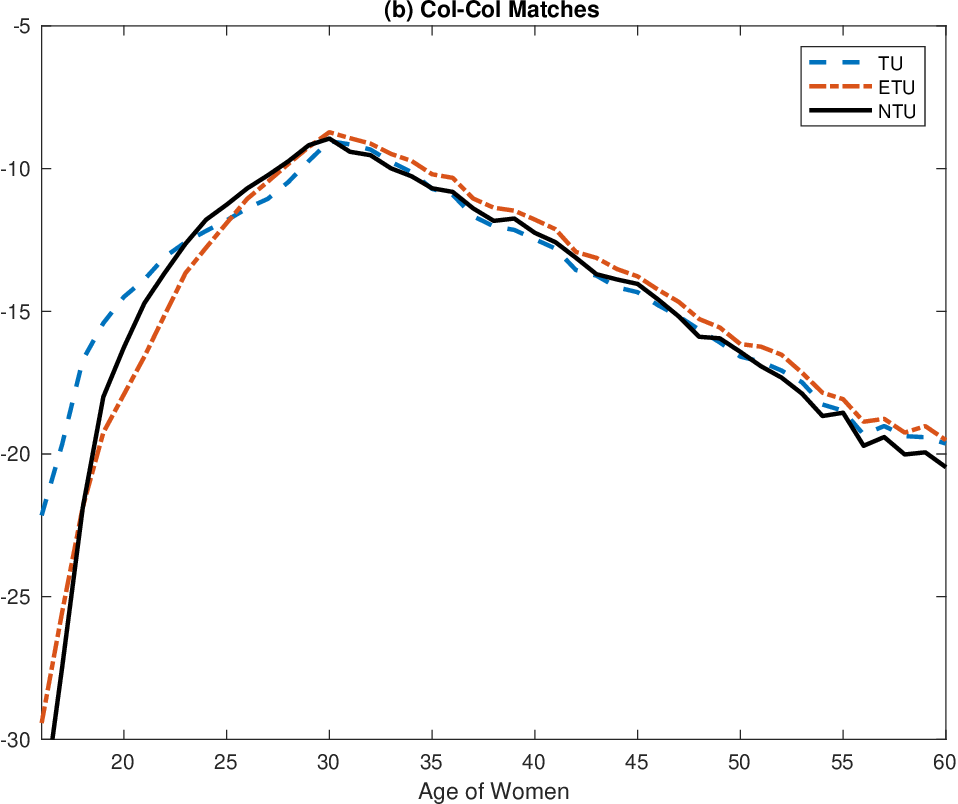}
%  \caption{A really Awesome Image}\label{fig:awesome_image2}
\endminipage\hfill
\minipage{0.31\textwidth}%
  \includegraphics[width=\linewidth]{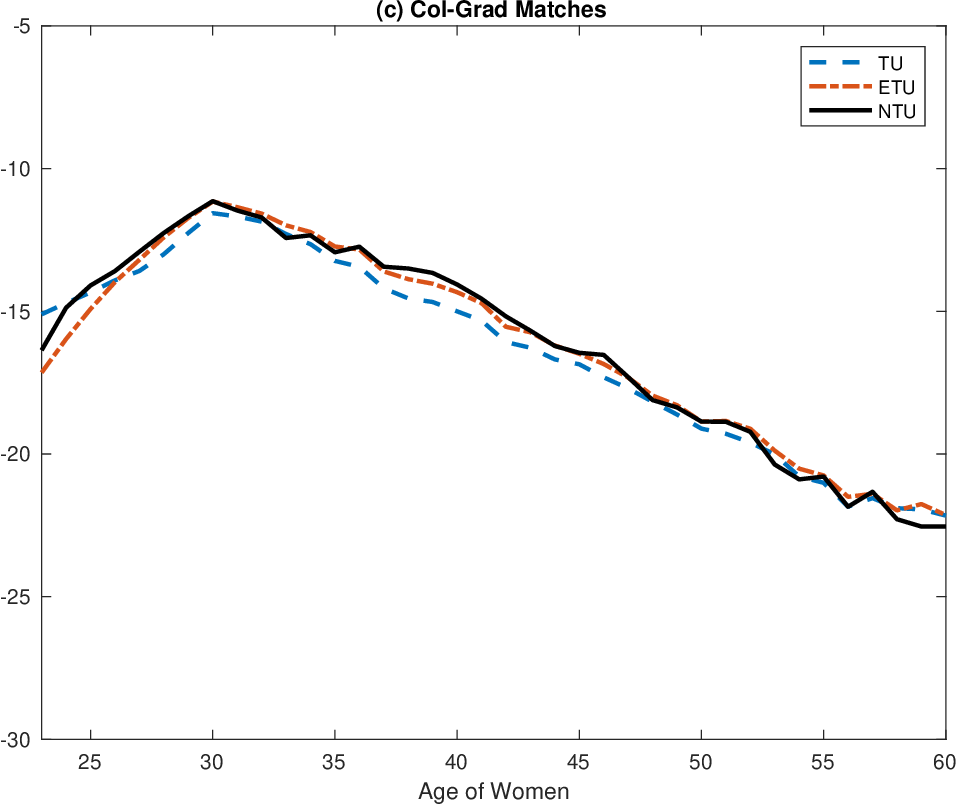}
%  \caption{A really Awesome Image}\label{fig:awesome_image3}
\endminipage\hfill
% \centering
\caption{Estimated $U_{xy}(\hat{\Lambdav})+V_{xy}(\hat{\Gammav})$ for college men of 30 years old  (Asymmetric Specification)}
\label{fig: joint_surplus_col_asy}
\end{figure}

\begin{figure}[!htbp]
\minipage{0.32\textwidth}
  \includegraphics[width=\linewidth]{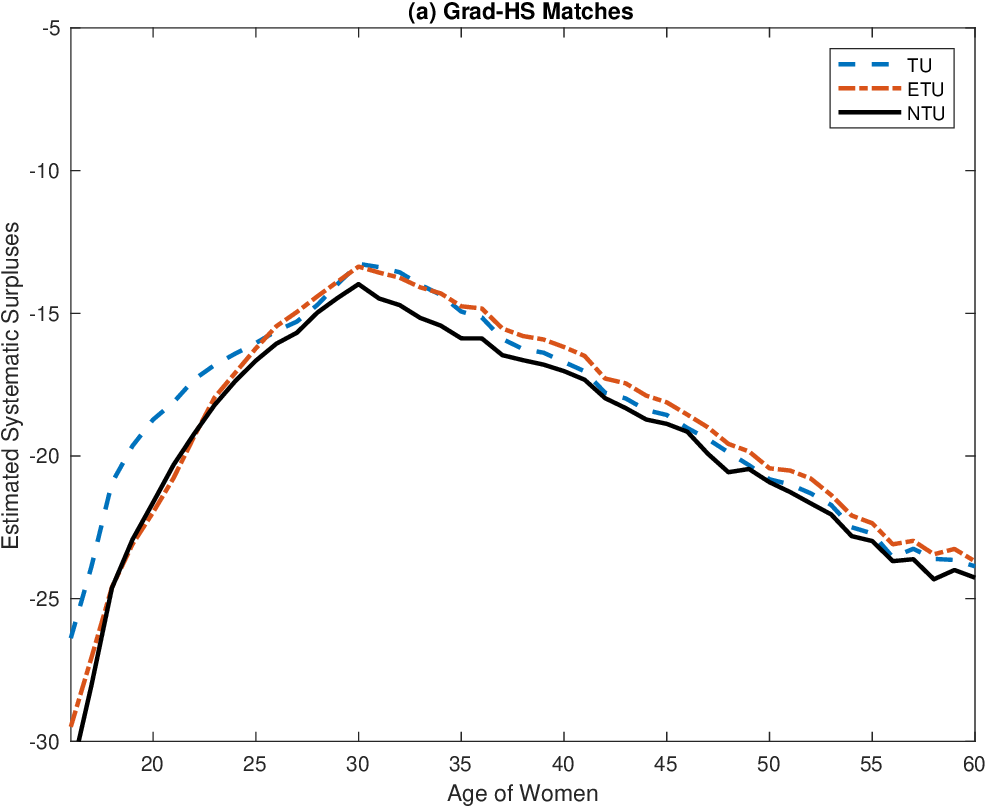}
%  \caption{A really Awesome Image}\label{fig:awesome_image1}
\endminipage\hfill
\minipage{0.31\textwidth}
  \includegraphics[width=\linewidth]{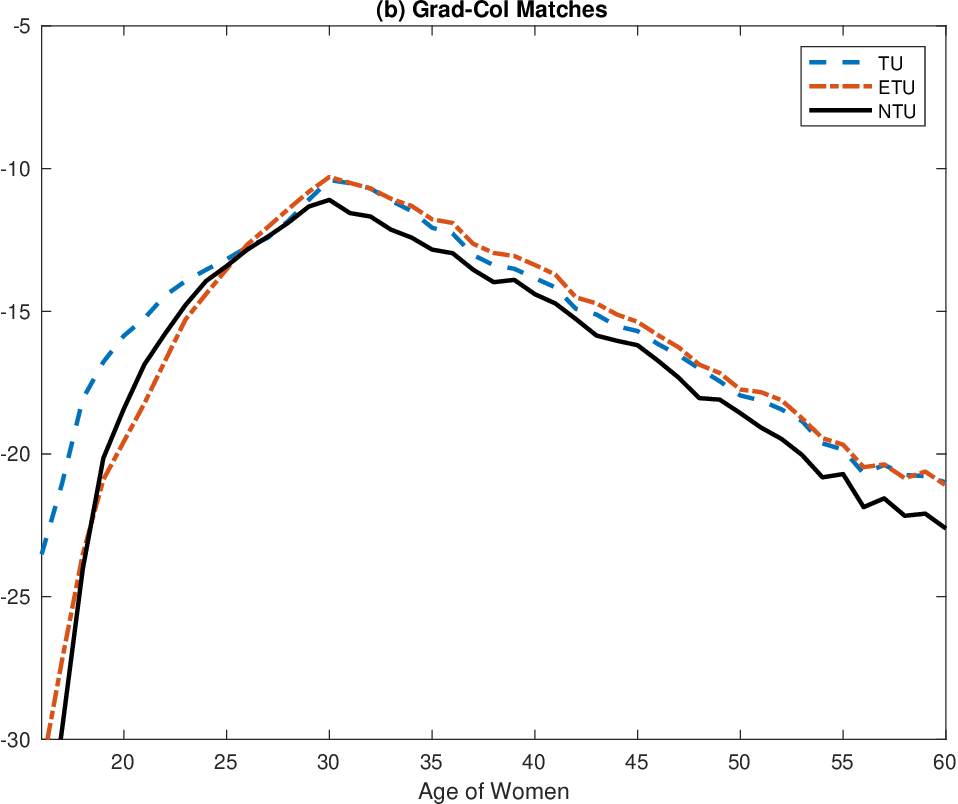}
%  \caption{A really Awesome Image}\label{fig:awesome_image2}
\endminipage\hfill
\minipage{0.31\textwidth}%
  \includegraphics[width=\linewidth]{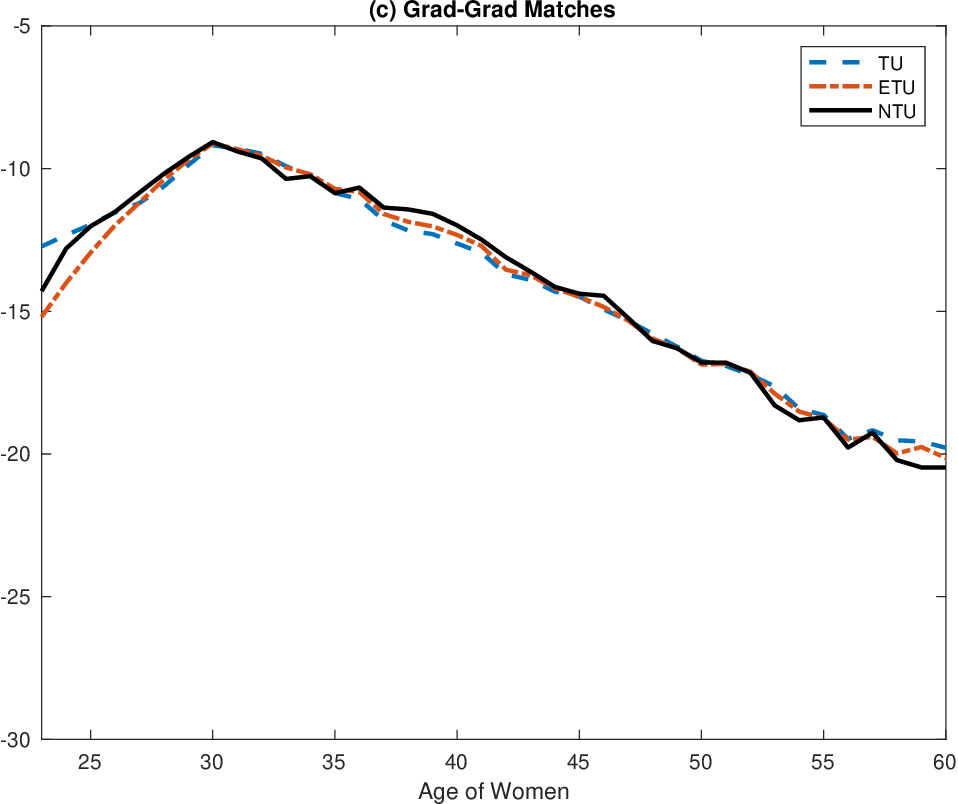}
%  \caption{A really Awesome Image}\label{fig:awesome_image3}
\endminipage\hfill
% \centering
\caption{Estimated $U_{xy}(\hat{\Lambdav})+V_{xy}(\hat{\Gammav})$ for graduate men of 30 years old  (Asymmetric Specification)}
\label{fig: joint_surplus_grad_asy}
\end{figure}

\clearpage

\subsection{Additional figures} \label{app:figures}

\begin{figure}[!htb]
\minipage{0.48\textwidth}
  \includegraphics[width=\linewidth]{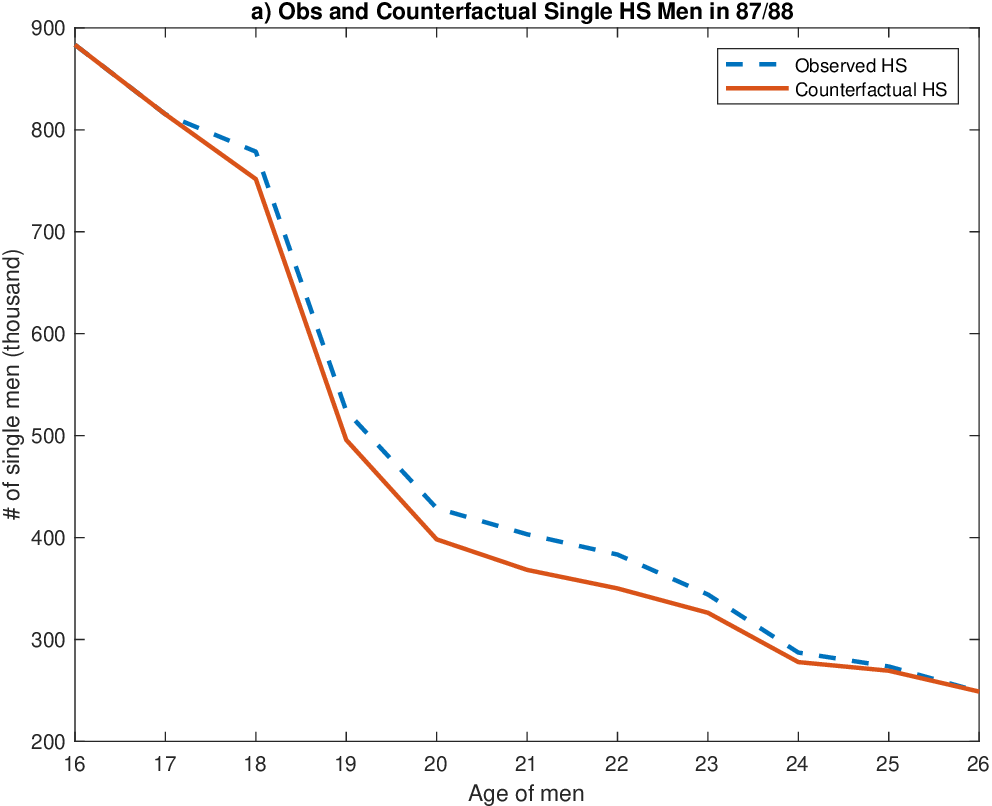}
%  \caption{A really Awesome Image}\label{fig:awesome_image1}
\endminipage\hfill
\minipage{0.48\textwidth}
  \includegraphics[width=\linewidth]{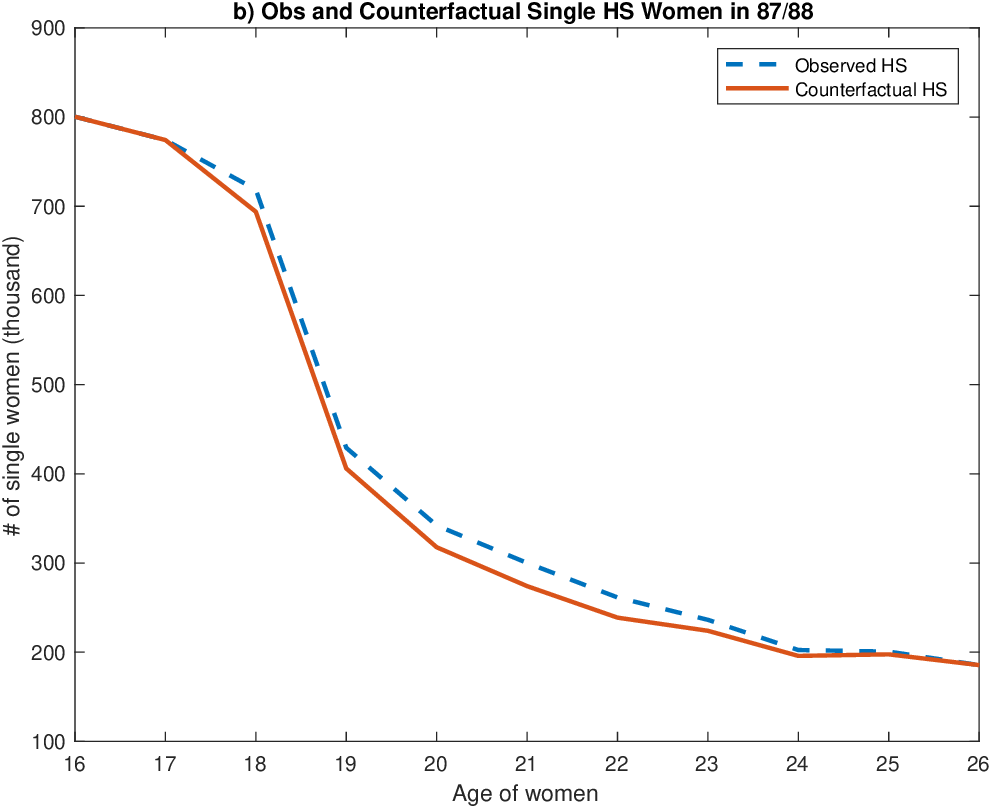}
%  \caption{A really Awesome Image}\label{fig:awesome_image2}
\endminipage\hfill
\minipage{0.48\textwidth}%
  \includegraphics[width=\linewidth]{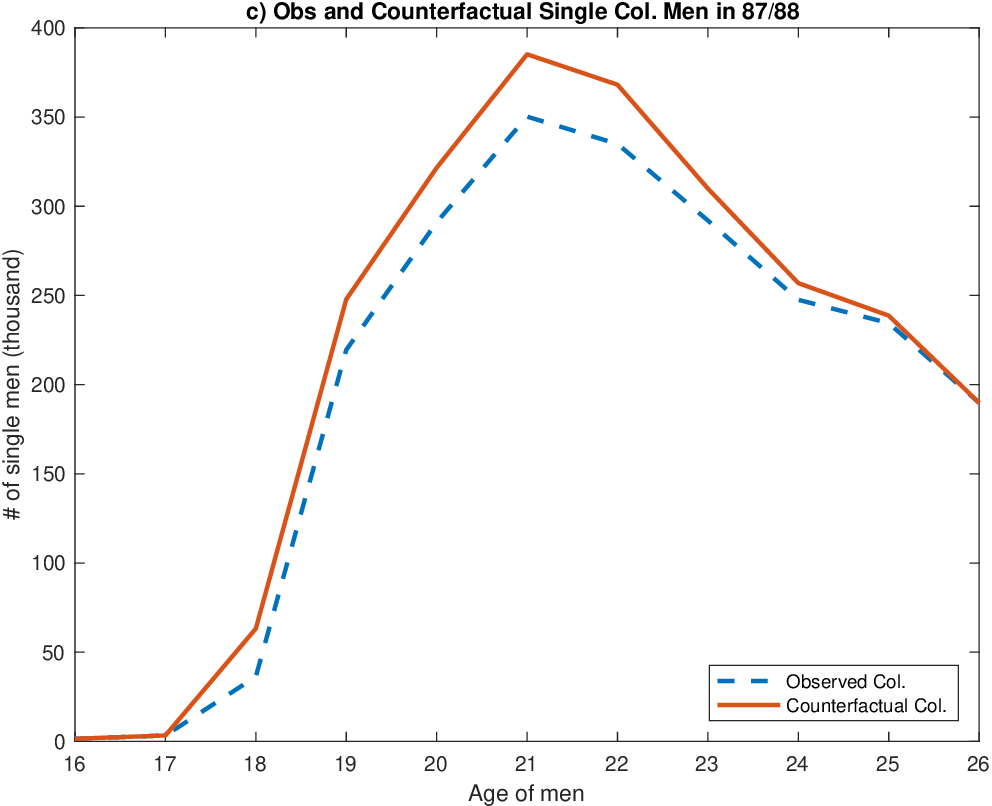}
%  \caption{A really Awesome Image}\label{fig:awesome_image3}
\endminipage\hfill
\minipage{0.48\textwidth}
  \includegraphics[width=\linewidth]{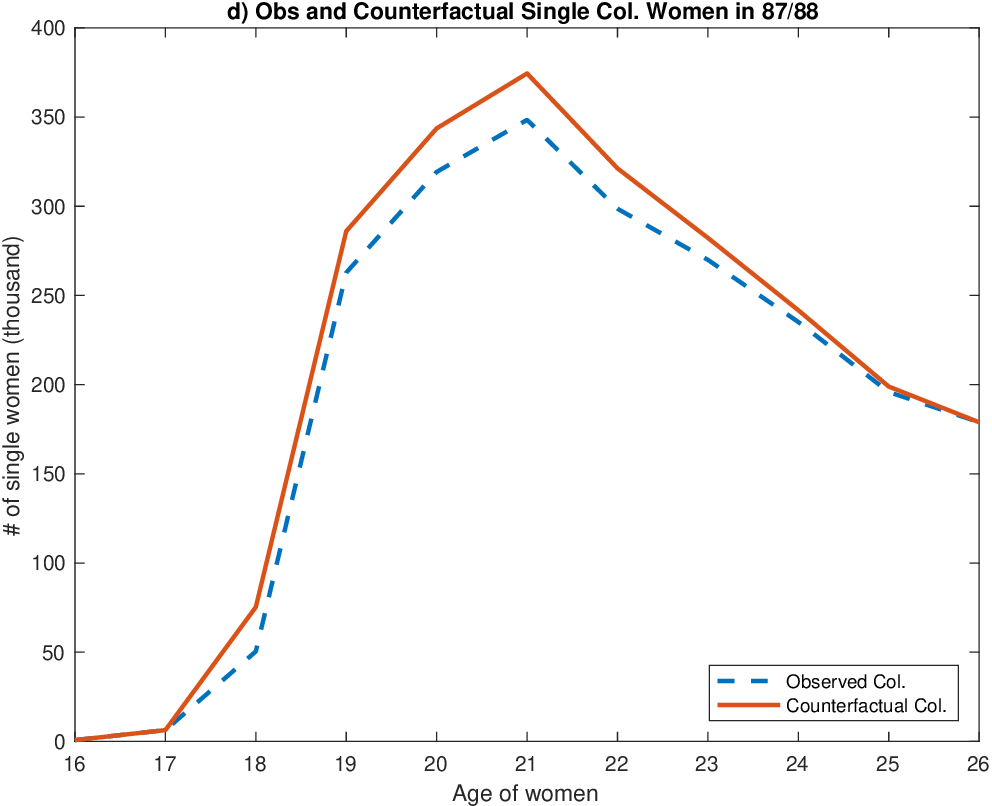}
%  \caption{A really Awesome Image}\label{fig:awesome_image1}
\endminipage\hfill
\centering
\caption{Changes in available HS and Col single men and women}
\label{fig:CompareS}
\floatfoot{These figures graph the observed and counterfactual available single men and women by ages between 16 and 22 in 1986. In the counterfactual scenario where the aid program was not eliminated, we expect there to be more available single college graduates and fewer available single high school graduates. As expected, our calculations create a wedge between the observed and counterfactual number of available single individuals between the ages of eighteen and twenty five.}
\end{figure}

\begin{figure}[!htbp]
\minipage{0.32\textwidth}
  \includegraphics[width=\linewidth]{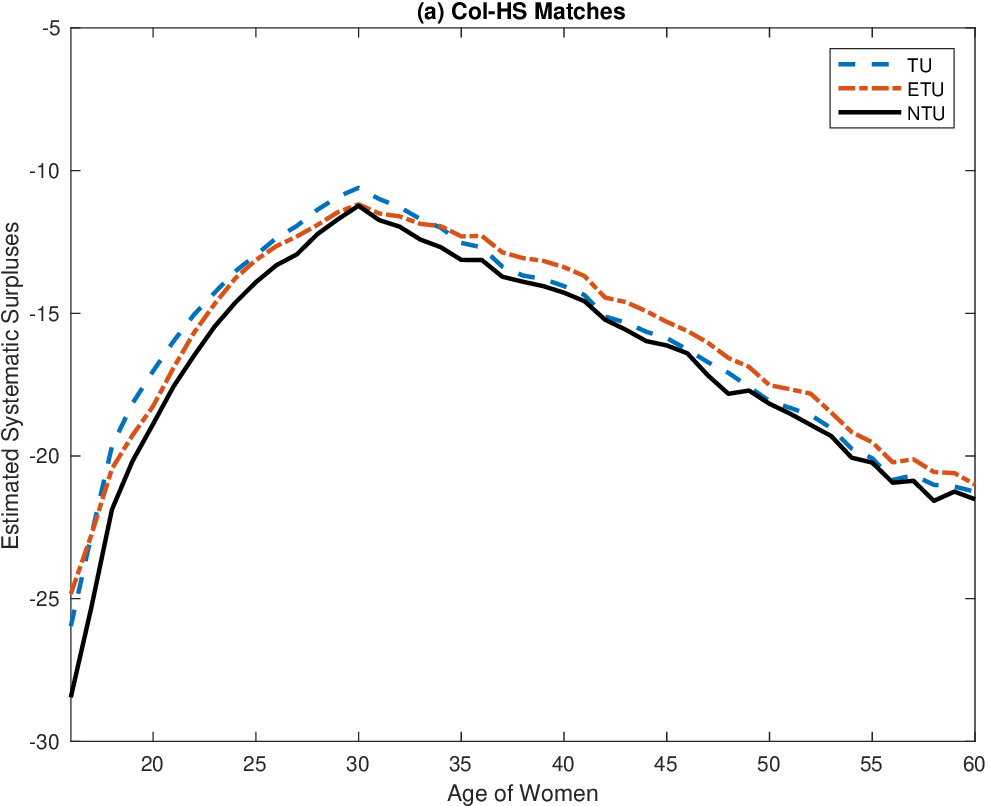}
%  \caption{A really Awesome Image}\label{fig:awesome_image1}
\endminipage\hfill
\minipage{0.31\textwidth}
  \includegraphics[width=\linewidth]{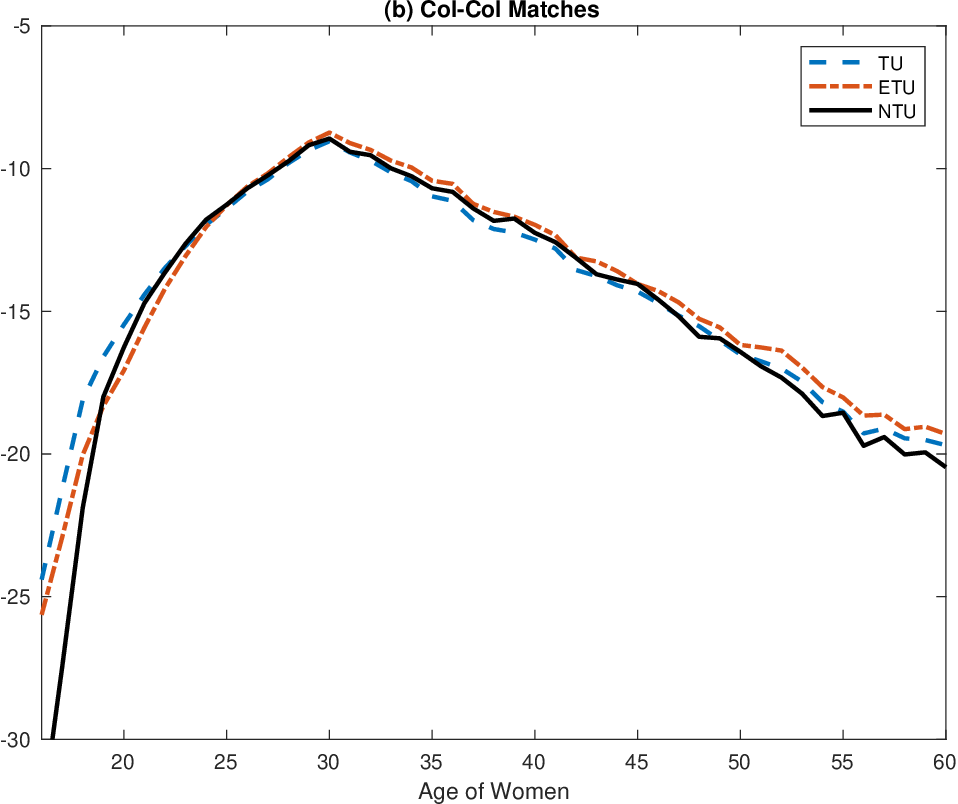}
%  \caption{A really Awesome Image}\label{fig:awesome_image2}
\endminipage\hfill
\minipage{0.31\textwidth}%
  \includegraphics[width=\linewidth]{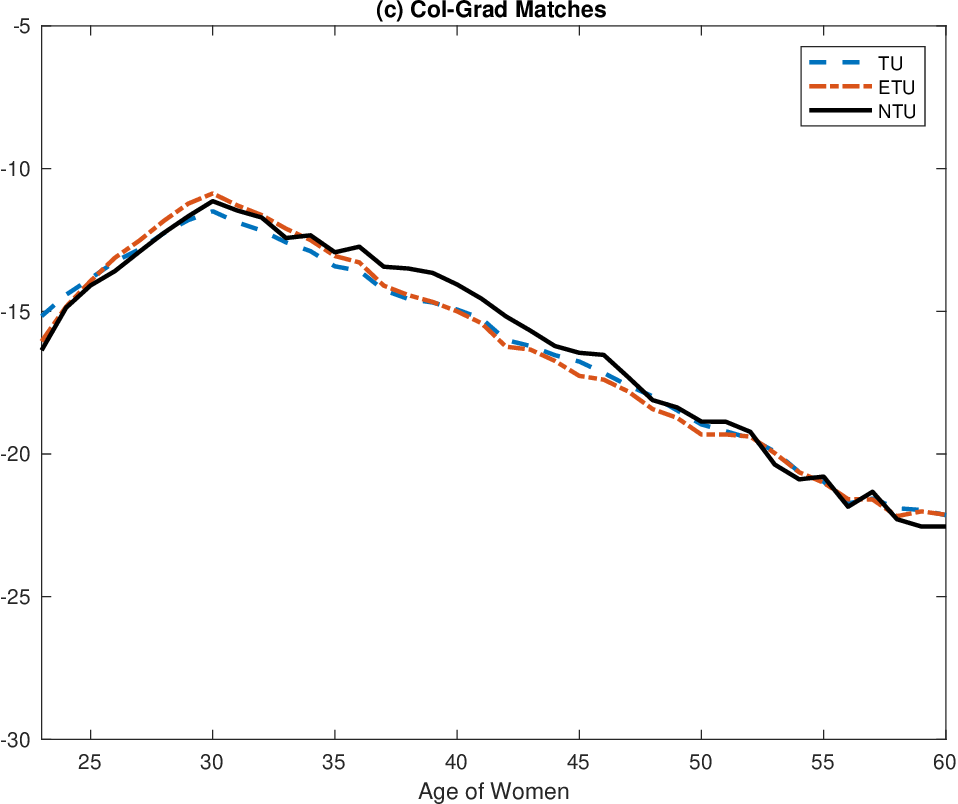}
%  \caption{A really Awesome Image}\label{fig:awesome_image3}
\endminipage\hfill
% \centering
\caption{Estimated $U_{xy}(\hat{\Lambdav})+V_{xy}(\hat{\Gammav})$ for college men of 30 years old}
\label{fig: joint_surplus_col}
\end{figure}

\begin{figure}[!htbp]
\minipage{0.32\textwidth}
  \includegraphics[width=\linewidth]{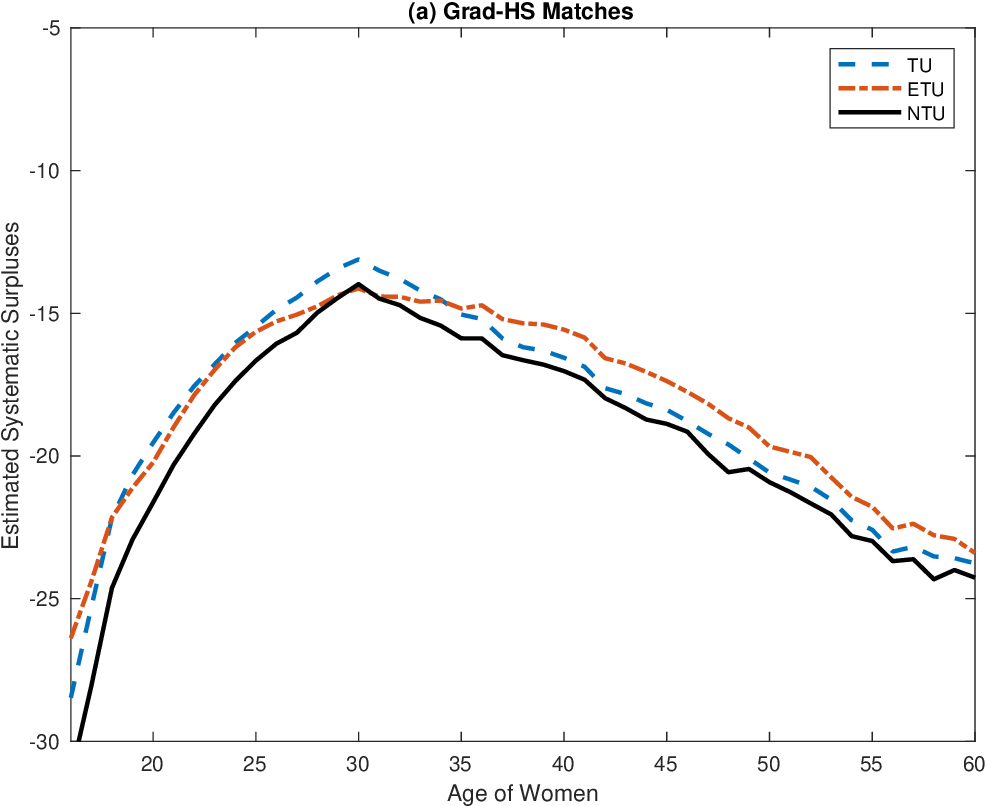}
%  \caption{A really Awesome Image}\label{fig:awesome_image1}
\endminipage\hfill
\minipage{0.31\textwidth}
  \includegraphics[width=\linewidth]{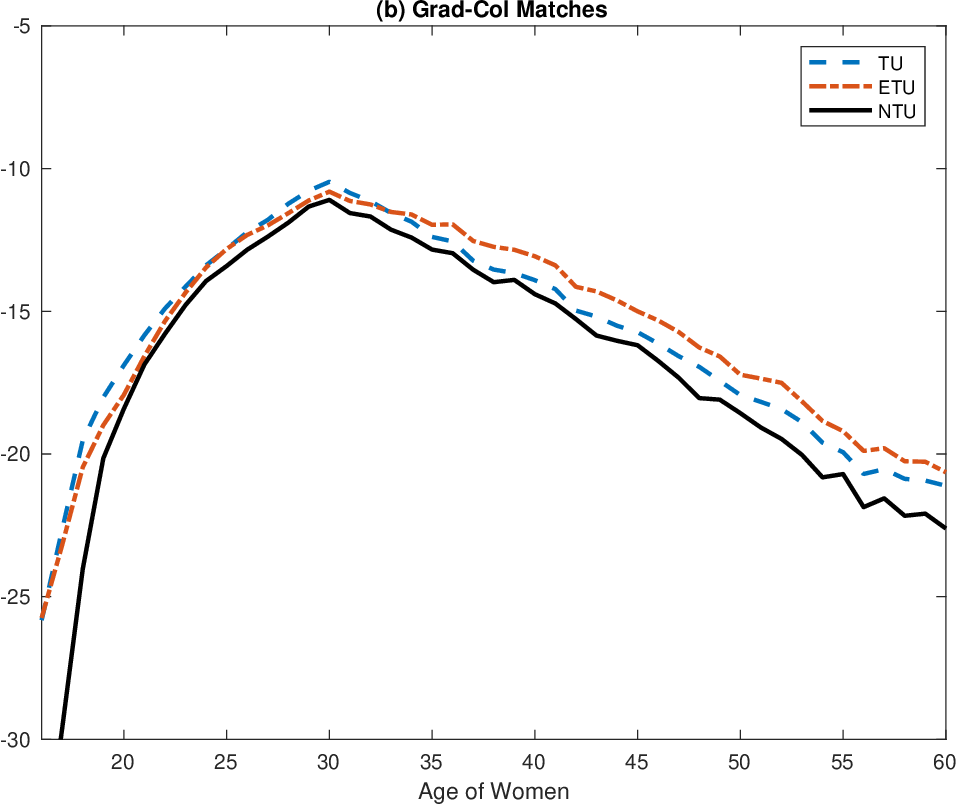}
%  \caption{A really Awesome Image}\label{fig:awesome_image2}
\endminipage\hfill
\minipage{0.31\textwidth}%
  \includegraphics[width=\linewidth]{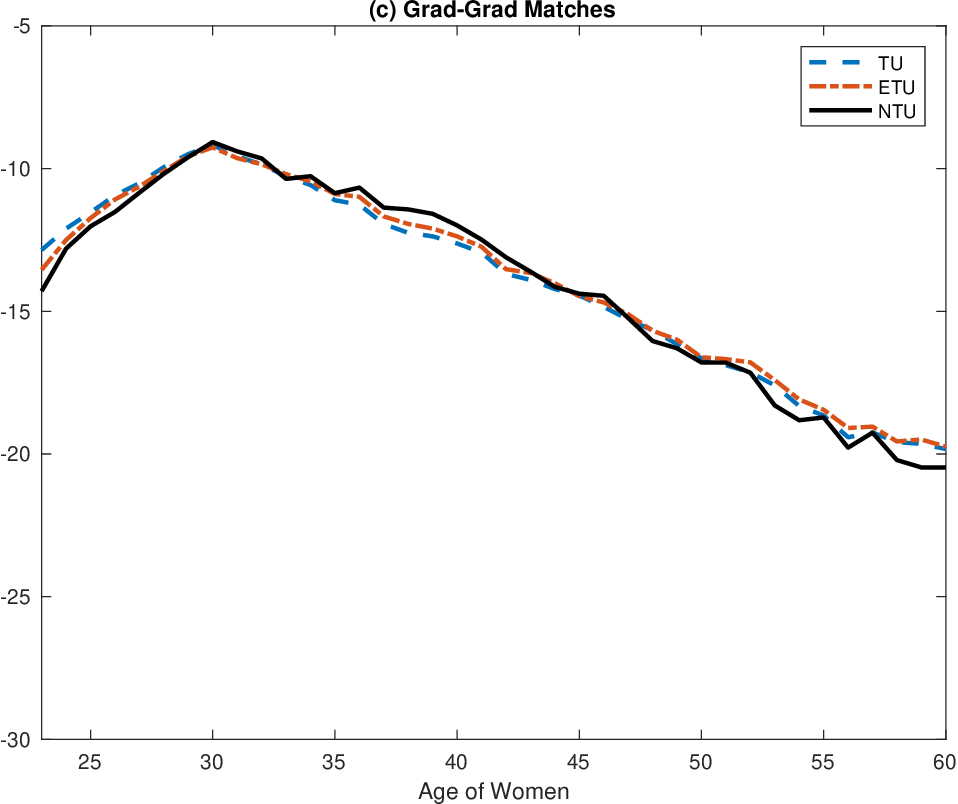}
%  \caption{A really Awesome Image}\label{fig:awesome_image3}
\endminipage\hfill
% \centering
\caption{Estimated $U_{xy}(\hat{\Lambdav})+V_{xy}(\hat{\Gammav})$ for graduate men of 30 years old}
\label{fig: joint_surplus_grad}
\end{figure}

\begin{figure}[!htbp]
\minipage{0.32\textwidth}
  \includegraphics[width=\linewidth]{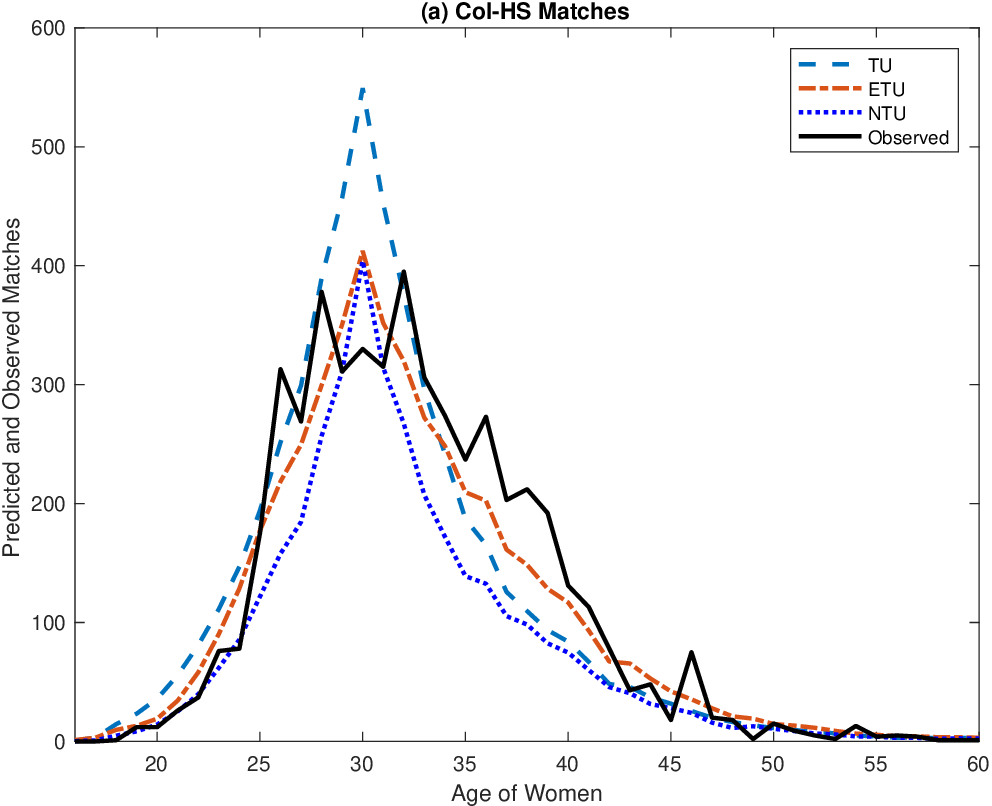}
%  \caption{A really Awesome Image}\label{fig:awesome_image1}
\endminipage\hfill
\minipage{0.31\textwidth}
  \includegraphics[width=\linewidth]{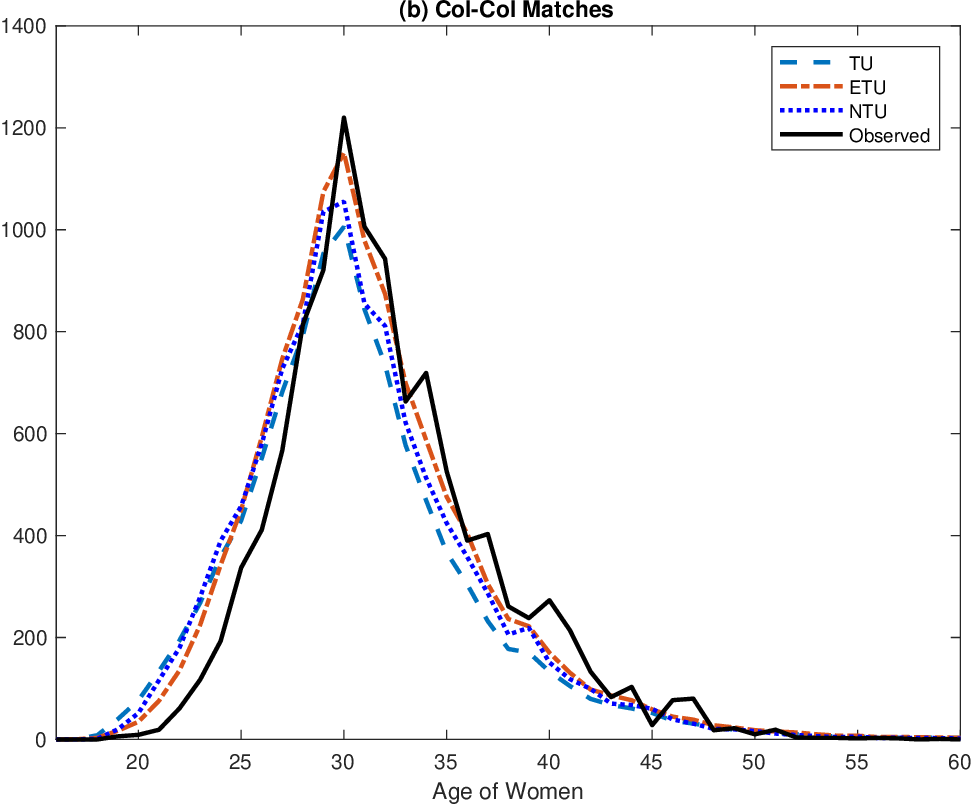}
%  \caption{A really Awesome Image}\label{fig:awesome_image2}
\endminipage\hfill
\minipage{0.31\textwidth}%
  \includegraphics[width=\linewidth]{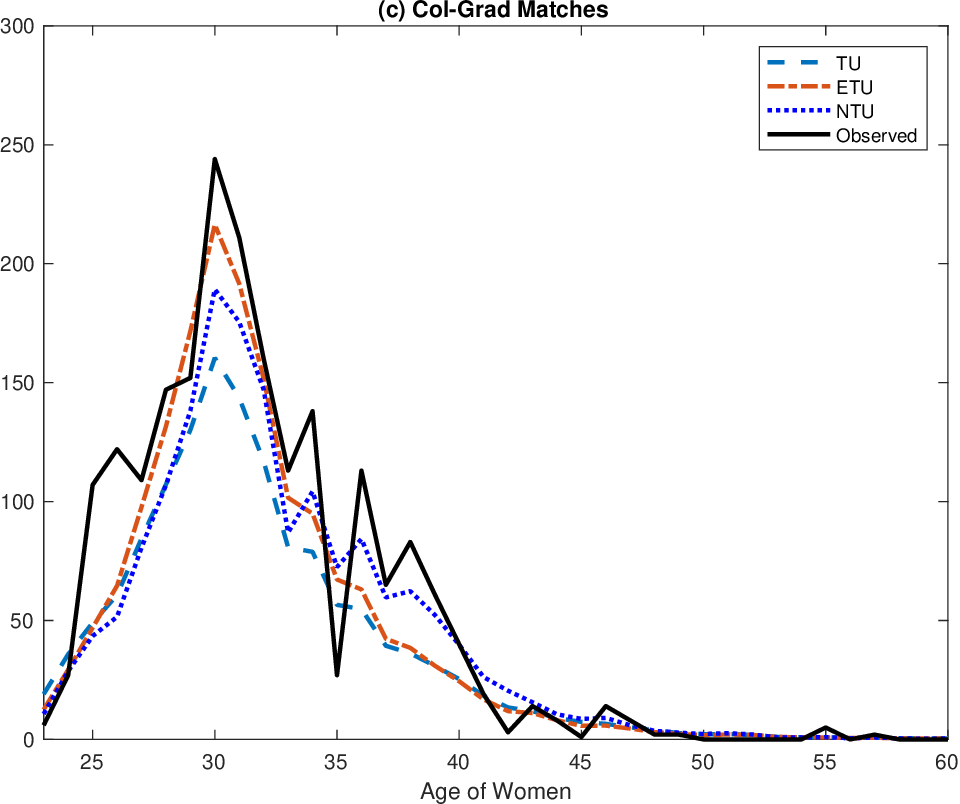}
%  \caption{A really Awesome Image}\label{fig:awesome_image3}
\endminipage\hfill
% \centering
\caption{Predicted and observed matches for college men of 30 years old}
\label{fig: matches_col}
\end{figure}

\begin{figure}[!htbp]
\minipage{0.32\textwidth}
  \includegraphics[width=\linewidth]{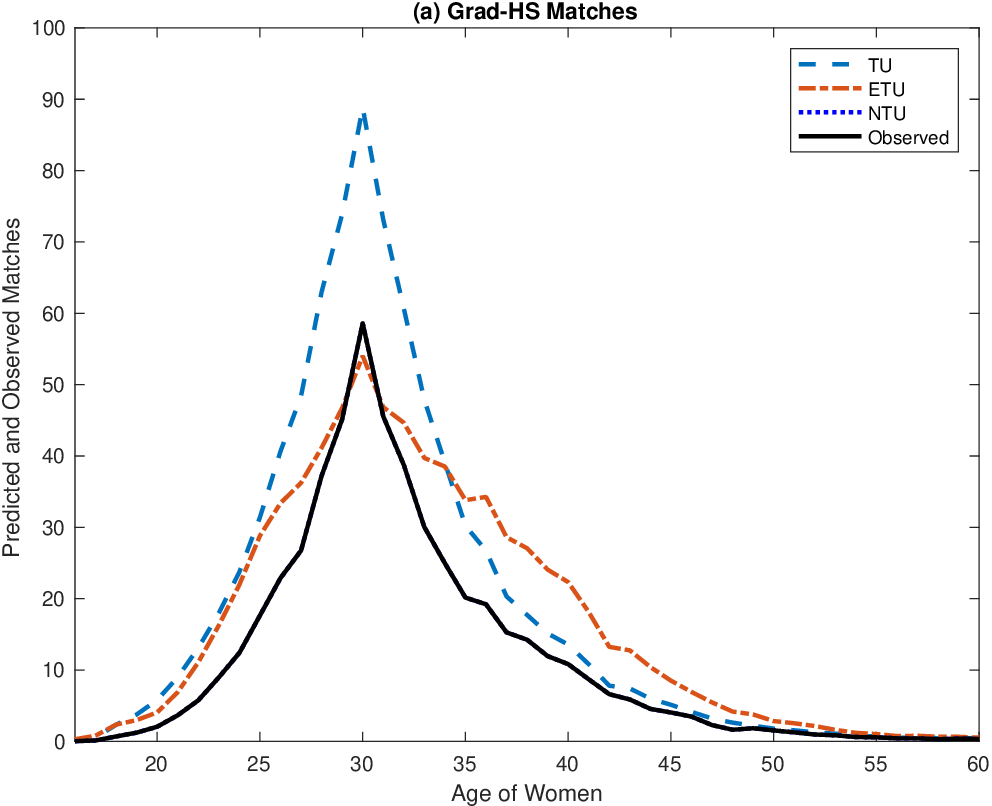}
%  \caption{A really Awesome Image}\label{fig:awesome_image1}
\endminipage\hfill
\minipage{0.31\textwidth}
  \includegraphics[width=\linewidth]{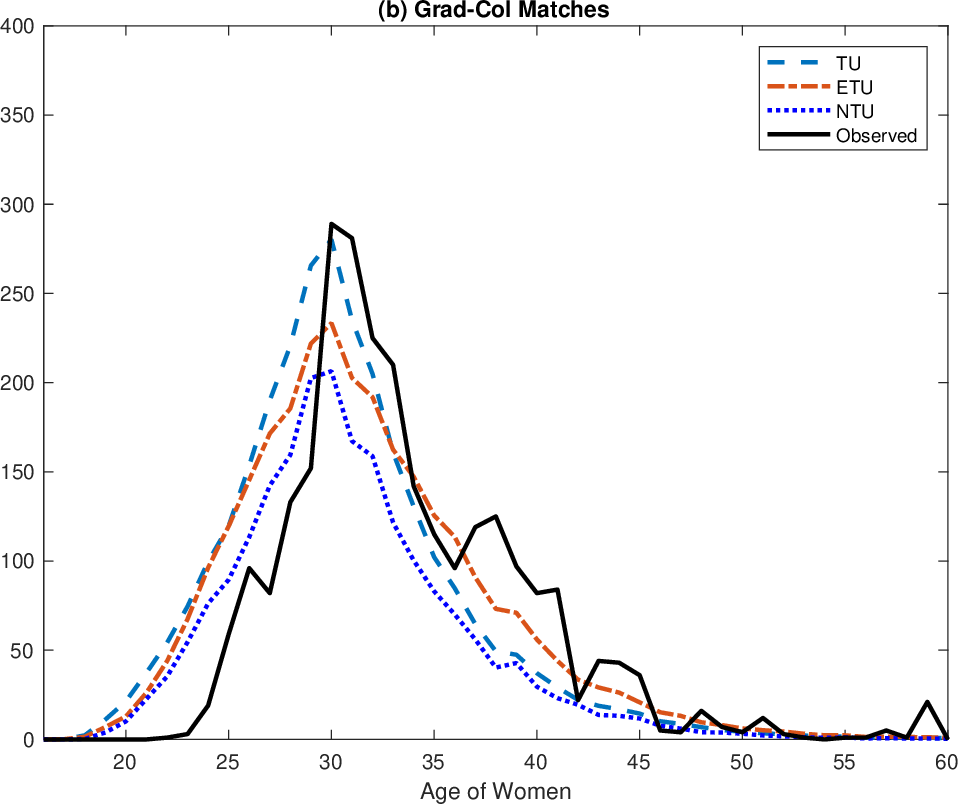}
%  \caption{A really Awesome Image}\label{fig:awesome_image2}
\endminipage\hfill
\minipage{0.31\textwidth}%
  \includegraphics[width=\linewidth]{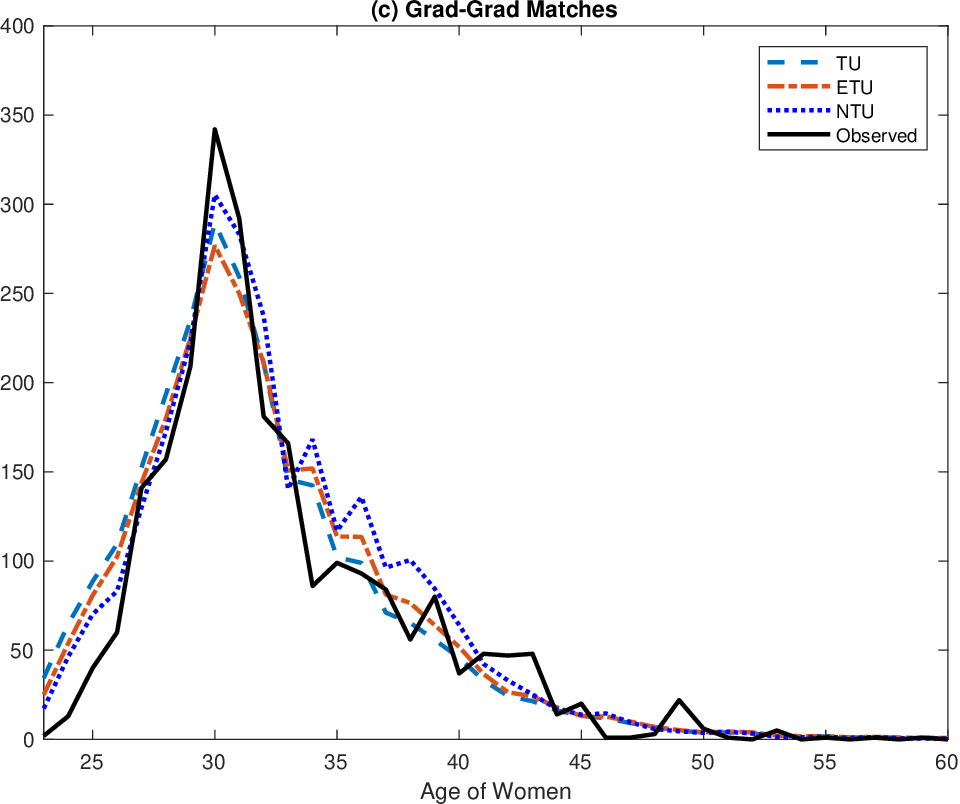}
%  \caption{A really Awesome Image}\label{fig:awesome_image3}
\endminipage\hfill
% \centering
\caption{Predicted and observed matches for graduate men of 30 years old}
\label{fig: matches_grad}
\end{figure}

\begin{figure}[!htb]
\minipage{0.32\textwidth}
  \includegraphics[width=\linewidth]{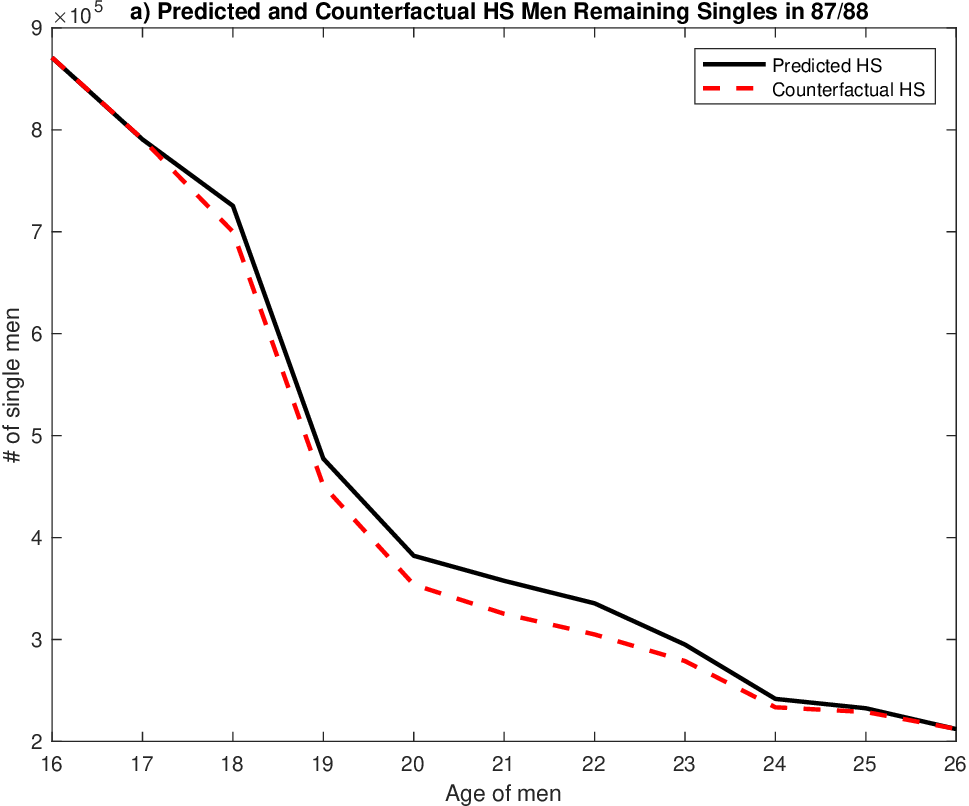}
%  \caption{A really Awesome Image}\label{fig:awesome_image1}
\endminipage\hfill
\minipage{0.32\textwidth}
  \includegraphics[width=\linewidth]{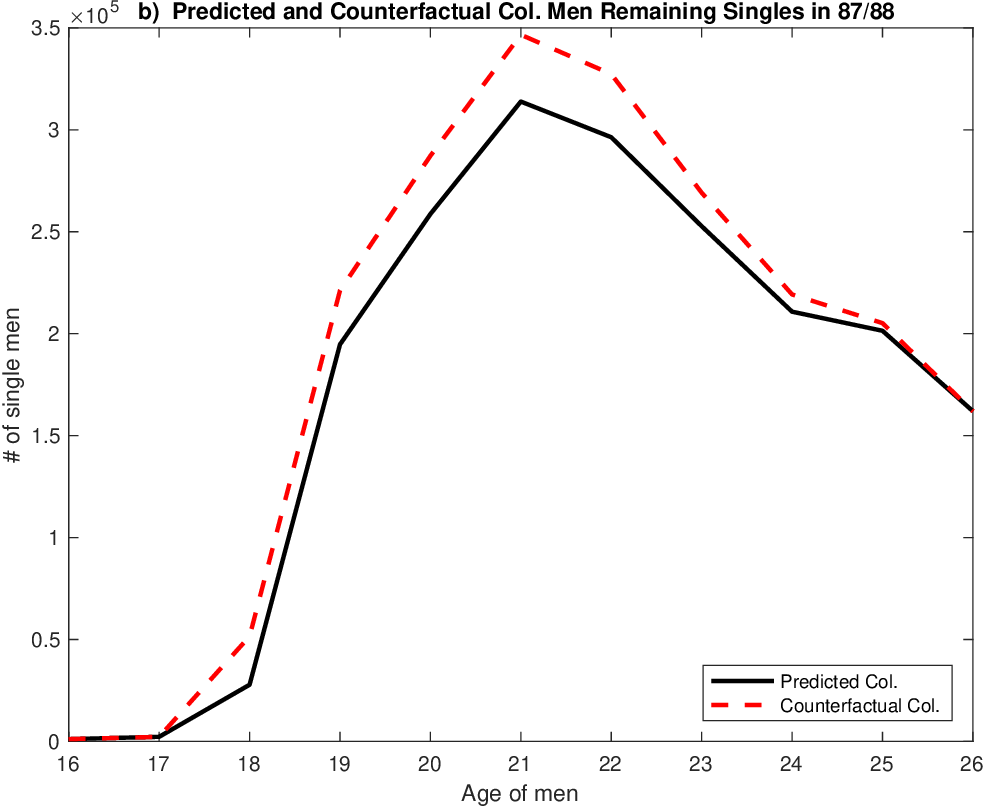}
%  \caption{A really Awesome Image}\label{fig:awesome_image2}
\endminipage\hfill
\minipage{0.32\textwidth}%
  \includegraphics[width=\linewidth]{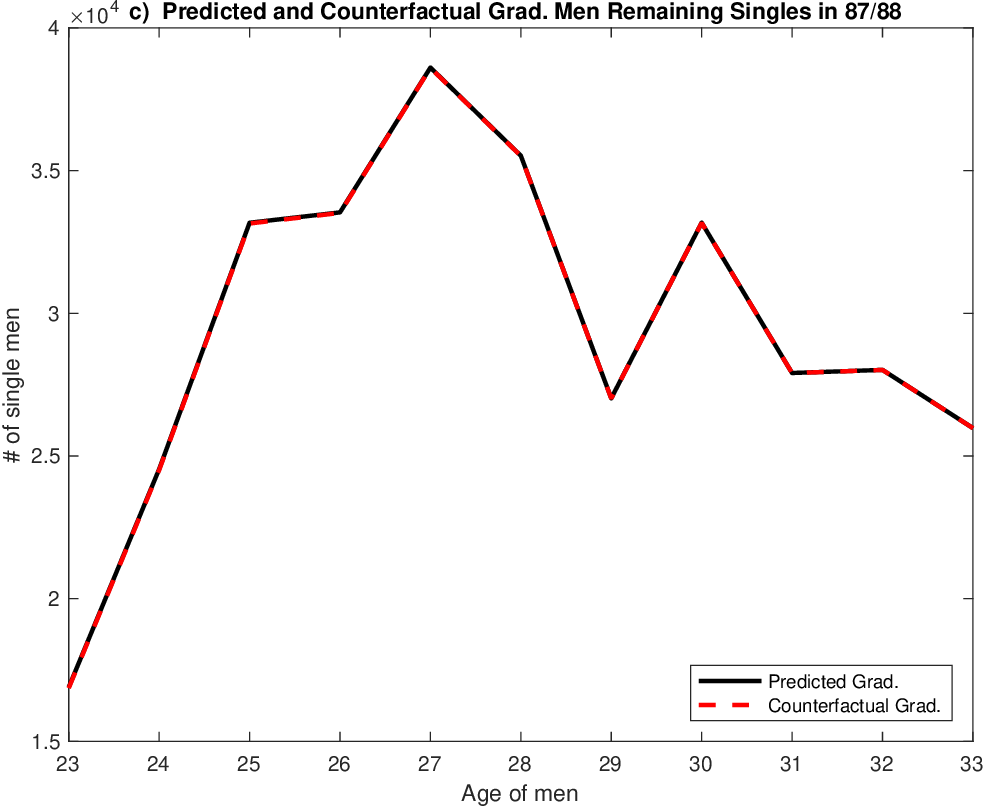}
%  \caption{A really Awesome Image}\label{fig:awesome_image3}
\endminipage\hfill
\minipage{0.32\textwidth}
  \includegraphics[width=\linewidth]{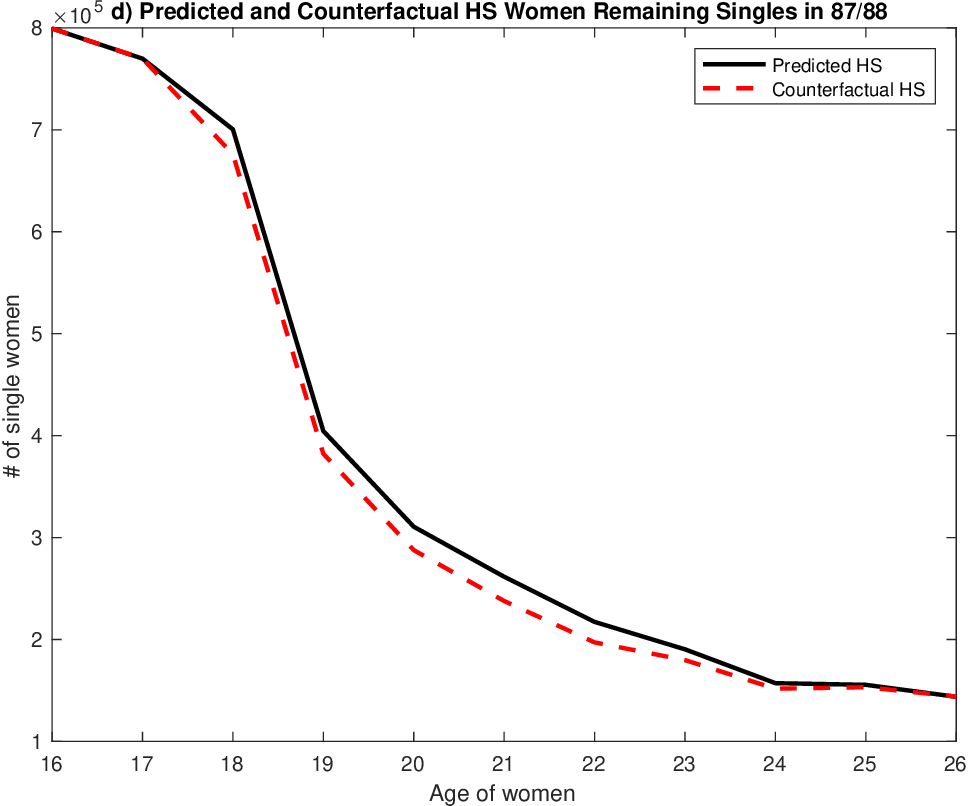}
%  \caption{A really Awesome Image}\label{fig:awesome_image1}
\endminipage\hfill
\minipage{0.32\textwidth}
  \includegraphics[width=\linewidth]{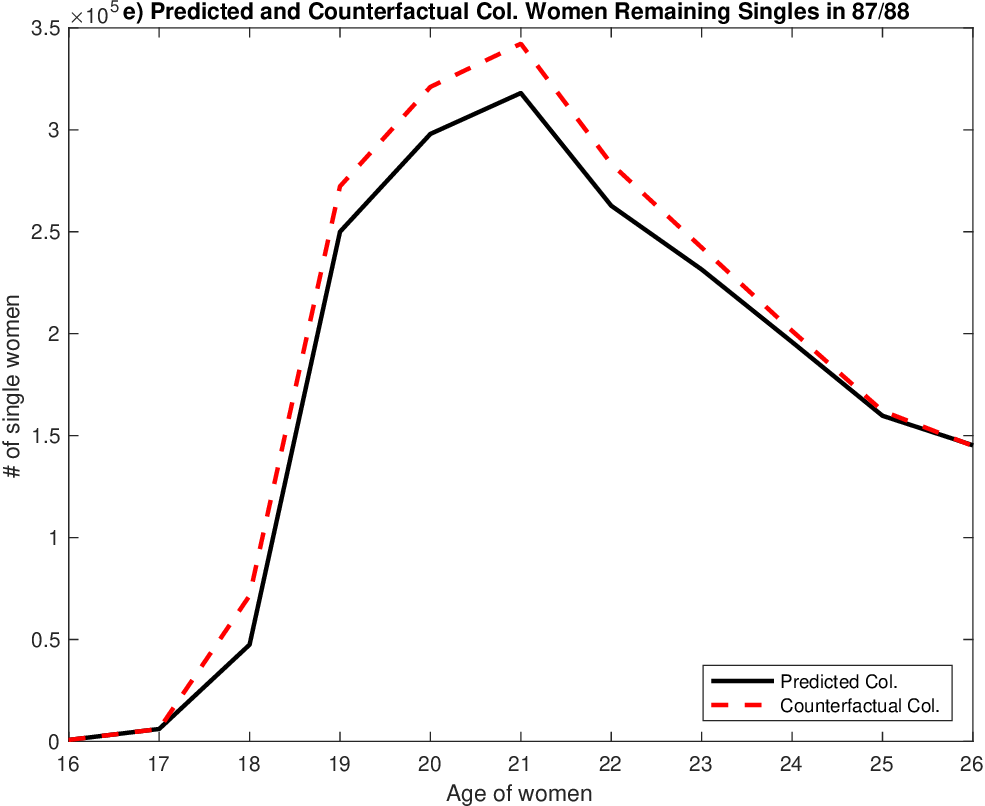}
%  \caption{A really Awesome Image}\label{fig:awesome_image2}
\endminipage\hfill
\minipage{0.32\textwidth}%
  \includegraphics[width=\linewidth]{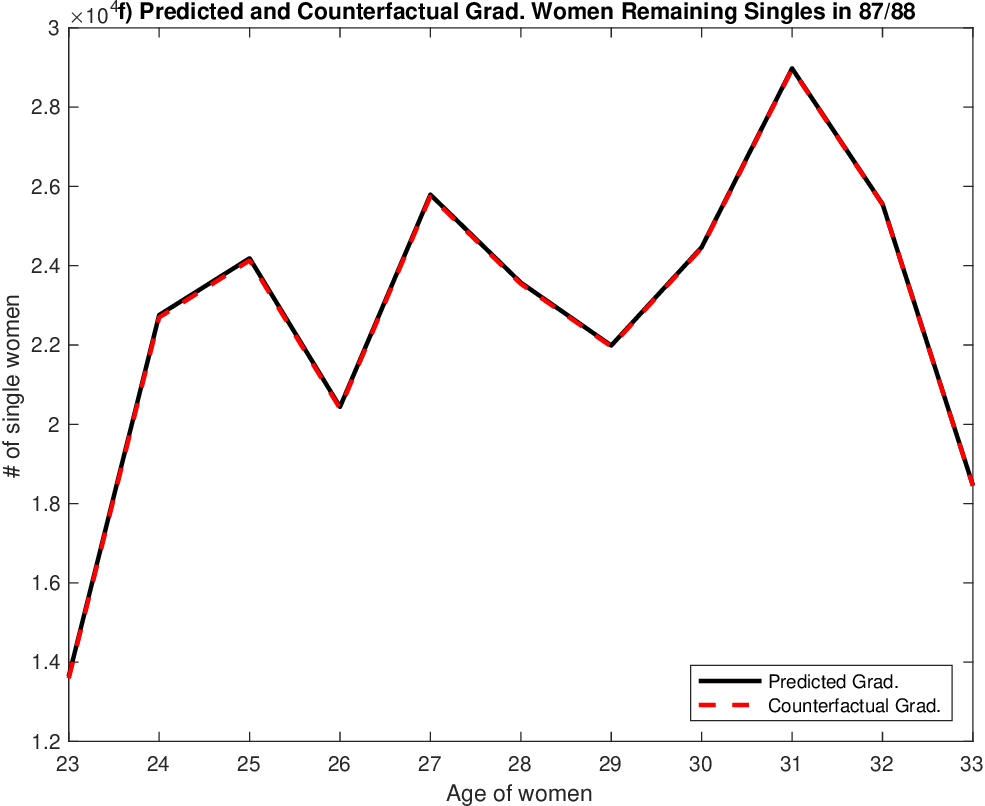}
%  \caption{A really Awesome Image}\label{fig:awesome_image3}
\endminipage
\centering
\caption{Predicted and counterfactual numbers of men and women remaining singles \\
\centering (TU-Optimal Model)}
\label{fig: pred_count_singles_TU}
\end{figure}

\begin{figure}[!htb]
\minipage{0.32\textwidth}
  \includegraphics[width=\linewidth]{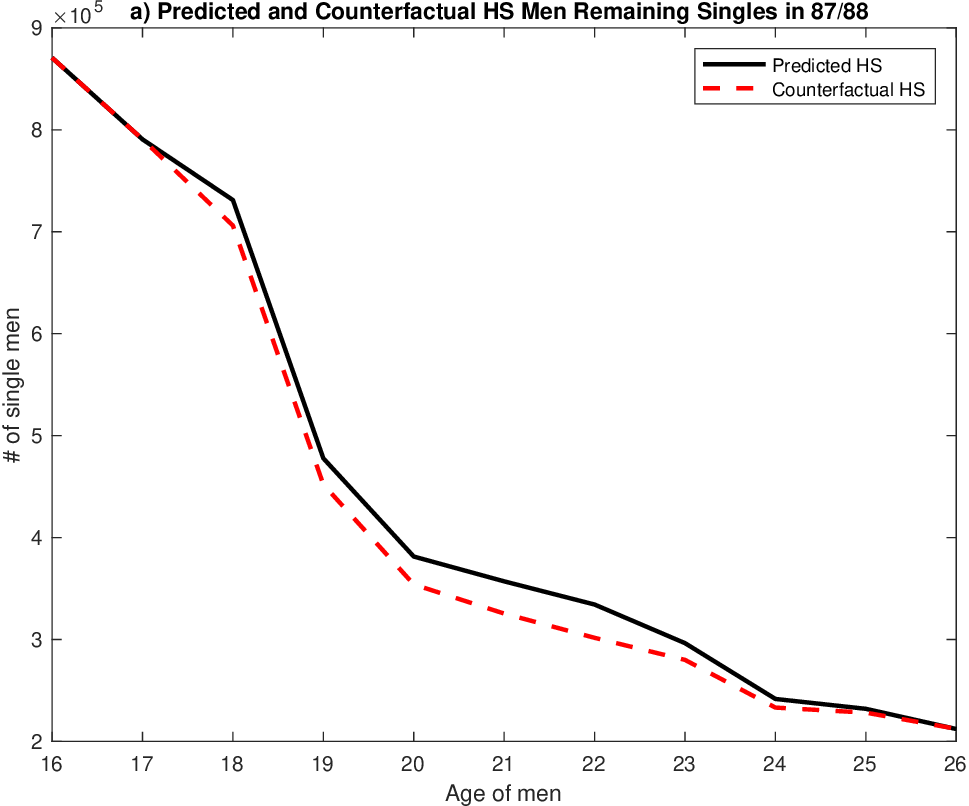}
%  \caption{A really Awesome Image}\label{fig:awesome_image1}
\endminipage\hfill
\minipage{0.32\textwidth}
  \includegraphics[width=\linewidth]{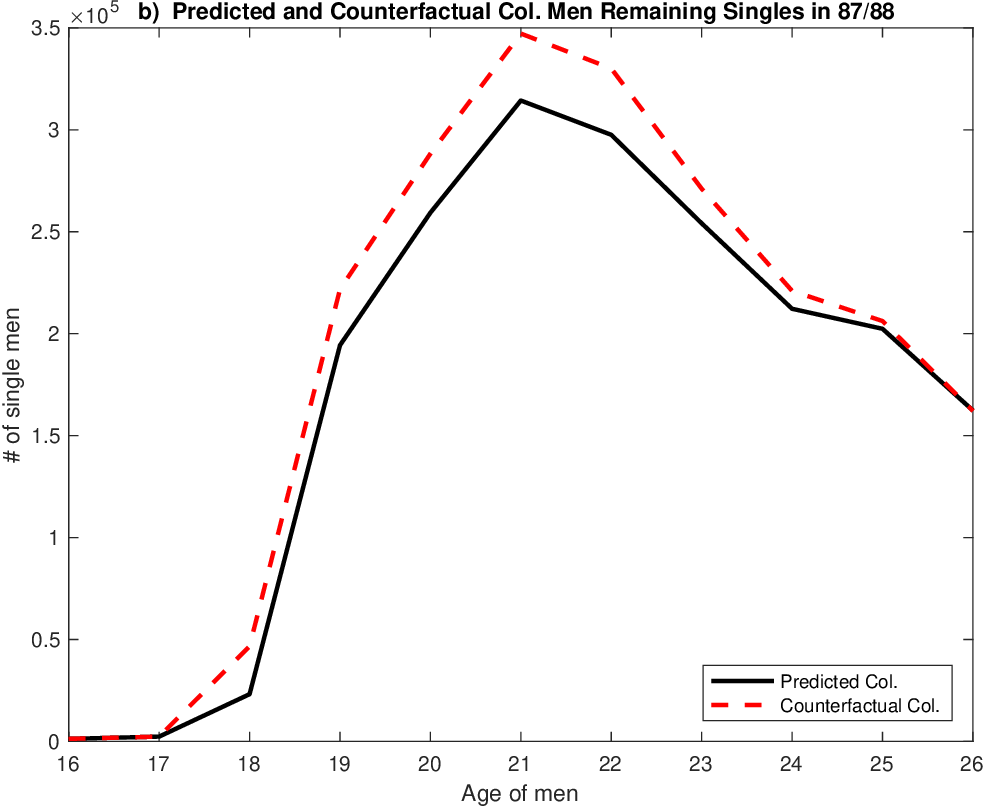}
%  \caption{A really Awesome Image}\label{fig:awesome_image2}
\endminipage\hfill
\minipage{0.32\textwidth}%
  \includegraphics[width=\linewidth]{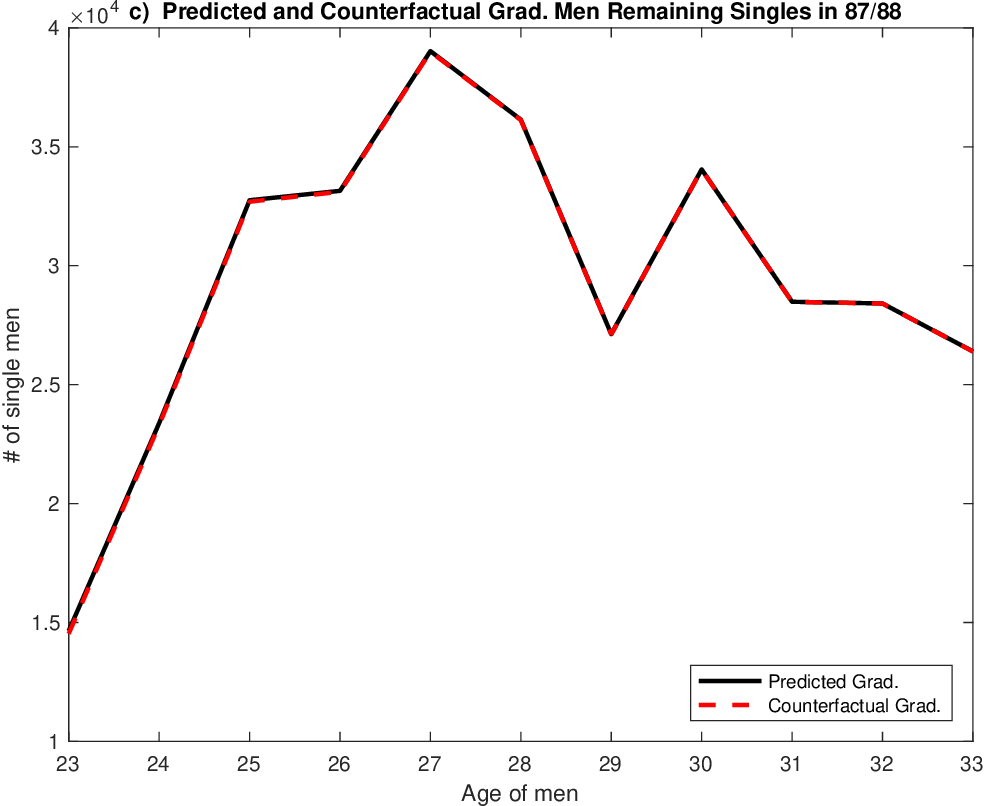}
%  \caption{A really Awesome Image}\label{fig:awesome_image3}
\endminipage\hfill
\minipage{0.32\textwidth}
  \includegraphics[width=\linewidth]{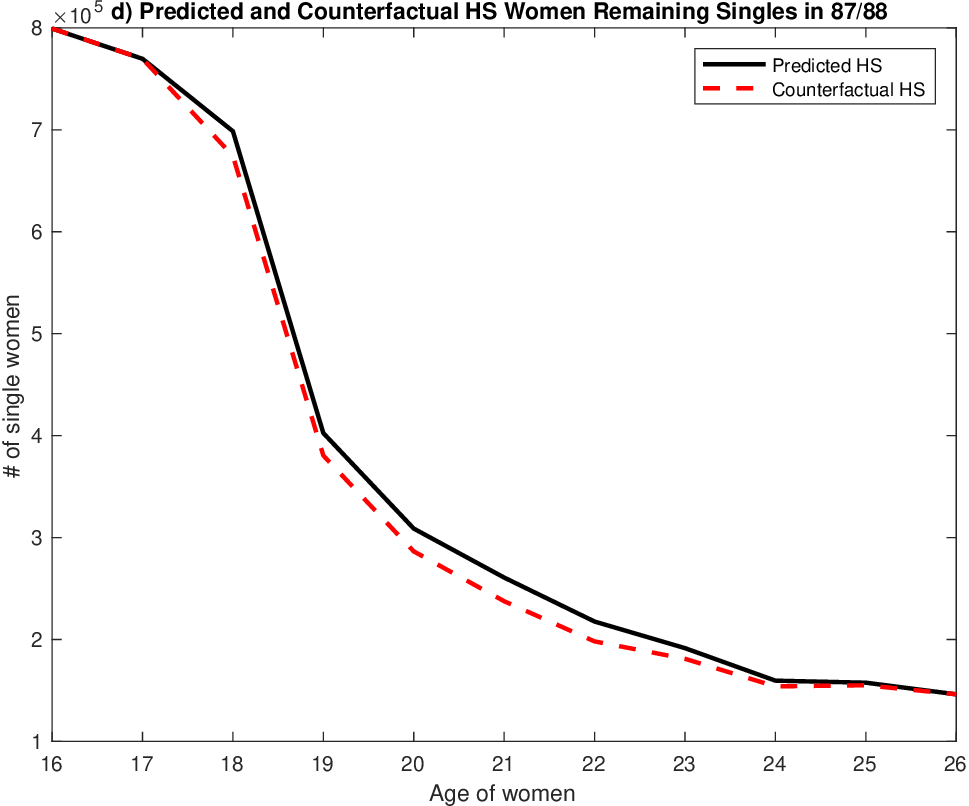}
%  \caption{A really Awesome Image}\label{fig:awesome_image1}
\endminipage\hfill
\minipage{0.32\textwidth}
  \includegraphics[width=\linewidth]{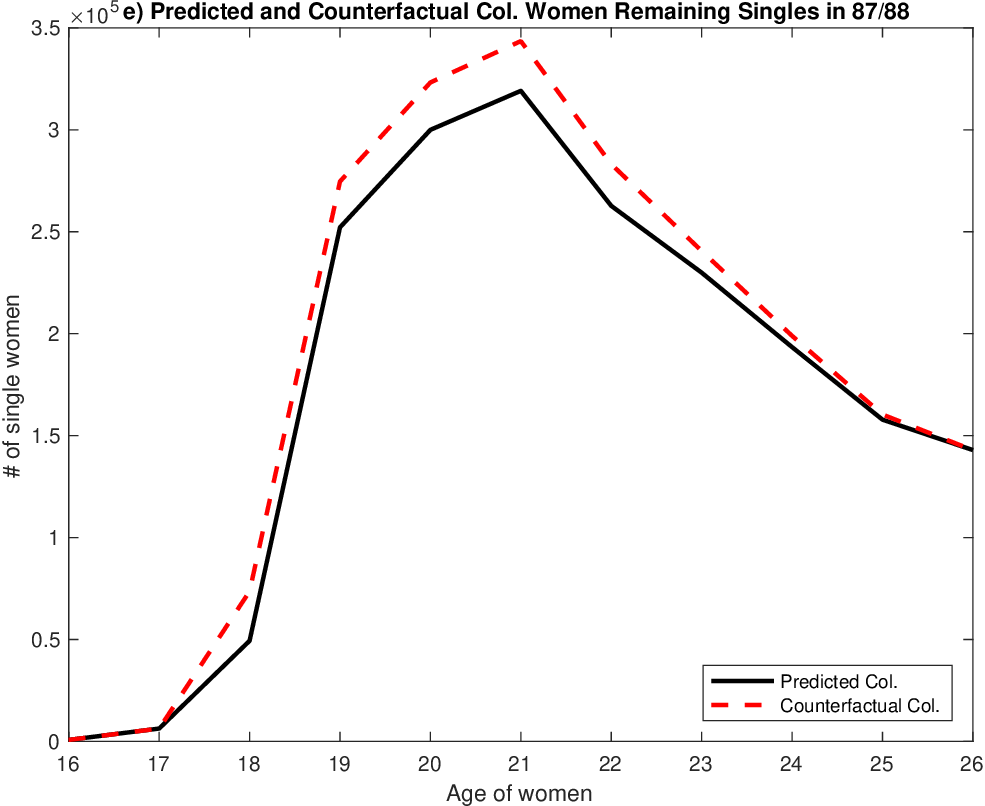}
%  \caption{A really Awesome Image}\label{fig:awesome_image2}
\endminipage\hfill
\minipage{0.32\textwidth}%
  \includegraphics[width=\linewidth]{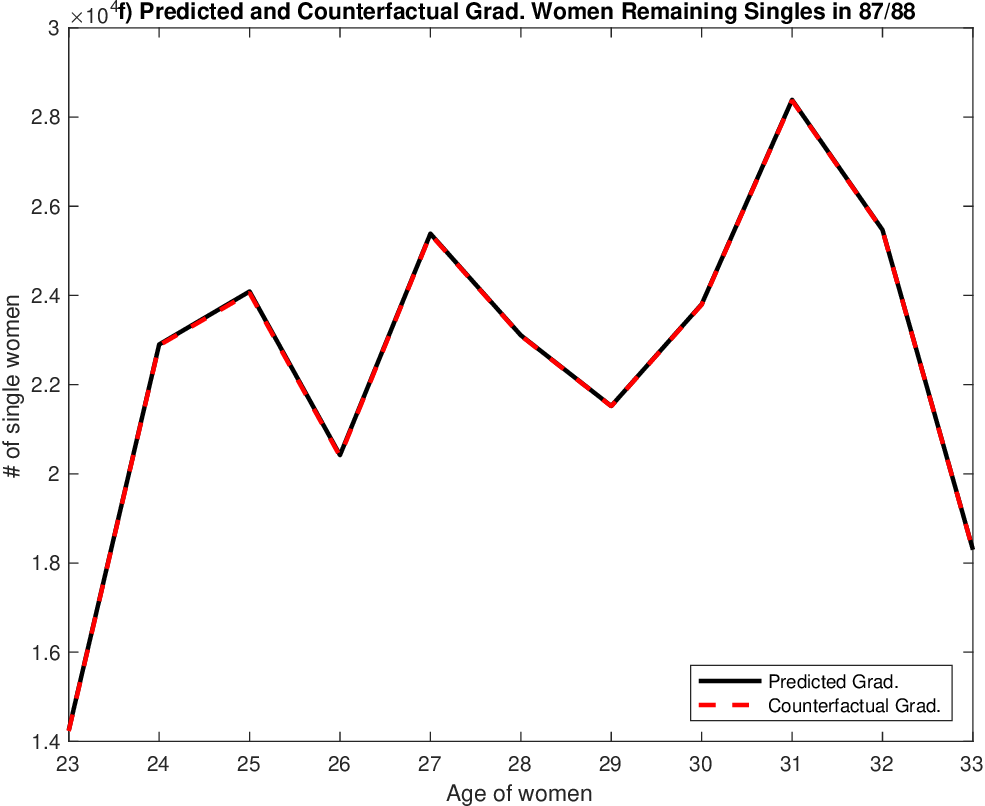}
%  \caption{A really Awesome Image}\label{fig:awesome_image3}
\endminipage
\centering
\caption{Predicted and counterfactual numbers of men and women remaining singles \\
\centering (NTU-Optimal Model)}
\label{fig: pred_count_singles_NTU}
\end{figure}

\end{document}